\documentclass[a4paper,fleqn]{cas-sc}

\usepackage[authoryear]{natbib}
\usepackage{soul}
\usepackage{caption} 
\usepackage{amsmath}  
\usepackage{bm}    
\usepackage{cases} 
\usepackage{threeparttable}  
\usepackage{siunitx}
\usepackage{bbding}
\usepackage{graphicx}
\usepackage{multirow}
\usepackage{booktabs}
\usepackage{caption}
\usepackage{subcaption}

\def\tsc#1{\csdef{#1}{\textsc{\lowercase{#1}}\xspace}}
\tsc{WGM}
\tsc{QE}

\begin{document}
\let\WriteBookmarks\relax
\def\floatpagepagefraction{1}
\def\textpagefraction{.001}

\shorttitle{MPM for simulating hypervelocity impact on asteroids}

\shortauthors{X. Yan et al.}

\title [mode = title]{The Material Point Method (MPM) for simulating hypervelocity impact on asteroids}



%

\author[1,2,3,4]{Xiaoran Yan}[orcid=0000-0001-6241-5744]
\cormark[1] 
\ead{yanxr159@gmail.com} 

\author[2,5]{Patrick Michel}[orcid=0000-0002-0884-1993]
\ead{michelp@oca.eu}


\author[1,6]{Ruichen Ni}[orcid=0000-0001-8218-3784]
\ead{nirc@synjoz.com}

\author[1,7]{Yifei Jiao}[orcid=0000-0003-1097-0521]
\ead{jiaoyf.thu@gmail.com}

\author[1]{Junfeng Li}
\cormark[2]
\ead{lijunf@tsinghua.edu.cn}

\address[1]{Tsinghua University, Beijing 100084, China}
\address[2]{Universit\'e C\^ote d'Azur, Observatoire de la C\^ote d'Azur, CNRS, Laboratoire Lagrange UMR7293, Nice 06304, France}
\address[3]{Shanghai Astronomical Observatory, Chinese Academy of Sciences, Shanghai 200030, China}
\address[4]{IFAC-CNR, Sesto Fiorentino (FI) 50019, Italy}
\address[5]{Department of Systems Innovation, School of Engineering, University of Tokyo, Tokyo 113-8656, Japan}
\address[6]{Hangzhou Synjoz Micro-Nano Technology Co., Ltd., Hangzhou 311227, China}
\address[7]{Department of Earth and Planetary Science, University of California, Santa Cruz, CA 95064, USA}

\cortext[1]{Corresponding author}
\cortext[2]{Co-corresponding author}



\begin{abstract}
Shock-physics numerical codes are essential tools for describing the short but extreme fragmentation stage of the hypervelocity impact process on asteroids.
However, accurately representing complex interior structures, surfaces, and contact mechanics in these events remains a significant challenge for traditional hydrocodes.
This study introduces and validates an innovative yet underutilized technique, i.e., the Material Point Method (MPM), to simulate hyper-velocity impacts on asteroids.
This approach offers new perspectives and solutions for capturing complex interfaces and handling the contact and boundary conditions in asteroid impact simulations.
Our MPM implementation incorporates critical improvements to material models, including a pressure-dependent $C^1$ continuous yield criterion with quantifiable plastic strain, and a resolution-independent Grady-Kipp fragmentation model, to capture the complex physics of geological materials under extreme conditions.
The framework is rigorously validated against laboratory impact experiments and benchmarked with smoothed particle hydrodynamics (SPH) simulations, confirming its robustness and precision.
Crucially, when applied to asteroid-scale collisions, our model successfully reproduces the formation of large, coherent fragments analogous to ($433$) Eros.
This work establishes MPM as a validated and powerful extension to the planetary scientist's toolkit, enabling the expansion of the parameter space and the treatment of complex contact and boundary conditions, which will enable more realistic simulations of asteroid evolution, family formation, and planetary defense scenarios.
\end{abstract}


\begin{highlights}
\item Developed a 3D Material Point Method (MPM) framework for asteroid impacts that explicitly tracks fragments and complex interfaces.
\item Implemented advanced material models, including a $C^1$-continuous yield criterion, quantifiable plastic strain and a resolution-independent damage model, to enhance physical realism.
\item Demonstrated that large, Eros-like remnants can survive catastrophic impacts, with survival critically dependent on the Weibull parameters governing the strength of the parent body.
\item Revealed the unique capability of MPM to model discontinuous structures, paving the way for future studies of rubble-pile asteroids and complex geologies.

\end{highlights}


\maketitle

\section{Introduction}\label{sec1}
From local-scale craters to global-scale catastrophic disruptions, asteroid impacts have played a crucial role in the formation and evolution of asteroids \citep{Michel2015asteroids, Zhang2021}.
As one of the significant mechanisms that reshape and restructure small celestial bodies, hypervelocity impacts sculpt the landscapes \citep{Richardson2004}, influence the internal structures \citep{Yu2017} and physical characteristics \citep{Jutzi2020, Guldemeister2022}, form specific asteroids and asteroid families \citep{Michel2015}, and may result in the orbital migration \citep{Bottke2023}.
Understanding impact processes helps trace the history of our solar system.

Space missions have intentionally created several impact experiments to gain insights into the composition of small celestial bodies and test asteroid impact defense capabilities.
The JAXA Hayabusa2 spacecraft produced an artificial impact crater on asteroid ($162173$) Ryugu, with a rim-to-rim diameter of $17.6 \pm \SI{0.7}{\metre}$, or $14.5 \pm \SI{0.8}{\metre}$ when measured from the original horizon \citep{Arakawa2020}.
The actual crater far exceeded its predicted size of $2$ to \SI{10}{\metre}, assuming that the surface was controlled by strength \citep{Arakawa2017}.
Similarly, the NASA DART mission, which aimed to use the kinetic impactor techniques to deflect the orbit of Dimorphos, the secondary component of the near-Earth binary asteroid system with the primary body ($65803$) Didymos \citep{Cheng2023}, also achieved a more substantial change in the orbital period of about $-33.0 \pm \SI{1.0}{\minute}$ than the anticipated \SI{7}{\minute} in a perfectly inelastic case \citep{Daly2023, Thomas2023}.
These surprising outcomes underscore the need for in-depth research into the physical properties of these bodies, and in-situ investigations, such as those of the ESA Hera mission to investigate Didymos' properties and DART impact outcome \citep{Michel2022}.
This highlights the importance of further study focusing on the hypervelocity impact process, to better guide mission implementation and extract valuable data from mission findings.

Due to scale limitations, experimental extrapolation alone is insufficient to cover all hypervelocity impact scenarios, while theoretical analysis lacks detail.
Thanks to the increasing performance of computers, shock-physics numerical codes have become commonly used in these large scale high-speed impact phenomena.
Numerical simulations refine theoretical investigations and effectively extend experimental results, providing a more comprehensive understanding of such complex impact events \citep{Jutzi2015modeling}.
The current state-of-the-art shock-physics codes include grid-based CTH \citep{McGlaun1990} and iSALE \citep{Wunnemann2006, Elbeshausen2009}, and mesh-free smoothed particle hydrodynamics (SPH) codes, such as Bern SPH \citep{Benz1994, Benz1995, Jutzi2008, Jutzi2015}, Spheral \citep{Owen1998} and Miluphcuda \citep{Schafer2016}.
Those hydrocodes are stable and extendable, with constitutive equations representing the dynamical response of materials \citep{Jutzi2015modeling}.
Each method also has its unique advantages. 
They have been collaboratively benchmarked and validated \citep{Pierazzo2008, Stickle2020}, and are widely used in the planetary science community \citep{Stickle2022, Luther2022}, bridging the gap between experimental results and the actual conditions of interest.

Recent research studies increasingly focus on accurately depicting the surface topography of asteroids and simulating their long-term evolution after impact.
For example, modeling Dimorphos as a weak rubble pile allowed impact simulations using the Bern SPH code to reproduce the data from the DART impact, suggesting a possibility that the impact might have globally reshaped the asteroid instead of merely leaving a localized crater \citep{Raducan2024}.
However, whether such extreme reshaping definitively occurred remains a subject of ongoing scientific debate \citep{Rivkin2026}.
Resolving this crucial question and validating these computational models are among the primary expected outcomes of the upcoming ESA Hera mission \citep{Michel2022}.

To accurately capture these complex post-impact phenomena, such as the global reshaping of the parent body or the fallback of ejecta in a microgravity field \citep{Jiao2023}, seamlessly bridging the short-term shock physics with long-term gravitational interactions has become a critical frontier.
Consequently, coupling continuum hydrocodes with discrete methods is increasingly necessary.
For instance, SPH results are frequently mapped to N-body or DEM codes for reaccumulation and ejecta evolution analysis \citep{Michel2001, Jiao2023}, and the Combined Finite-Discrete Element Method (FDEM) has been applied to naturally transition from continuum fracturing to discrete fragment interactions \citep{Mnjiuza1995}.
Despite these advancements, the complex geology of small asteroids, as revealed by recent space missions, poses new challenges for the current leading shock-physics codes regarding computational efficiency, contact simulation between objects, and the precise shape extraction of fragments post-breakup for subsequent discrete simulations.
SPH methods, while powerful, can face challenges such as tensile instability \citep{Monaghan2000}, require significant computational resources for finding the neighbor particles, and despite recent progress, can still struggle with applying specific boundary conditions (such as symmetric or transmitting boundaries used to reduce computational domain and costs) \citep{ElMir2019}.
Grid-based methods, on the other hand, contend with common grid-related disadvantages, such as grid distortion in the Lagrangian approach, and difficulties in solving the convection term and tracking boundaries and fragmentation in the Eulerian approach \citep{Ma2009}.
These challenges necessitate advancements in methodologies to meet the evolving needs of asteroid impact simulations and celestial body analysis.

Another method in the field of shock dynamics -- the material point method, or MPM for short -- may offer new approaches and solutions to the above challenges as a complement to the existing methods.

MPM is an extension of the FLIP particle-in-cell method, combining the advantage of both Lagrangian and Eulerian descriptions \citep{Sulsky1994}.
It has been widely applied in the mechanical industry, including investigations into the debris clouds produced by impacted thin lead plates \citep{huang2008}, the spalling of Armco steel during impact \citep{Chen2012}, the perforation of metal targets \citep{Lian2011}, etc.
However, its potential in the field of planetary science remains under-explored.
\citet{Tonge2016} designed the Tonge–Ramesh model for brittle materials under the Uintah MPM framework \citep{Germain2000}.
Based on this material model, \citet{Tonge2016a} further investigated the impact events forming the largest craters on asteroid ($433$) Eros.
Then \citet{ElMir2019} benchmarked the model by a set of dynamic Brazilian disk experiments on basalt samples, and built an end-to-end MPM-DEM approach that simulates both the fragmentation and gravitational reaccumulation phase of hypervelocity impacts on asteroids. 
These studies showcase the capability of MPM to handle complex impact events.
However, no significant progress has been made in applying MPM to study impacts on small celestial bodies after that.
More importantly, the MPM framework has yet to be validated against experimental data and benchmarked with classical scenarios simulated by other hydrocodes, which is essential and crucial.
Benchmarking is necessary to distinguish the differences brought by numerical methods and material models, as well as quantify them separately \citep{Pierazzo2008}.
Additionally, it helps understand how to compare simulation results from different methods \citep{Stickle2022}.
Only by accomplishing the validation will the strengths of each approach be leveraged, thereby enhancing the credibility and reliability of simulation results.

This paper, therefore, introduces and validates a 3D MPM framework for simulating hypervelocity impacts on small celestial bodies, culminating in key scientific insights such as the formation of large, Eros-like fragments.
The paper is structured as follows.

Section~\ref{sec2} provides the theoretical foundations of the Material Point Method.
Building on this, Section~\ref{sec3} details the implementation of advanced material models essential for capturing impact physics.
It presents three key enhancements: an improved strength model with a smoothed yield surface, a modified Tillotson equation of state, and a comprehensive damage formulation to track material failure. 
Section~\ref{sec4} is dedicated to the validation and application of MPM.
First, it is rigorously benchmarked against laboratory impact experiments and established SPH simulations.
Second, the framework is applied to asteroid-scale collisions, demonstrating its capability to produce large, coherent remnants analogous to ($433$) Eros.
Section~\ref{sec5} discusses the broader implications of these results, highlighting the unique advantages of MPM and outlining its future potential as a powerful tool in planetary science.
Finally, Section~\ref{sec6} summarizes the principal conclusions of this work.
Supplementary details on the numerical algorithms and plasticity theory are provided in Appendices~\ref{secA} and \ref{secB} to ensure reproducibility and completeness.

\section{Material Point Method}\label{sec2}
\begin{figure}[pos=htbp]
\centering
  \includegraphics[width = \textwidth]{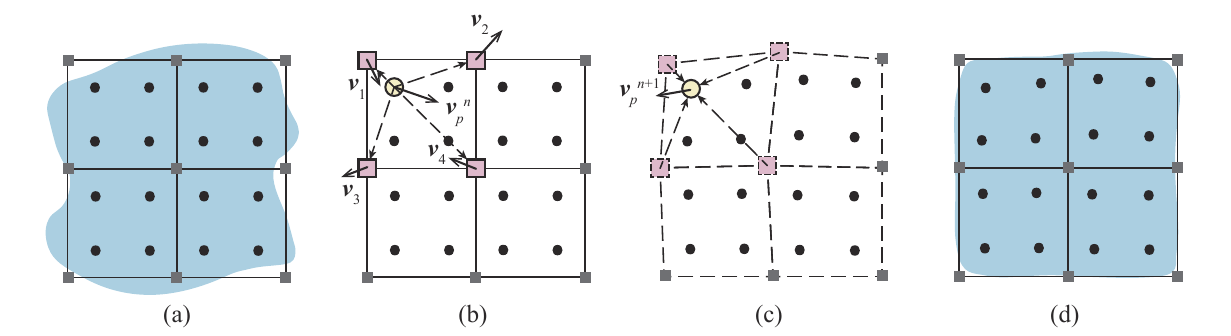}
  \caption{Illustration of MPM framework.
  \textit{a}. Initially discretizing the material domain into a group of Lagrangian material points and predefining a regular background grid.
  \textit{b}. In each timestep, mapping the mass and momentum of the material points to the background grid nodes, to solve the momentum equations as well as the boundary conditions.
  \textit{c}. The material points moving along the deformation of the grid, allowing interpolation of the kinetic solution to update the positions and velocities of material points (Lagrangian step). Also updating the density and stress of each material point (at different stages during a time step, based on the stress updating format).
  \textit{d}. Discarding the deformed grid, resetting a new one as predefined (Eulerian step), and going to step (b) to calculate the next timestep.}\label{fig:MPMillu}
\end{figure}

The fundamental methodology of material point method (MPM) is to discretize the continuum body into a group of Lagrangian material points (also called particles), and utilize a rigidly attached Eulerian background grid in each time step which the material points move with \citep{Sulsky1994, zhang2016}, as shown in Fig.~\ref{fig:MPMillu}.

\begin{enumerate}[(a)]

\item The material points carry all the physical properties during the entire loading process, including mass, density, position, velocity, stress and strain, internal energy, as well as other state variables. 
This is a typical Lagrangian description, making it simple to trace the material interfaces, and it is able to implement history-dependent constitutive models.

\item In each timestep, a predefined background grid is always regenerated at the beginning.
Then, the kinematic quantities of the material points are mapped to the background grid nodes, constructing a Lagrangian finite element discretization of the material domain.

\item By using the standard finite element formulation with the grid, the momentum equations can be solved numerically.
Boundary conditions can also naturally be imposed on the grid nodes.
While the positions and velocities of the grid nodes are updated for one timestep, the material points are bound with the deforming grid, allowing the update of the positions and velocities by interpolating from the grid nodes based on the initial configuration.
And other states are updated based on the material point itself. 
This is the Lagrangian phase, where the usual convection term associated with Eulerian formulations does not appear.

\item At the end of a timestep, the deformed background grid is discarded, and the simulation proceeds to the next time step with a renewed regular grid. 
The reset of the background grid is regarded as the Eulerian phase that avoids the mesh distortion and element entanglement.

\end{enumerate}

The core renovations of the MPM lie in the material point discretization
\begin{equation}
    \rho (\bm{x}) = \sum_{p=1}^{n_{\mathrm{p}}} m_p \delta (\bm{x} - \bm{x}_p)  \, \text{,}
    \label{eq:rho}
\end{equation}
and the interpolation bridging the mesh nodal variables and the variables on the material point (taking the displacement $\bm{u}$ as an example)
\begin{equation}
    \bm{u}_p = \sum_{I=1}^{n_{\mathrm{g}}} N_I(\bm{x}_p) \bm{u}_I  \,  \text{,}
    \label{eq:u}
\end{equation}
which adopts the same form as the standard finite element interpolation.
In Eqs.~(\ref{eq:rho}) and (\ref{eq:u}), the subscripts $p$ and $I$ are used as indices for the material points and the grid nodes, respectively.
For a specific material point $p$, $n_{\mathrm{p}}$ denotes the total number of material points in the entire computational domain, while $n_{\mathrm{g}}$ represents the number of grid nodes whose shape functions $N_I$ have non-zero values at $\bm{x}_p$ (i.e., the nodes that contribute to the interpolation at point $p$).
Moreover, $\bm{x}_p$ refers to the coordinates of point $p$, and $\delta(\bm{x})$ is the Dirac delta function.

Unlike traditional Eulerian or Arbitrary Lagrangian-Eulerian (ALE) methods, where state variables are stored on the mesh and require advective remapping that can severely smear sharp interfaces and history-dependent variables, MPM fundamentally differs by permanently storing all mass, momentum, and internal state variables (such as stress, damage, and plastic strain) on the Lagrangian material points \citep{Sulsky1994, zhang2016}.
The background grid serves merely as a temporary computational scratchpad to solve the momentum equations and calculate spatial gradients.
Consequently, the Eulerian phase in MPM simply involves discarding the distorted grid and resetting a new one, entirely bypassing the advection equations.
While the interpolation between particles and the grid can introduce some numerical dissipation, the history variables themselves are strictly conserved on the material points without advective smearing.
To facilitate this particle-grid mapping in standard three-dimensional implementations, it is common practice to initially populate $2 \times 2 \times 2 = 8$ uniformly spaced material points within each active background grid cell.
This configuration establishes a robust numerical integration domain for the shape functions.

By virtue of this particle-grid mapping, MPM effectively merges the Lagrangian and Eulerian descriptions, which minimizes numerical dissipation and speeds up the neighbor searching, making it an efficient, extendable, and robust code for modeling extreme events involving large deformations and complex material history.
However, the method is not without inherent numerical challenges.
Specifically, as particles traverse grid boundaries, they can generate cell-crossing noise, resulting in localized stress oscillations.
Our framework mitigates these effects using the Generalized Interpolation Material Point (GIMP) method \citep{Bardenhagen2004}. The interplay between this numerical noise and discretization precision is evaluated in Section~\ref{sec4_1_1}, while broader methodological limitations and future improvements are discussed in Section~\ref{sec5_2}.

This section provides a brief introduction to the basic concepts of MPM.
A detailed description of the governing equations, discretization, and the explicit solution scheme is provided in Appendix~\ref{secA}.
For a comprehensive treatment of the method, readers are referred to \citet{zhang2016}.

In this work, we adapt and enhance the open-source MPM3D-F90 code\footnote{https://github.com/xzhang66/MPM3D-F90}.
Originally designed for engineering applications, we have tailored it for the specific challenges of planetary science.
Our key contributions to the framework include:
(1) implementing an adaptive time-stepping scheme for the explicit leapfrog integrator to ensure stability in high-velocity scenarios (Appendix~\ref{secA_2});
(2) developing and incorporating advanced constitutive models tailored for the brittle failure of geological materials (Section~\ref{sec3}); and
(3) performing extensive code optimization to improve computational performance and accuracy.
These enhancements transform MPM3D-F90 into a powerful tool for simulating hypervelocity impacts on small celestial bodies.

\section{Material model}\label{sec3}
The material model, also known as the constitutive equation, describes the inherent dynamic properties of various materials and how they respond to specific mechanical loads.
This supplements the governing equations with stress-strain relationships.
For the study of hypervelocity impacts on asteroids, it is essential to have a comprehensive description that covers the mechanical behaviors of carbonaceous, siliceous, and metallic materials under extreme loading, such as significant plastic deformation, compaction, fracture, etc \citep{Flynn2018, Binzel2019, Demeo2014}.

To model these diverse responses, the stress tensor $\bm{\sigma}$ is decomposed into its deviatoric (shear) component $\bm{s}$ and its isotropic (hydrostatic pressure) component $p$
\begin{equation}
    \bm{\sigma} = \bm{s} - p\bm{I}  \,  \text{.}
\end{equation}
This decomposition allows us to treat the material's resistance to shear and compression separately. The deviatoric stress $\bm{s}$ is governed by a strength model, which defines the yield limit and the post-yield behavior of a elastoplastic material. The pressure $p$ is determined by an equation of state (EOS), which relates pressure to density and internal energy. Furthermore, a damage model is used to describe material weakening and failure under tensile loading.

When the elastic trial stress exceeds the material's yield limit, permanent (plastic) deformation occurs.
To capture this, our framework implements a plastic correction procedure that returns the stress state to a physically admissible yield surface while accumulating plastic strain.
This procedure is based on the principles of plastic flow theory, which are detailed in Appendix~\ref{secB_1}.
Building upon this established theory, Section~\ref{sec3_1} introduces our novel, modified Lundborg strength model, specifically designed for the pressure-dependent behavior of brittle rocks.

Section~\ref{sec3_2} presents the Tillotson EOS, a widely-used model for impact simulations.
Our implementation includes its extension to the cold expansion stage and a formulation for the sound speed correction.

Additionally, to capture fracture, Section~\ref{sec3_3} presents our improved damage model, which features a resolution-independent method for initializing Weibull-distributed flaws, crucial for obtaining consistent results across different simulation scales.
A brittle fracture failure mode is also introduced.

\subsection{A modified Lundborg strength model}\label{sec3_1}

Classical strength models for geological materials, such as the Drucker-Prager \citep{Drucker1952} and the original Lundborg models \citep{Lundborg1968} (detailed in Appendix~\ref{secB_3}), successfully capture pressure-dependent strengthening.
While plastic flow theory is an essential component of modern shock-physics codes, the numerical implementation for pressure-dependent geological materials often relies on simplified radial return algorithms \citep{Jutzi2015modeling}.
In such simplified approaches, if the trial effective shear stress $\tau^*$ exceeds the corresponding yield strength $\sigma_{\mathrm{Y}}$, the deviatoric stress components are simply reduced to the yield envelope by multiplying a factor $\sigma_{\mathrm{Y}} / \tau^*$.
This purely deviatoric scaling fails to strictly follow the yield surface normal, thereby decoupling the volumetric and deviatoric plastic strain increments.
Additionally, classical multi-surface models introduce non-smooth corners on the yield surface, complicating the analytical determination of the true plastic flow direction.

\begin{figure}[pos=htbp]
\centering
  \includegraphics[width = \textwidth]{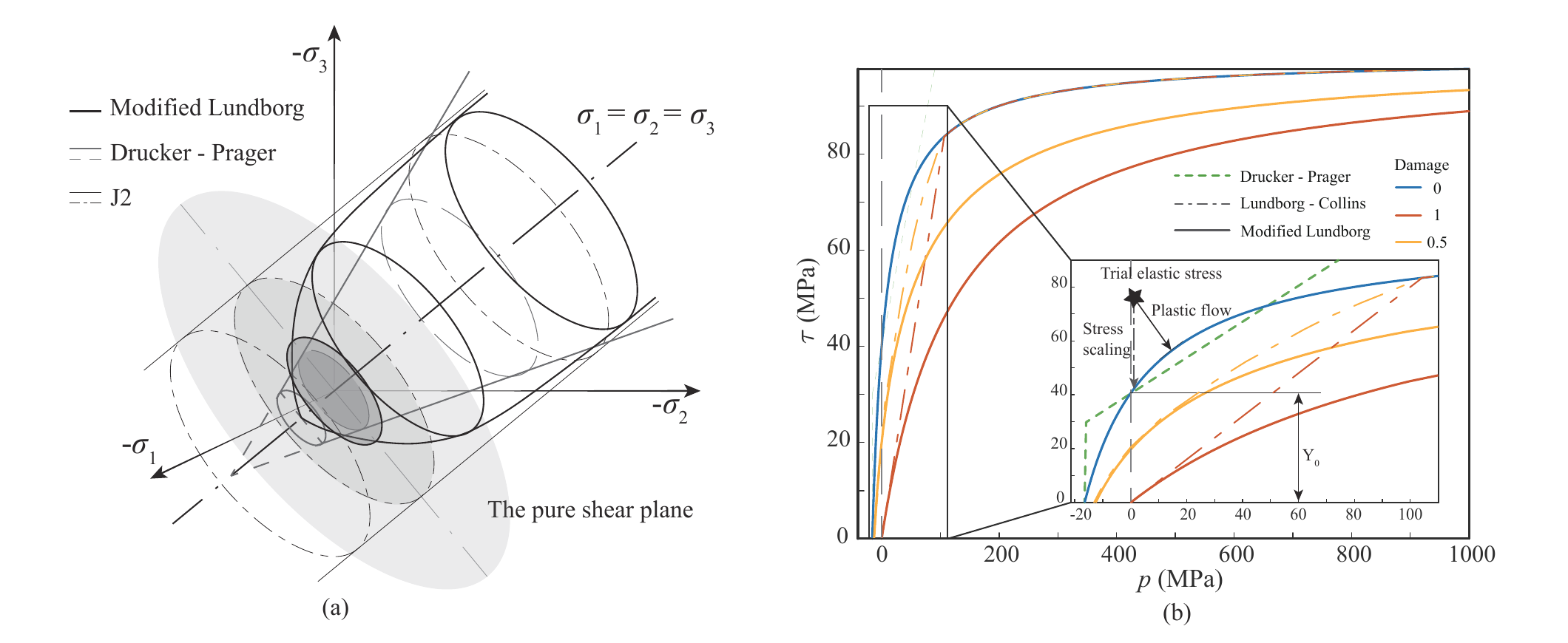}
  \caption{Illustration of the yield surfaces for various strength models: 
  \textit{a}. in the principal stress space, and
  \textit{b}. in the $p$--$\tau$ plane.
  The enlarged view in \textit{b} also compares two algorithms that return the elastic stress trial solution to the yield surface: the plasticity correction based on the flow rules, and the deviatoric stress scaling by the factor $\sigma_{\mathrm{Y}} / \tau^*$.}
  \label{fig:yield}
\end{figure}

To address these limitations, we propose a modified Lundborg yield condition with a single, smooth ($C^1$ continuous) yield surface:
\begin{equation}
    f (\bm{\sigma}) = \tau - 
    \left[(1 - D) \frac {\mu_{\mathrm{i}}p} {1 + \mu_{\mathrm{i}}p/(Y_{\mathrm{m}}-Y_{0})} + 
    (1 - D) Y_{0} + 
    D \frac {\mu_{\mathrm{d}} p} {1 + \mu_{\mathrm{d}} p / Y_{\mathrm{m}}}\right]
    \left(1 - \frac {Q} {Q_{\mathrm{melt}}}\right)  \, \text{,}
\end{equation}
where $D$ is the damage variable (0 for intact, 1 for failed), $\mu_{\mathrm{i}}$ and $\mu_{\mathrm{d}}$ are the internal friction coefficients for intact and damaged material, $Y_0$ is the cohesion (zero-pressure shear strength), and $Y_{\mathrm{m}}$ is the ultimate shear strength limit.
As the internal energy due to elastic deformation does not contribute to thermal effects, here the melting coefficient is quantified solely based on heat energy.
$Q = Q_0 + \Delta Q$, where $Q_0$ represents the initial heat energy associated with the initial temperature, and $\Delta Q$ is derived from Eq.~(\ref{eq:heatenergy}).
$Q_{\mathrm{melt}}$ is defined as the heat energy at the point of material melting.
For damaged materials without softening, the expression of shear strength becomes an asymptotic curve, also with $Y_{\mathrm{m}}$ as its limit.

Based on the plastic flow theory explained in Appendix~\ref{secB_1}, with this smoothed expression and the plastic potential function Eq.~(\ref{eq:potential}), the calculation of the plastic corrector can be derived as follows
\begin{subequations}
  \begin{numcases} {}
    \Delta\lambda = \frac {f(\bm{\sigma}^{*})} {G + Kq_{\psi}H} \label{eq:dlambda} \\
    p = p^{*} + K q_{\psi} \Delta\lambda \label{eq:pcor} \\
    \tau = \tau^* - G \Delta\lambda \\
    \bm{s} = \bm{s}^* - G \Delta\lambda \frac{\bm{s}}{\tau} = \frac{\tau}{\tau^*} \bm{s}^*
  \end{numcases}
\end{subequations}
with the abbreviated form
\begin{equation*}
    H = \left[\frac {(1 - D)\mu_{\mathrm{i}}} {(1 + \frac {\mu_{\mathrm{i}} p} {Y_{\mathrm{m}} - Y_{0}})^{2}} + 
    \frac {D \mu_{\mathrm{d}}} {(1 + \frac {\mu_{\mathrm{d}} p} {Y_{\mathrm{m}}})^{2}}\right] \left(1 - \frac {Q} {Q_{\mathrm{melt}}}\right)  \, \text{.}
\end{equation*}
$K$ represents the bulk modulus that could be obtained from $K = E_{\mathrm{Y}} / (3 (1 - 2\nu))$.
By substituting Eq.~(\ref{eq:dlambda}) into Eq.~(\ref{eq:pcor}) and iterating, the pressure $p$ is obtained.
Subsequently, substituting this result back into Eq.~(\ref{eq:dlambda}) yields the value of $\Delta\lambda$, and then $\tau$ and $\bm{s}$.
The enlarged view of Fig.~\ref{fig:yield}(b) shows the difference between the plasticity correction based on the flow rules and the deviatoric stress scaling.
Moreover, the increment of plastic strain can be rewritten as
\begin{subequations}
  \begin{align}
    & \Delta \bm{\varepsilon}^{\prime \mathrm{p}} = \Delta\lambda \frac {\bm{s}} {2 \tau} \\
    & \Delta \varepsilon^{\mathrm{p}}_{\mathrm{k}} = \Delta\lambda q_{\psi} / 3 \\
    & \Delta \varepsilon^{\mathrm{p}} = \Delta\lambda \sqrt{\frac{1}{3} + \frac{2}{9} q_{\psi}^2}
  \end{align}
\end{subequations}
where $\bm{\varepsilon}^{\prime \mathrm{p}}$ means the deviatoric part of plastic strain, and $\varepsilon^{\mathrm{p}}_{\mathrm{k}}$ means the spheric part.

With the methodology outlined above, the proposed enhancements refine the Lundborg strength model by smoothing the yield surface and providing an analytical plastic correction factor.
These improvements retain the characteristic of pressure-dependence with an upper limit.
Such modifications bolster the model's precision and reliability in simulating hypervelocity impacts on asteroids.
The specific outcomes and benefits of these enhancements will be elaborated in Sections~\ref{sec4_1_5} and~\ref{sec4_2}, demonstrating the enhanced predictive capabilities and their implications for understanding asteroid impact dynamics.

\subsection{The equation of state}\label{sec3_2}
The pressure component $p$ is determined by a function of density $\rho$ and specific internal energy $E$
\begin{equation}
    p = p (\rho, \, E)  \,  \text{.}
\end{equation}
This relationship is known as the equation of state (EOS).

\begin{figure}[pos=htbp]
\centering
  \includegraphics[width = 0.9\textwidth]{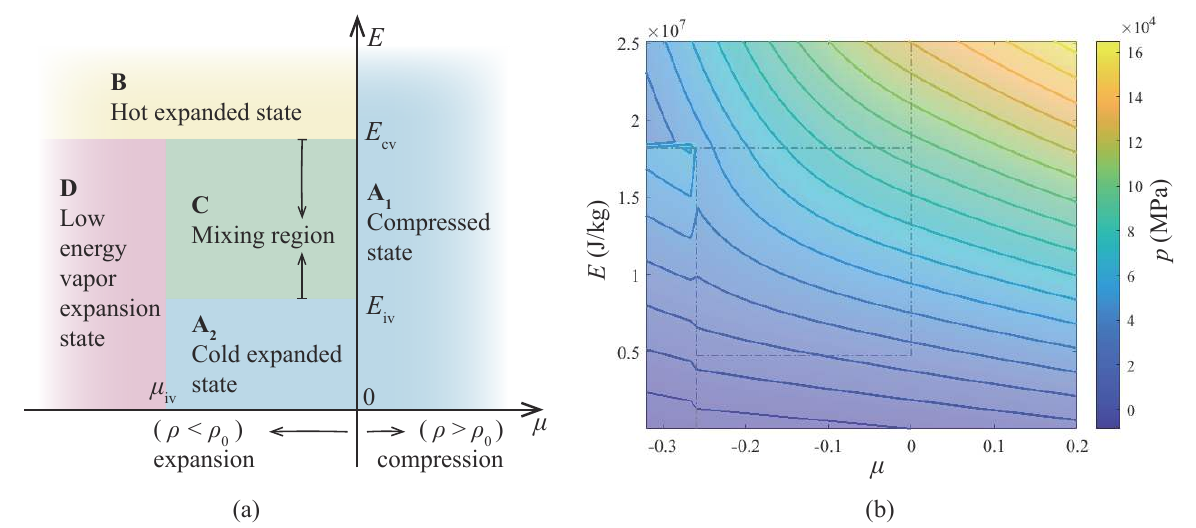}
  \caption{Representation of the Tillotson EOS in the $\mu$--$E$ space: 
  \textit{a}. the regions of the states, and
  \textit{b}. gradation and contours of pressure $p$, with the dashed lines correlating with the divisions of regions from (\textit{a}).
  The horizontal axis is labeled as $\mu = \rho / \rho _0 - 1$.
  The value $E_{\mathrm{iv}}$ refers to the specific energy when the material is incipiently vaporized, while $E_{\mathrm{cv}}$ corresponds to the state of complete vaporization.
  The quantity $\mu_{\mathrm{iv}} = \rho_{\mathrm{iv}} / \rho _0 - 1$ is defined by $\rho_{\mathrm{iv}}$, the density at which incipient vaporization occurs.}
  \label{fig:Tillotson}
\end{figure}

The Tillotson EOS is the main equation of state used in this paper. 
The classical Tillotson equations are a simplified amalgamation that combines the Hugoniot curve with the Thomas-Fermi model, representing the solid(or liquid) or gaseous phase of the material \citep{Tillotson1962}. 
Subsequent research implements a mixing region between these two phases, as well as the low energy vapor expansion states, whereupon the Tillotson EOS could provide full coverage of the density-energy space in a concise form for the hypervelocity impact simulation \citep{Brundage2013} (see Fig.~\ref{fig:Tillotson}).

Let us predefine $\eta  = \rho / \rho _0$, $\mu  = \eta - 1$ and $\nu  = 1/\eta - 1$, where $\rho _0$ denotes the initial density.
For compressed and cold expanded states, i.e. $\rho \geqslant \rho_0$, or $\rho_{\mathrm{iv}} \leqslant \rho < \rho_0 \,\&\, E \leqslant E_{\mathrm{iv}}$, the equation takes the form:
\begin{equation}
    p = P_1(\rho,E)
    =\left[ a + \frac{b} {E/(E_0\eta ^2) + 1} \right] \rho E + A \mu + B \mu^2  \, \text{.}
\end{equation}
For hot expanded states where $\rho < \rho_0 \,\&\, E \geqslant E_{\mathrm{cv}}$,
\begin{equation}
    p = P_2(\rho,E)
    =a \rho E + \left[ \frac{b\rho E} {E/(E_0\eta ^2) + 1} + A \mu e^{-\beta \nu } \right]{e^{ - \alpha \nu^2}}  \, \text{.}
\end{equation}
And the mixing region in between $\rho_{\mathrm{iv}} \leqslant \rho < \rho_0 \,\&\, E_{\mathrm{iv}} < E < E_{\mathrm{iv}}$ could be defined as the interpolation of energy
\begin{equation}
    p = P_3(\rho,E)
    =\frac {P_2 \left( E - E_{\mathrm{iv}} \right) + P_1 \left( E_{\mathrm{cv}} - E \right)} {E_{\mathrm{cv}} - E_{\mathrm{iv}}}  \, \text{.}
\end{equation}
The formula of low energy vapor expansion states, with $\rho < \rho_{\mathrm{iv}} \,\&\, E < E_{\mathrm{cv}}$, is given by
\begin{equation}
    p = P_4(\rho,E)
    =\left[ a + \frac{b} {E/(E_0\eta ^2) + 1} \right] \rho E + A \mu  \, \text{.}
\end{equation}
Here, 'incipient vaporization' is denoted by the subscript 'iv', and 'complete vaporization' by 'cv'.
In the equation $A = \rho_0 C^2$, $C$ represents the speed of sound at low pressure.
The sum $a + b$ corresponds to the Gruneisen coefficient at zero pressure, with $a$ consistently set to $0.5$.
The Tillotson parameters $E_0$ and $B$ are defined to 'provide the best over-all $P$--$V$--$E$ surface', as described by \citet{Tillotson1962}.

The speed of sound, a critical parameter derived from the EOS, is also dependent on the current state of $p$, $\rho$, and $E$.
For a one-dimensional solid or a fluid, the bulk sound speed $c_{\mathrm{K}}$ is given by
\begin{equation}
    c_{\mathrm{K}}^2 =\left. \frac{\partial p} {\partial \rho} \right|_S
    = \left. \frac{\partial p} {\partial \rho} \right|_E + \frac{p} {\rho^2} \left. \frac{\partial p} {\partial E} \right|_{\rho}   \, \text{,}
\end{equation}
where the derivation and specific expressions for each state are omitted for brevity.
For three-dimensional solids, compressional (longitudinal) waves propagate at a higher speed $c_{\mathrm{p}}$ than shear (transverse) waves
\begin{equation}
    c_{\mathrm{p}} = \sqrt{\frac{4G}{3\rho} + c_{\mathrm{K}}^2}  \, \text{,}
    \label{eq:cp}
\end{equation}
neglecting changes in the shear modulus $G$.
Therefore, the longitudinal wave speed $c_{\mathrm{p}}$ is used as the reference sound speed in our calculations.
This updated sound speed is essential, as it plays a pivotal role in the plastic flow correction (Eq.(\ref{eq:dlambda})), the determination of the critical time step (Eq.(\ref{eq:dt})), the evaluation of damage accumulation (Eq.~(\ref{eq:damage})), and the calculation of artificial viscosity used to capture shock fronts \citep{Neumman1950}.

Other EOSs are also included in this MPM code, including elastic relation, linear polynomial EOS, Mie--Gr\"uneisen EOS \citep{zhang2016}, simplified Tillotson EOS \citep{Jutzi2008}, and could be further extended.

\subsection{The damage model}\label{sec3_3}
Extreme loading causes damage, leading to a reduction in strength or even fracture of the material
\begin{equation}
    \bm{\sigma} = (1 - D) \bm{s}_{\mathrm{E}} - p^* \bm{I}  \, \text{,}
\end{equation}
where $\bm{s}_{\mathrm{E}}$ is the elastic stress without considering damage, $p^* = (1 - D)p$ for tensile cases and $p^* = p$ for pressure \citep{Benz1994}. For brittle solids, \citet{Grady1980} introduced a fracture model beginning with an isotropic Weibull distribution of incipient flaws whose number activated at strain $\varepsilon$ obeys:
\begin{equation}
    \frac{n(\varepsilon)} {V_{\mathrm{s}}} = k \varepsilon^{m}
\end{equation}
\citep{Weibull1939, Jaeger1979}, where $k$ and $m$ are Weibull parameters, and $V_{\mathrm{s}}$ is the volume of the solid phase of the target ($V_{\mathrm{s}} = V$ for non-porous material).
Hence, the average activation strain of the $n_{\mathrm{tot}}$ flaws inside the target is derived to be
\begin{equation}
    \bar{\varepsilon}_{\mathrm{act}} = \frac{m}{m+1} \left( \frac{n_{\mathrm{tot}}}{kV} \right) ^\frac{1}{m}  \, \text{.}
\end{equation}

To implement the Weibull distribution in a numerical code, the total number of flaws is always set to $n_{\mathrm{tot}} = n_{\mathrm{p}} \ln(n_{\mathrm{p}})$ which could ensure a uniformly random distribution as well as controlled computational complexity \citep{Benz1995}, and flaws are sequentially assigned randomly to each computational subregion (particle) \citep{Benz1994}.
Then the explicit smallest activation strain $\varepsilon_{\mathrm{min},p}$, the largest activation strain $\varepsilon_{\mathrm{max},p}$ and the total number of flaws $n_{\mathrm{f},p}$ in each particle must be recorded.

Presuming the flaws in each particle still follow a Weibull distribution, the number of flaws activated by strain $\varepsilon_{\mathrm{eff},p}$ can be estimated as
\begin{equation}
    n_{\mathrm{act},p} = \left( \frac {\varepsilon_{\mathrm{eff},p}} {\varepsilon_{\mathrm{min},p}} \right)^{m^p}  \, \text{,}
    \label{eq:nact}
\end{equation}
with the Weibull parameter $m^p$ equals to $\ln(n_{\mathrm{f},p}) / \ln(\varepsilon_{\mathrm{max},p}/\varepsilon_{\mathrm{min},p})$, where the effective local tensile strain equals to
\begin{equation}
    \varepsilon_{\mathrm{eff},p} = \frac {\sigma_{\mathrm{max},p}} {(1-D_p)E_{\mathrm{Y}}}  \, \text{.}
    \label{eq:dmgsoft}
\end{equation}
$\sigma_{\mathrm{max}}$ is the maximum tensile stress after a principal axis transformation \citep{Grady1980}.
Assuming that a single crack grows at a constant velocity $c_{\mathrm{g}} = 0.4 c_{\mathrm{p}}$ related to Eq.~(\ref{eq:cp}), based on the assumption that a crack relieves stresses in a volume approximately equal to its circumscribing sphere \citep{Walsh1965}, \citet{Melosh1992} generalized the dynamical propagation of damage to higher dimensions, and \citet{Benz1994} implemented this computation into SPH codes.
Then \citet{Schafer2016} modified it to a multi--flaw subvolume (particle)
\begin{equation}
    \frac {\mathrm{d} D^{1/3}} {\mathrm{d}t} = n_{\mathrm{act}} \cdot \frac {c_{\mathrm{g}}} {R_{\mathrm{s}}}
    \label{eq:damage}
\end{equation}
where $R_{\mathrm{s}}$ represents the effective radius of the subvolume when fully damaged.
Note that damage is not allowed to exceed its upper limit $D_p^{\mathrm{max}} = (n_{\mathrm{act},p} / n_{\mathrm{f},p})^{(1/3)}$.

One problem is that the total number of flaws in the above-mentioned approach relies on precision, similar to the average activation strain.
Besides, \citet{Sevecek2021} evaluates that this approach may not be efficient enough for sets of large numbers of particles.
The assignment of flaws into particles follows a binomial distribution, so in this paper, a Poisson statistics estimation could be utilized to directly obtain the total number of flaws in each particle.
Based on these two points, a new approach is proposed to initialize the flaw distribution. 
\begin{itemize}

\item $n_{\mathrm{tot}}$ can be determined in a straightforward way based on the sizes of objects constructed for the target, such as pebbles, grains composing the regolith, or crystal substructure.
Note that $n_{\mathrm{tot}}$ should be sufficiently large to ensure that each particle contains multiple flaws, usually $n_{\mathrm{tot}} > 100 n_{\mathrm{p}}$.

\item To capture the highly dynamical events during hypervelocity impacts, a wide spread of fracture strains is required, normally at least $\varepsilon_{\mathrm{max}} / \varepsilon_{\mathrm{min,V}} = 10$ where $\varepsilon_{\mathrm{min,V}} = (kV)^{-1/m}$ \citep{Benz1994}.
Therefore the smallest activation strain of each particle $\varepsilon_{\mathrm{min},p}$, regarded as the active strain, is set to vary by an order of magnitude in this approach, i.e., $\left[\varepsilon_{\mathrm{min},p}\right]_{\mathrm{max}} = 10 \left[\varepsilon_{\mathrm{min},p}\right]_{\mathrm{min}} = (10^m/kV)^{-1/m}$.
In other words, the corresponding smallest number of flaws of each particle ranges from $1$ to $10^m$ ($10^m < n_{\mathrm{tot}}$).
Assuming these flaws are uniformly dispersed in a numbered sequence for simplicity, by producing a random seed, the smallest number of flaws as well as $\varepsilon_{\mathrm{min},p}$ of each particle is determined.
And in this way, the average active strain goes to $\bar{\varepsilon}_{\mathrm{min}} = 10m / (m+1) \cdot \varepsilon_{\mathrm{min,V}}$.

\item The assignment of flaws into particles follows a binomial distribution, i.e., a sequence of $n_{\mathrm{tot}}$ independent experiments with the probability of success $P_{\mathrm{s}} = 1/n_{\mathrm{p}}$.
Since $n_{\mathrm{tot}}$ is large enough, and $P_{\mathrm{s}} n_{\mathrm{tot}} > 100$, the normal distribution $\mathcal{N}(P_{\mathrm{s}} n_{\mathrm{tot}}, P_{\mathrm{s}} n_{\mathrm{tot}}(1-P_{\mathrm{s}}))$ is a reasonable approximation to the result of this binomial experiment.
Generate another random seed and find the corresponding variable $x$ in the standard normal distribution table.
By mapping $x$ to the normal distribution $\mathcal{N}(P_{\mathrm{s}} n_{\mathrm{tot}}, P_{\mathrm{s}} n_{\mathrm{tot}}(1-P_{\mathrm{s}}))$, $n_{\mathrm{f},p}$ is obtained.

\item The biggest number $N_{\mathrm{max},p}$ of flaws of particle $p$ (corresponding to $\varepsilon_{\mathrm{max},p}$) follows an exponential distribution, with the probability density function $f(N_{\mathrm{max},p}) = (1 - 1/n_{\mathrm{p}})^{n_{\mathrm{tot}}-N_{\mathrm{max},p}} / n_{\mathrm{p}}$, and the cumulative distribution function is denoted as $F(N_{\mathrm{max},p}) = \begin{matrix} \sum_{i=1}^{N_{\mathrm{max},p}} f(i) \end{matrix} \approx (1 - 1/n_{\mathrm{p}})^{n_{\mathrm{tot}}-N_{\mathrm{max},p}}$.
Giving a random seed $x$ again, $N_{\mathrm{max},p}$ is evaluated by $n_{\mathrm{tot}} - \ln(x) / \ln(1-1/n_{\mathrm{p}})$, and $\varepsilon_{\mathrm{max},p} = ( N_{\mathrm{max},p}/kV ) ^{1/m}$.

\end{itemize}
With this method, the distribution of flaws in each particle can be efficiently provided, which has no effect on the discretization precision, as discussed in Section~\ref{sec4_1_1}. 

Using numerous state variables, this MPM code also incorporates several extra damage models.
The brittle fracture could be initiated based on the maximum principal stress ($D = 1$ when $\sigma_{\mathrm{max}} \geqslant \sigma_0$), and fatigue damage may accumulate based on the accumulated plastic strain ($D = \sum \Delta \varepsilon^{\mathrm{p}} / \varepsilon_0$).
These damage models aim to enhance the characterization of diverse dynamic properties of materials.

\section{Benchmarks and validations}\label{sec4}
Almost all aspects of the MPM framework have been thoroughly checked and analyzed \citep{Ma2009}.
This section mainly focuses on the validation of the MPM code under the scenario of hypervelocity impacts between asteroids, or targets composed of asteroidal materials, at different spatial scales.
Since both are particle-based methods, classical SPH simulations were chosen for comparison.

\subsection{Laboratory impact experiments}\label{sec4_1}
Laboratory impact experiments on spherical basalt targets carried out in 1991 in Japan fully measured the kinetic aspects of the fragments, including the cumulative mass distribution and mass-velocity distribution \citep{Nakamura1991}. They have thus been an ideal benchmark to validate a shock-physics code and determine material parameters \citep{Benz1994, Remington2020}.
The first version of the widely used Bern SPH code was developed and validated by comparing with these experiments \citep{Benz1994, Benz1995}. This code extended the application of SPH to solids with the implementation of material models \citep{Pierazzo2008}.
Therefore, these experiments and the SPH simulations are both testbeds for our MPM code.

\begin{figure}[pos=h]
\centering
  \includegraphics[width = 0.2\textwidth]{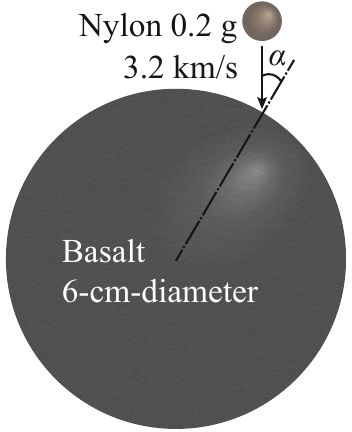}
  \caption{Schematic of the laboratory impact experiments on spherical basalt targets performed by \citet{Nakamura1991}.
  The nominal case is annotated, where the impact angle $\alpha = 30^{\circ}$.
  Further simulations encompass the influence of measurement errors, such as the mass of impactors, and the angle or relative velocity of the impact.}
  \label{fig:experiment}
\end{figure}

In the reference benchmarking experiment performed by \citet{Nakamura1991}, a spherical nylon projectile shoots a basalt sphere target off-axis at around $3.2$ km/s.
The projectile has a diameter of $7$ mm and a weight of $0.213$ g, the target has a diameter of $6$ cm under a density of $2.7$ g/cm$^3$, and the point of impact is offset by half of the radius perpendicular to the direction of velocity, resulting in an impact angle of $\alpha = 30^{\circ}$ (defined as the angle between the impact velocity vector and the surface normal at the point of impact), as shown in Fig.~\ref{fig:experiment}.
This impact experiment yielded a core-shaped major remnant, and statistically revealed a power-law distribution approximation in the mass-velocity relationship for fragments larger than a few millimeters.

The simulations by \citet{Benz1994} were in close agreement with these experimental results, especially the unprecedented accuracy of the mass and velocity of the largest fragment.
According to their numerical modeling, the elastic-perfectly plastic model and the Tillotson EOS were used for both the basalt target and the nylon projectile (it is possible to introduce a slight plastic hardening to guarantee the numeric stability, i.e., a non-zero but relatively small plastic modulus $E^{\mathrm{p}}$).
The basalt sphere was further described by the Grady-Kipp fracture model.
Given its lesser importance and the lack of material parameters, the nylon projectile was characterized without considering damage.
All relevant material characteristics of this nominal case can be found in Table~\ref{tab:mpmpara}.

\begin{table}
\begin{threeparttable}
\caption{Material parameters for laboratory impact experiment modeling}
\label{tab:mpmpara}
\begin{tabular}{lrr}
\toprule 
Description & Target & Projectile \\
\midrule
Material$^{\rm a}$ & Basalt & Nylon \\
Radius (\si{\milli\metre}) & $30.0$ & $3.5$ \\
Density (\si{\kilogram\per\cubic\metre}) & $2700$ & $1180$ \\
Mass (\si{\gram}) & \num{303} & \num{0.213} \\
Young's modulus $E_{\mathrm{Y}}$ (\si{MPa}) & $53100$ & $219$ \\
Poisson's ratio $\nu$ & $0.15$ & $0.45$ \\
\midrule
Strength model & linear hardening & linear hardening \\
Yield stress $\sigma_{\mathrm{Y}}$ (\si{MPa}) & $3500$ & $10$ \\
Plastic modulus $E^{\mathrm{p}}$ (\si{MPa}) & $5.31$ & $2.19$ \\
\midrule
Equation of state$^{\rm b}$ & Tillotson & Tillotson \\
$E_0$ (\si{MJ/kg}) & $487$ & $7$ \\
$E_{\mathrm{iv}}$ (\si{MJ/kg}) & $4.72$ & $2$ \\
$E_{\mathrm{cv}}$ (\si{MJ/kg}) & $18.2$ & $2.4$ \\
$\rho_{\mathrm{iv}}$ (\si{kg.m^{-3}}) & $2000$ & $1100$ \\
$A$ (\si{GPa}) & $26.7$ & $10.1$ \\
$B$ (\si{GPa}) & $26.7$ & $33.8$ \\
$a$ & $0.5$ & $0.6$ \\
$b$ & $1.5$ & $2.0$ \\
$\alpha$ & $5$ & $10$ \\
$\beta$ & $5$ & $5$ \\
\midrule
Fracture model & Grady--Kipp & - \\
Weibull $m$ & 8.5 & - \\
Weibull $k$ (\si{m^{-3}}) & $3.0\times10^{39}$ & - \\
Total number of flaws $n_{\mathrm{tot}}$ & $1.0\times10^{9}$ & - \\
\bottomrule
\end{tabular}
\begin{tablenotes}
\item[a] Parameters except that of EOS are extract from \citet{Benz1994}.
\item[b] \citet{Buruchenko2017}.
\end{tablenotes}
\end{threeparttable}
\end{table}

\subsubsection{Discretization settings}\label{sec4_1_1}
\begin{figure}[pos=h]
\centering
  \includegraphics[width = 0.45\textwidth]{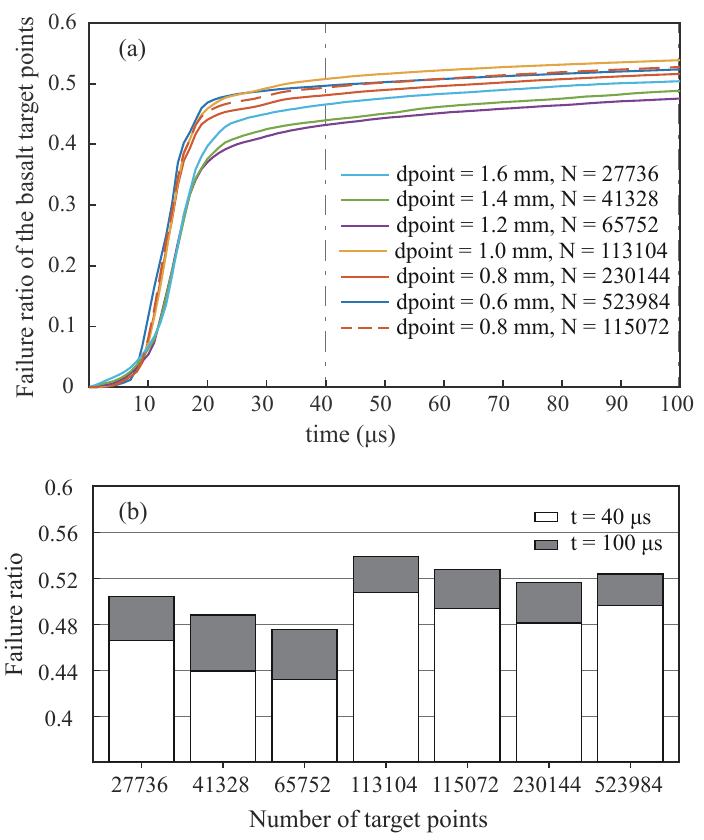}
  \caption{Development of the failure ratio of the basalt target with different precision.
  \textit{a}. Increase in the ratio of failure points over time. Each color corresponds to a specific discretization precision, indicated by the spacing between material points (dpoint).
  The total number of material points (N) discretizing the target basalt sphere is also noted.
  The orange dashed line corresponds to the half-particle-number scenario, taking advantage of the symmetric condition that will be discussed in Section~\ref{sec5_2}. 
  \textit{b}. Extracted failure ratio at time instants of \SI{40}{\micro\second} and \SI{100}{\micro\second} for different discretization precisions.}
  \label{fig:precision1}
\end{figure}

\begin{figure}[pos=htbp]
\centering
  \includegraphics[width = \textwidth]{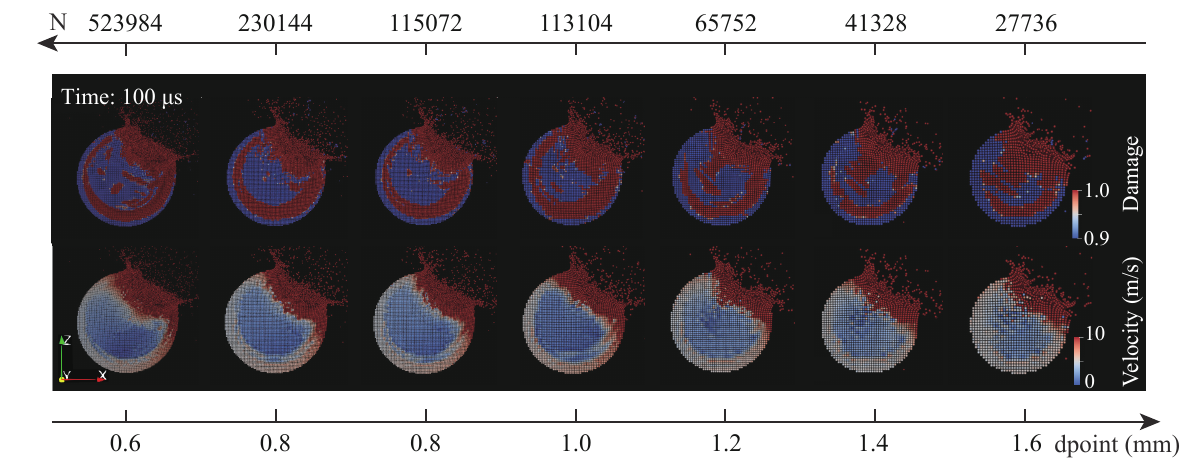}
  \caption{Cross-sectional views along the symmetry plane at \SI{100}{\micro\second} for different discretization precisions.
  The top row displays the degree of damage greater than $0.9$ for each point (considered as failed), while the bottom row shows point velocity distributions.
  Discretization precision decreases from left to right, and the third column corresponds to the half-particle number scenario, taking advantage of the symmetric condition.}
  \label{fig:precision2}
\end{figure}

Initially, different levels of discretization precision, or in other words, different interparticle spacings, are tested to confirm the robustness of our MPM code.
Accordingly, a regular cubic background grid with a cell edge length of twice the initial spacing between material points is employed, initially populating $8$ ($2 \times 2 \times 2$) material points within each active grid cell covering the material domain.
The total energy is monitored in real-time to ensure energy conservation, with fluctuations in our simulation that do not exceed \SI{0.6}{\percent}.
Meanwhile, an appropriate total number of material points and simulation duration are found.

Fig.~\ref{fig:precision1} illustrates how the proportion of failed particles (damage greater than $0.9$) developed under different discretization precision.
As depicted in Subfigure~\ref{fig:precision1}(a), a consistent trend in the temporal evolution of failure ratio is observed across varying levels of discretization precision.
Damage initially increases rapidly within the first \SI{20}{\micro\second}, and then transitions to a slower rate of growth.
With improved discretization precision, the failure ratio at \SI{100}{\micro\second} gradually converges to \SI{52}{\percent}, while the discrepancy in failure ratio for simulations with lower precision also does not exceed \SI{10}{\percent}.
Subfigure~\ref{fig:precision1}(b) extracts the failure ratio at \SI{40}{\micro\second} and \SI{100}{\micro\second} in each cases, showing an increment ranging from \SI{2.7}{\percent} to \SI{4.9}{\percent}.

The damage and velocity along the symmetrical cross-section of the target also display similar distributions with different discretization precisions at \SI{100}{\micro\second} in Fig.~\ref{fig:precision2}.
Spallation takes place in all cases, leading to a comparatively lower central velocity and higher shell velocity.
However, the core-shaped fragment only appears with higher discretization resolution.
If the total number of material points describing the target basalt ball decreases to $65752$, or the distance between material points exceeds \SI{1.2}{\milli\metre}, the failure zone is unable to develop into a closed shell due to the reduced capacity of capturing detailed features.
As a result, unreleased stress converges in the central portion, causing additional damage around the core.
Conversely, pushing to extremely high resolutions can introduce minor, localized numerical artifacts.
In MPM, finer grids dictate that material points cross cell boundaries more frequently during large deformations, leading to the accumulation of cell-crossing noise (the mitigation and limitations of which are discussed further in Section~\ref{sec5_2}).
For instance, when the number of material points increases to $523984$, this accumulated noise causes a slight expansion of the failure region inside the core-shaped fragment.

However, this microscopic noise does not represent a divergence.
Numerical convergence in this framework is assessed macroscopically by observing the plateauing of key statistical variables, such as the total failure ratio (which converges toward \SI{52}{\percent} in Fig.~\ref{fig:precision1}), and the stabilization of the fragment morphologies.
In the absence of reference results, an appropriate resolution is determined by identifying this converged plateau.
Ultimately, these macroscopically consistent performances indicate computational stability, and explicitly support the resolution-independent nature of our proposed approach for initializing Weibull flaws.

While the simulation duration in \citet{Benz1994} was limited to \SI{40}{\micro\second}, our simulations that go further in time show that the follow-up accumulation of damage is not negligible.
Some material points have activated flaws but with damage not fully developed, and the subsequently attenuated stress wave continues to reflect within the target. 
Therefore, the more prolonged the simulation duration, the more damage accumulated, although at a reduced rate.
Considering the computational costs, \SI{100}{\micro\second} is adopted as the nominal duration.
And the failure threshold is set to damage reaching $0.9$, to compensate for the insufficient development of damage due to time truncation.
Besides, an initial spacing of \SI{0.8}{\milli\metre} between material points is chosen as the nominal precision. The target basalt sphere is thus discretized into $230144$ material points.
This selection ensures both computational accuracy and efficiency and also reduces the likelihood of excessive computational errors.

\subsubsection{Searching for fragments}\label{sec4_1_2}
A post-processing algorithm for fragment identification is then implemented to search for and statistically analyze these impact residuals.
A fragment is defined as a region consisting of continuous unfailed material points bounded by a perimeter of failed points \citep{Benz1994}.
An unfailed region with two parts connected by only one or a few unfailed points would have a connection neck too weak to withstand future deformation, potentially leading to the region splitting into two separate parts. 
Therefore, based on the number of its unfailed neighbors referring to the density-based spatial clustering theory, unfailed points are labeled as either extendable or non-extendable \citep{Ester1996}.

Although the fragment identification process occurs post-impact simulation, the principles of the Material Point Method (MPM) still offer valuable insights for the algorithm.
A critical application is the employment of the background grid, which significantly enhances the quick localization of material points and the efficient identification of their neighbors.
Furthermore, material points are distinctly maintained without spatial overlap, each carrying physical quantities unaffected by others.
This characteristic, in conjunction with the background grid, facilitates a seamless implementation of the density-based spatial clustering method.
Rather than computing the coordination number, the algorithm efficiently assesses the ratio of unfailed material points within each grid cell, streamlining the process of bulk labeling extendable points.
Therefore, starting from a single extendable free (without regard to any fragments) point, a new fragment is recognized using a grid-based friends-of-friends neighborhood search algorithm.
The algorithm first classifies all the adjacent extendable points, then adds the related non-extendable points, and finally the failed layer.
Fragments are sequentially numbered as they are recognized.
And the shape of each identified fragment is inherently determined by the collective arrangement of its material points, negating the need for external contouring algorithms.
After all extendable points are registered, the remaining non-extendable or failed points, which are not incorporated into any fragments, are considered to be finely fractured to dust (with sizes below the resolution threshold) and are thus not assigned fragment numbers.

Assuming rigid body motion, kinematic quantities of each fragment can be deduced as
\begin{equation}
    \begin{cases}
    m_f = \sum {m_p} \\
    \bm{x}_{f \mathrm{c}} = \left( \sum {m_p \bm{x}_p} \right) / m_f \\
    \bm{P}_f = \sum {m_p \bm{v}_p} \\
    \bm{L}_{f \mathrm{c}} = \sum \bm{x}_p \times m_p \bm{v}_p - \bm{x}_{f \mathrm{c}} \times \bm{P}_f \\
    \bm{\omega}_f = \bm{J}_{\mathrm{c}}^{-1} \bm{L}_{f \mathrm{c}}
    \end{cases}
    \, \text{,}
\end{equation}
where $\bm{L}_{f \mathrm{c}}$ represents a specific angular momentum, the subscript $f$ is the identification of one fragment, and $\mathrm{c}$ denotes the relative to the center of mass.
And $\bm{J}_{\mathrm{c}}$ is the inertia matrix of the fragment referred to the center of the mass coordinate system.
For the nominal case, the core mass in our simulation is quantified at \SI{24.6}{\percent} of the target's mass, possessing a velocity of \SI{2.056}{\metre\per\second}.

\subsubsection{Nominal case}\label{sec4_1_3}
\begin{figure}[pos=htbp]
\centering
  \includegraphics[width = \textwidth]{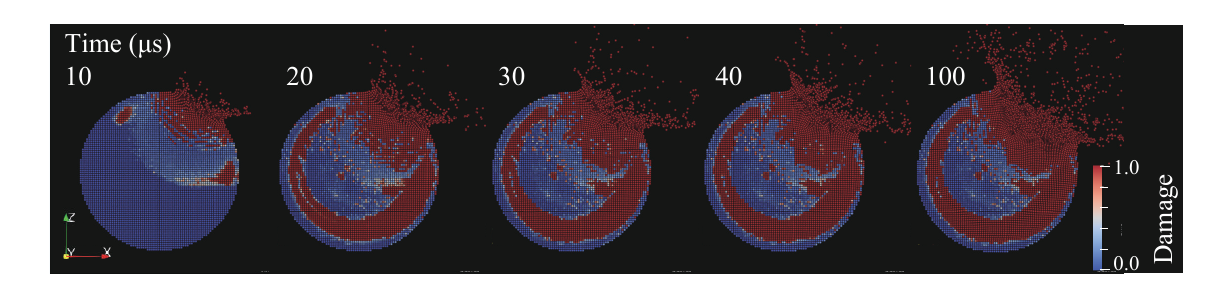}
  \caption{Snapshots at $10$, $20$, $30$, $40$, and \SI{100}{\micro\second} of the nominal impact case under the discretization precision of \SI{0.8}{\milli\metre} inter--particle spacing.
  Shown are the degrees of damage ranging from $0$ to $1$ in the central symmetry plane.}
  \label{fig:snapshots}
\end{figure}

The damage growth of this nominal case is shown in Fig.~\ref{fig:snapshots}.
The snapshot taken at \SI{10}{\micro\second} demonstrates two mechanisms at the origin of material damage induced by the shock wave.
Firstly, the shearing and compressive effect of the powerful shock wave, which produces linear weakening at the point of impact while spreading spherically, activates flaws in the material points located along its path.
Subsequently, upon reflection from the target's surface, the weakened compressive wave transforms into a tensile wave that leads to spalling underneath the surface.
During \SI{10}{\micro\second} to \SI{20}{\micro\second}, this spall runs along a shell, forming an almost intact core.
And the damage continues to accumulate thereafter.
Notice that the damage distribution at times $10$, $20$, $30$ and \SI{40}{\micro\second} and the formation of the core-shaped unfailed region correspond very well to Fig.~$7$ in \citet{Benz1994}, and the ejection in the vicinity of the impact point positions is closely aligned with the experimental records reported in \citet{Nakamura1991} Fig.~$1$b.

\begin{figure}[pos=h]
\centering
  \includegraphics[width = 0.4\textwidth]{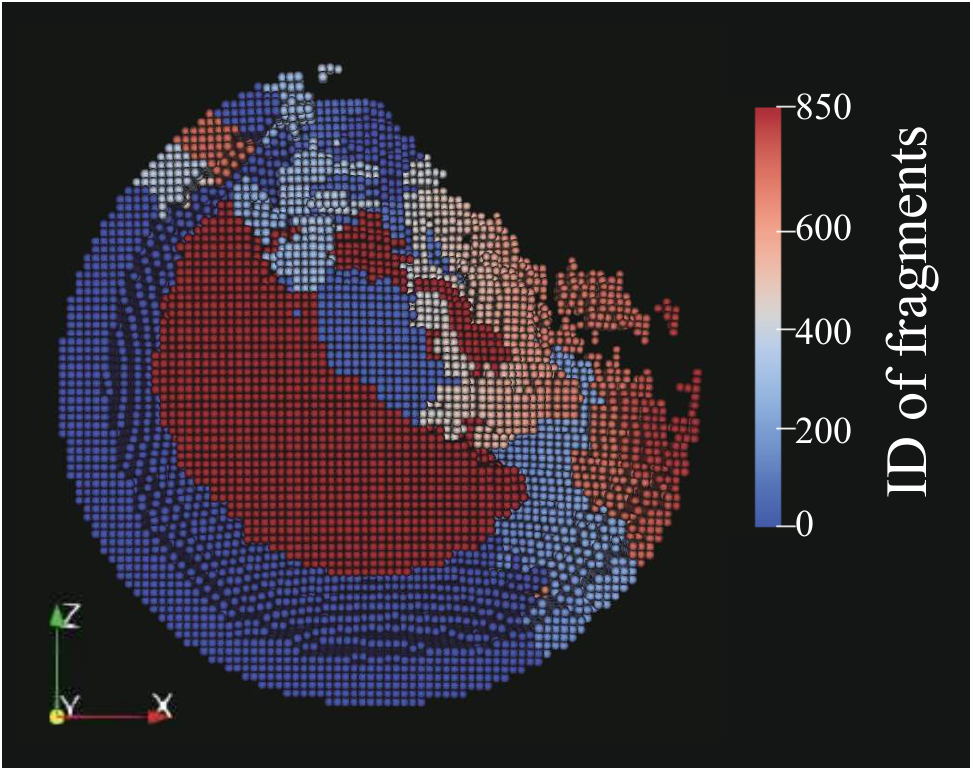}
  \caption{Shape and position of the fragments in a central clip, \SI{100}{\micro\second} after the impact.
  The core-shaped major fragment is painted in red, and others are colored by their identifying number generated during the fragment search.}
  \label{fig:frag}
\end{figure}

This fragment searching algorithm generates $826$ fragments in the nominal case, which are presented in Fig.~\ref{fig:frag}.
It is worth noting that because the target forms a highly fractured outer spallation shell, a full 3D visualization would be visually occluded by the external debris.
Therefore, the 2D central cross-section is presented here as it most clearly reveals the internal fragmentation hierarchy and the core fragment.
The core fragment colored in red is the largest one.
And the second largest one is a piece of spallation, detached near the impact site's antipode.
For other fragments, generally, the closer a fragment is to the point of impact, the smaller it tends to be.

\begin{figure}[pos=htbp]
\centering
  \includegraphics[width = 0.4\textwidth]{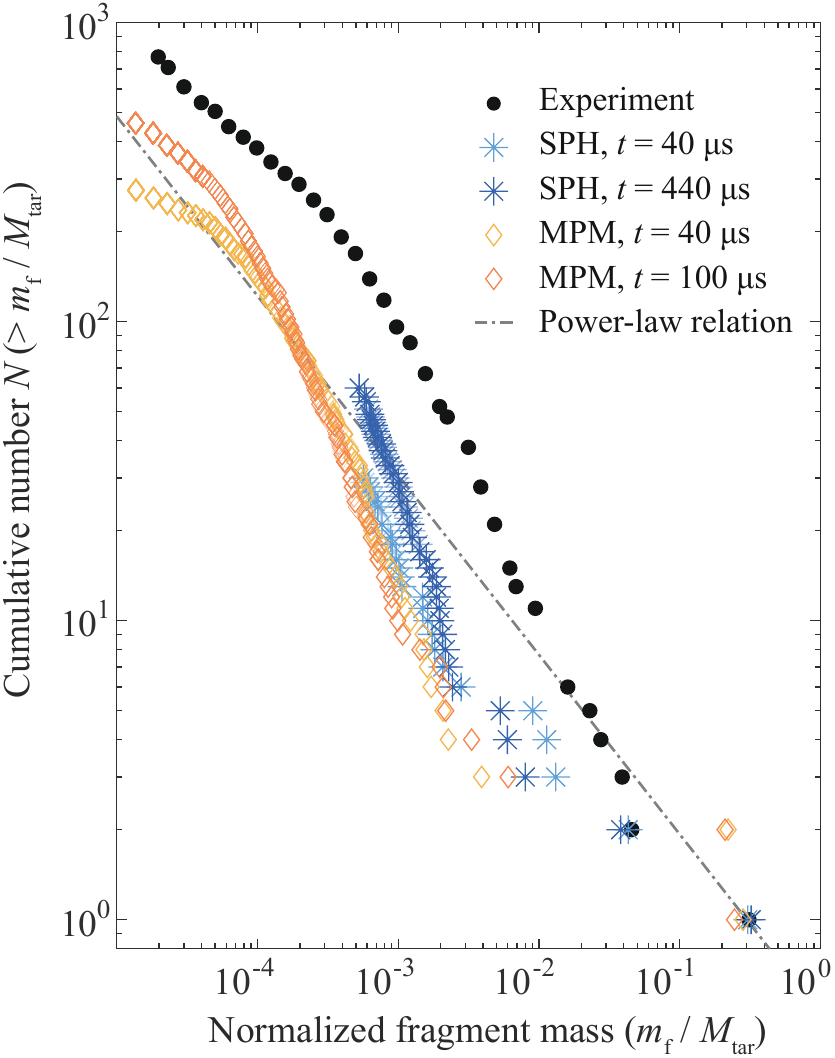}
  \caption{The cumulative mass distribution of fragments obtained from MPM simulation (orange diamonds), in comparison with the experimental results (black circles), the SPH calculation (blue asterisks, lighter shade at \SI{40}{\micro\second} while darker at \SI{440}{\micro\second}), and the predicted power-law mass distribution of comminuted fragments (gray dashed line).}
  \label{fig:compmn}
\end{figure}

\begin{figure}[pos=htbp]
\centering
  \includegraphics[width = 0.4\textwidth]{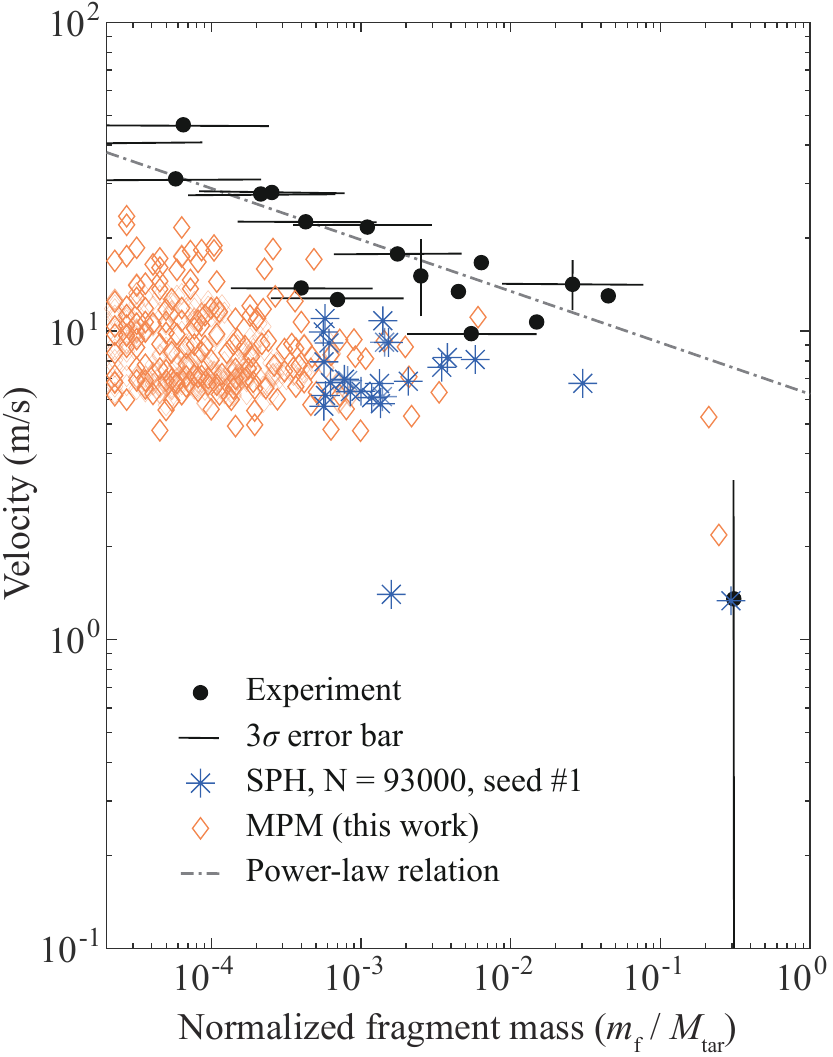}
  \caption{The mass-velocity distribution of fragments for the experimentally-measured results (black circles) associated with $3 \mathrm{\sigma}$ error bars (solid lines), the SPH calculation (blue asterisks), the MPM simulation (orange diamonds), and the predicted power-law relation (gray dashed line).}
  \label{fig:compmv}
\end{figure}

An exhaustive statistical analysis is conducted on the mass and velocity of all fragments, facilitating a comparison with experimental data \citep{Nakamura1991} and SPH simulation results \citep{Benz1994}. 

Fig.~\ref{fig:compmn} depicts the cumulative unit mass $m_{\mathrm{f}} / M_{\mathrm{tar}}$ distribution of fragments, with a dashed line portraying the predicted power-law relationship of $n(m) \propto m^{-3/5}$ for fine fractured fragments \citep{Nakamura1991}, here $m$ denotes the unit mass.
The mass of the largest fragment obtained in the MPM simulation shows reasonable agreement with the findings of prior research (unit mass $0.31$, $0.30$, and $0.246$ in the experiment, SPH simulation, and our MPM simulation, respectively).
Although this represents a relative difference of approximately \SI{21}{\percent} compared to the experimental value, such variances are consistent with the typical \SI{20}{\percent} to \SI{30}{\percent} accepted error margins for physical quantities in planetary impact hydrocode studies \citep{Pierazzo2008}.
Meanwhile, the second one (envelope) is much heavier, consequently leading to the lower mass of intermediate fragments.
Following the idea mentioned by \citet{Benz1994}, the comparison between fragments generated at various times verifies the accuracy of the postprocessing scheme.
A slight shrinkage of the first and second fragments, and a slight lifting of the slope of the intermediate mass fragments' distribution appear, as a result of the damage growth that breaks the narrow connection linking protrusions and their main body.
Besides, the normalized mass distribution of comminuted fragments ($10^{-5} < m_{\mathrm{f}} / M_{\mathrm{tar}} < 10^{-3.3}$) generally corresponds to the predicted power-law relation.

The comparison of unit mass versus $3$-D velocity distribution is shown in Fig.~\ref{fig:compmv}.
The experimental results obtained by \citet{Nakamura1991} are plotted as solid black circles.
Due to recognition difficulty, fragments from the inner part of the target with a smaller size or lower velocity were omitted from the analysis.
The captured fragments are associated with $3 \mathrm{\sigma}$ error bars, which quantify the uncertainty of mass or the $3$-D velocity estimated from side-view films, arising from the fragment outlining of images.
Then, a power-law relation $V(m_{\mathrm{f}}) = 6.4 (m_{\mathrm{f}} / M_{\mathrm{tar}})^{-0.155}$ is fitted according to the second largest fragment and the smaller ones \citep{Nakamura1991}.
Therefore, these experimental results portray the upper-velocity boundary for fragments that exclude the core.
Upon examining the figure, it is observed that the MPM simulation captures the general downward trend of the velocity distribution for fragments of lower mass, roughly aligning with the predicted power-law relation despite the expected numerical scatter.
However, the MPM simulation overestimates the velocity of our core fragment (\SI{2.056}{\metre\per\second}) when compared to the experimental and SPH simulation results (\SI{1.35}{\metre\per\second} and \SI{1.352}{\metre\per\second}, respectively).
Despite this deviation, the observed velocity of the core fragment remains within the $3 \mathrm{\sigma}$ confidence interval, underscoring the simulation's overall alignment with empirical observations.

However, the method of measuring angular velocity in the referenced literature is not explicitly stated, rendering it challenging to include a comparative statistical analysis for this specific kinematic quantity in our study. Therefore, statistics on angular velocity have been omitted in this paper.

The above analysis substantiates the reliability of our MPM simulation to reproduce laboratory impact experiments, as it is congruent with the referenced experimental and SPH-simulated data.

\subsubsection{Sensitivity on impact conditions}\label{sec4_1_4}

\begin{table}
\begin{threeparttable}
\caption{Influence of impact conditions}
\label{tab:initial}
\setlength{\tabcolsep}{2.5pt}
\begin{tabular}{lccccccc}
\toprule 
\multirow{2}{*}{Initial condition} &
\multirow{2}{0.08\columnwidth}{Nominal \\ case$^{\mathrm{a}}$} &
\multicolumn{2}{c}{Impact angle (\si{\degree})} &
\multicolumn{2}{c}{Velocity (\si{\kilo\metre\per\second})} &
\multicolumn{2}{c}{Mass of projectile (\si{\gram})} \\
\cmidrule(lr){3-4} \cmidrule(lr){5-6} \cmidrule(lr){7-8}
  & & $0$ (vertical) & $45$ & $2.7$ & $3.7$ & $0.175$ & $0.225$ \\
\midrule
\parbox{3cm}{Momentum of\\projectile (\si{\kilogram\metre\per\second})} 
& $0.64$ & $0.64$ & $0.64$ & $0.54$ & $0.74$ & $0.56$ & $0.72$ \\
Core-shaped fragment
& \Checkmark & \Checkmark & \XSolid & \XSolid & \Checkmark & \Checkmark & \Checkmark \\
$m_{\mathrm{f,max}} / M_{\mathrm{tar}}$ 
& $0.246$ & $0.057$ & $0.544$ & $0.605$ & $0.121$ & $0.276$ & $0.213$ \\
$v_{\mathrm{f,max}}$ (\si{\metre\per\second})
& $2.056$ & $0.727$ & $3.032$ & $3.092$ & $2.468$ & $1.973$ & $2.408$ \\
Damage at \SI{100}{\micro\second}$^{\mathrm{b}}$
&
\begin{minipage}[c]{0.1\columnwidth}
    \centering
    {\includegraphics[width=\linewidth]{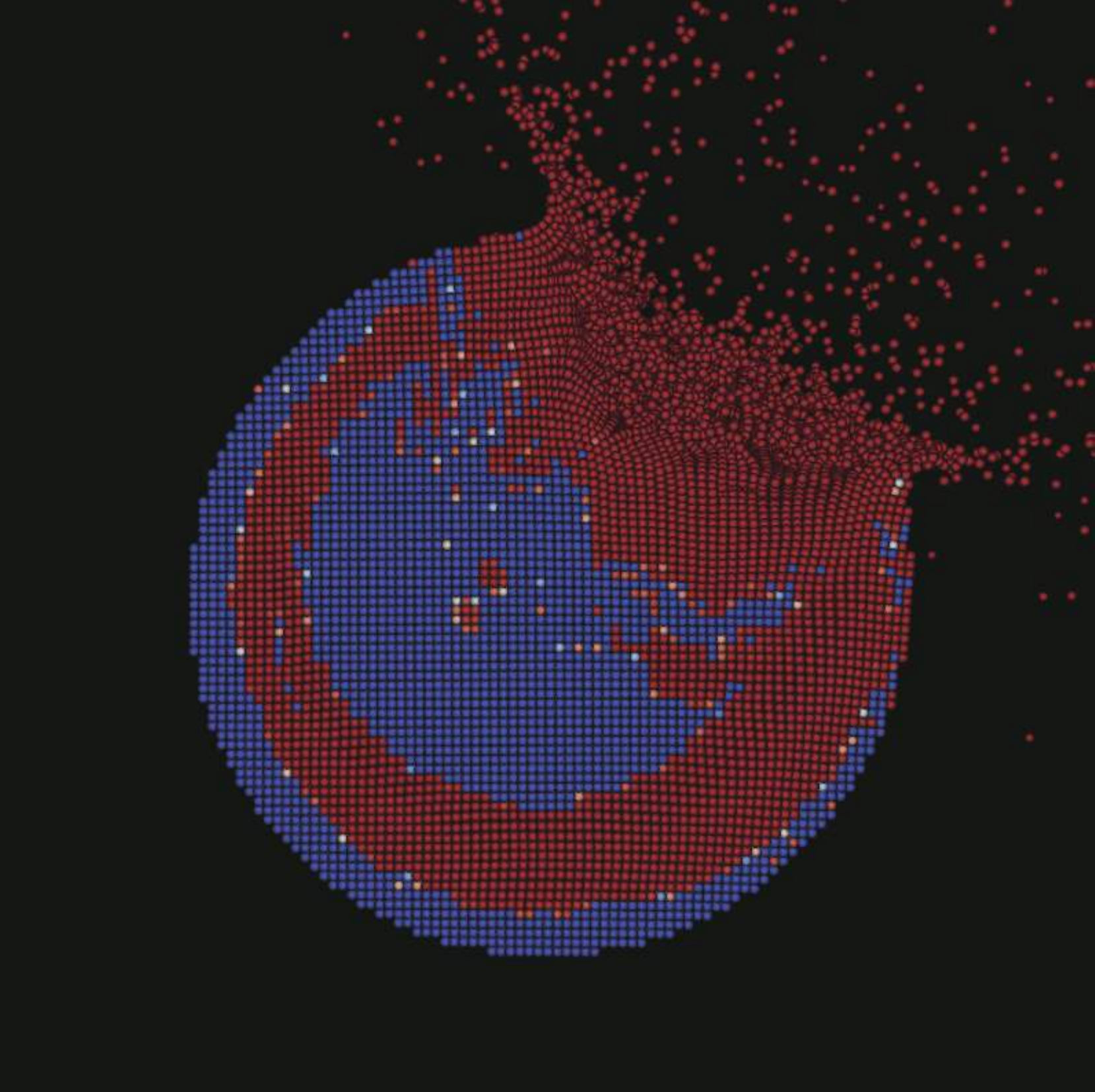}}
\end{minipage}
&
\begin{minipage}[c]{0.1\columnwidth}
    \centering
    {\includegraphics[width=\linewidth]{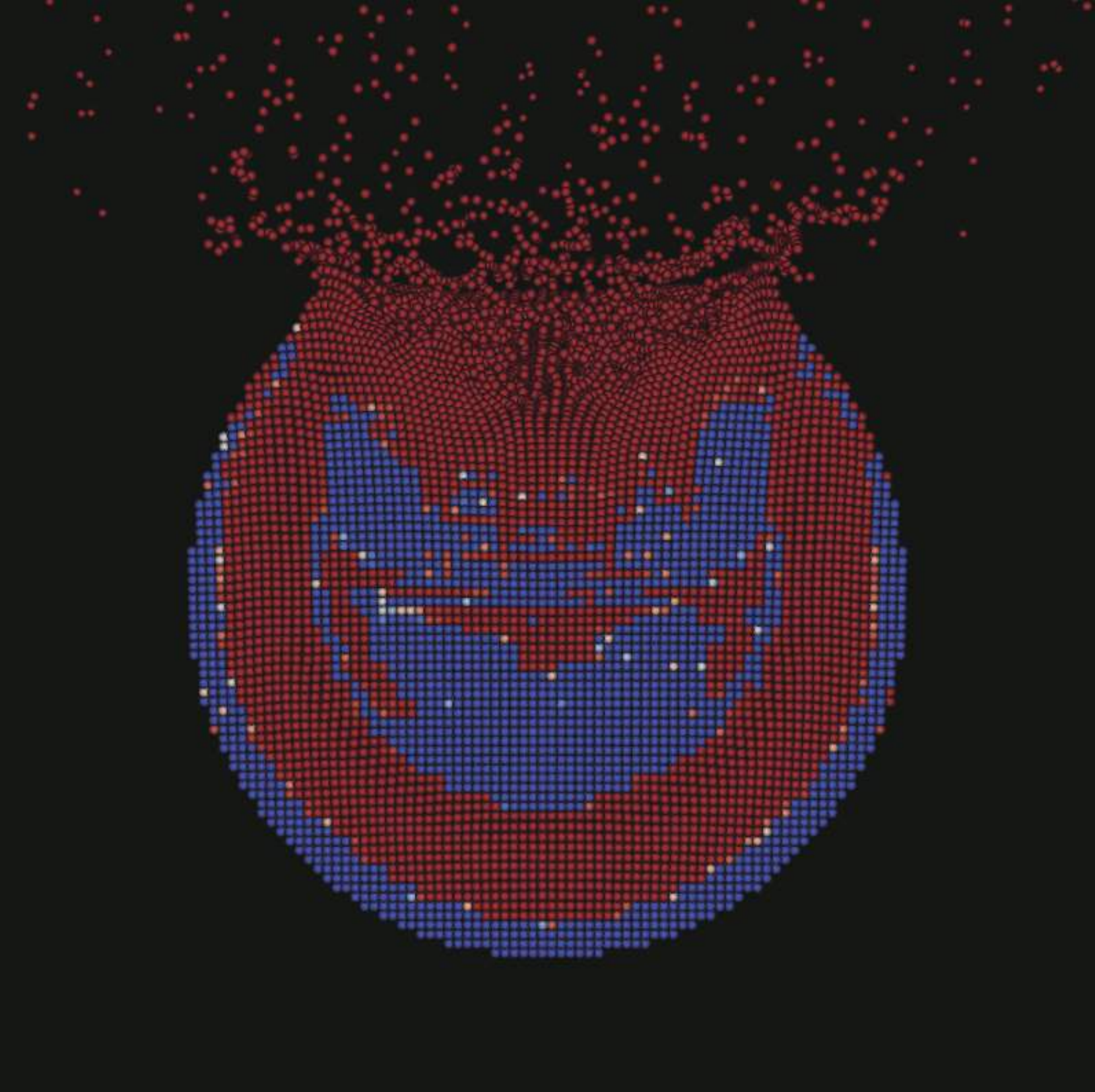}}
\end{minipage}
&
\begin{minipage}[c]{0.1\columnwidth}
    \centering
    {\includegraphics[width=\linewidth]{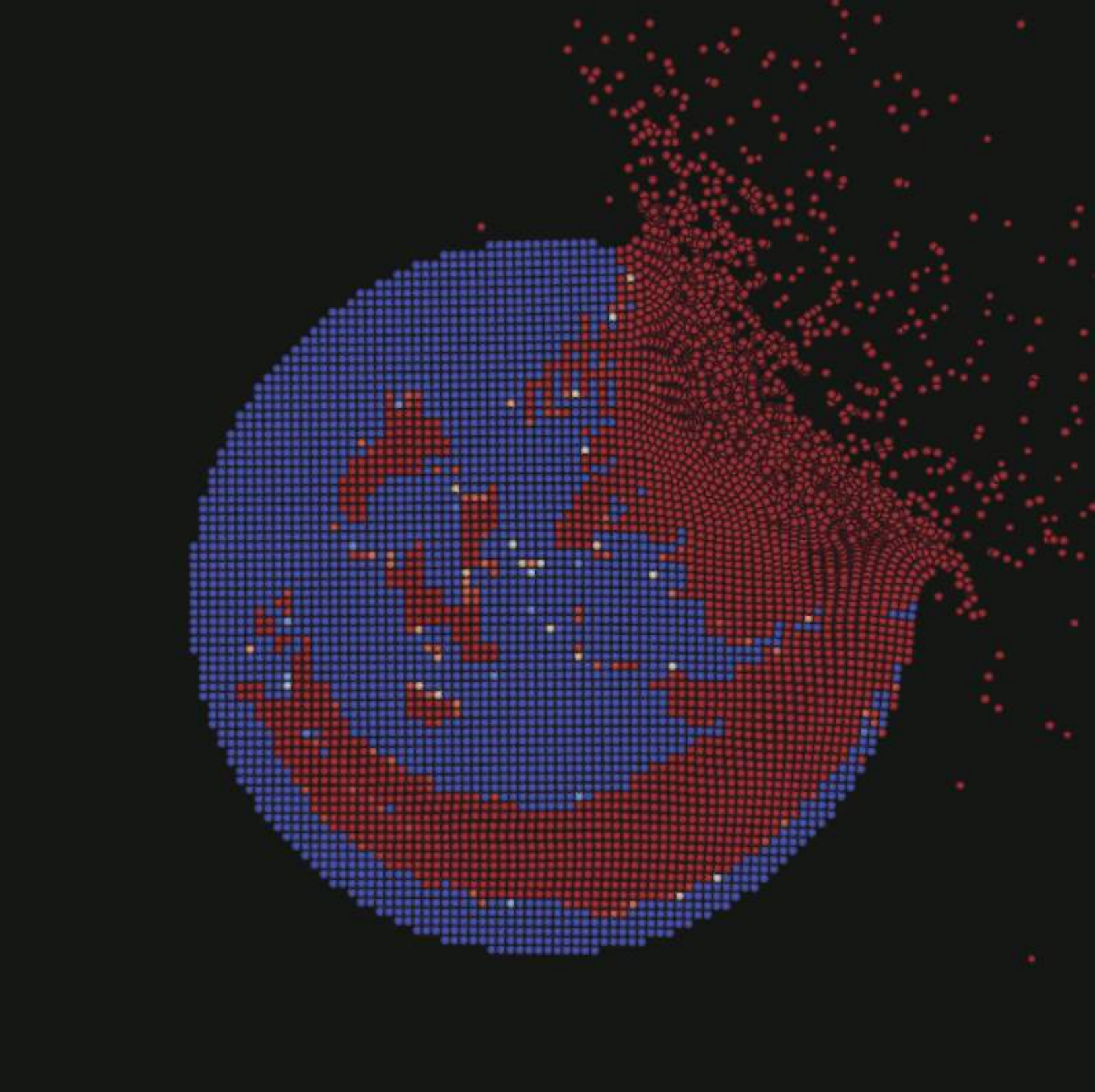}}
\end{minipage}
&
\begin{minipage}[c]{0.1\columnwidth}
    \centering
    {\includegraphics[width=\linewidth]{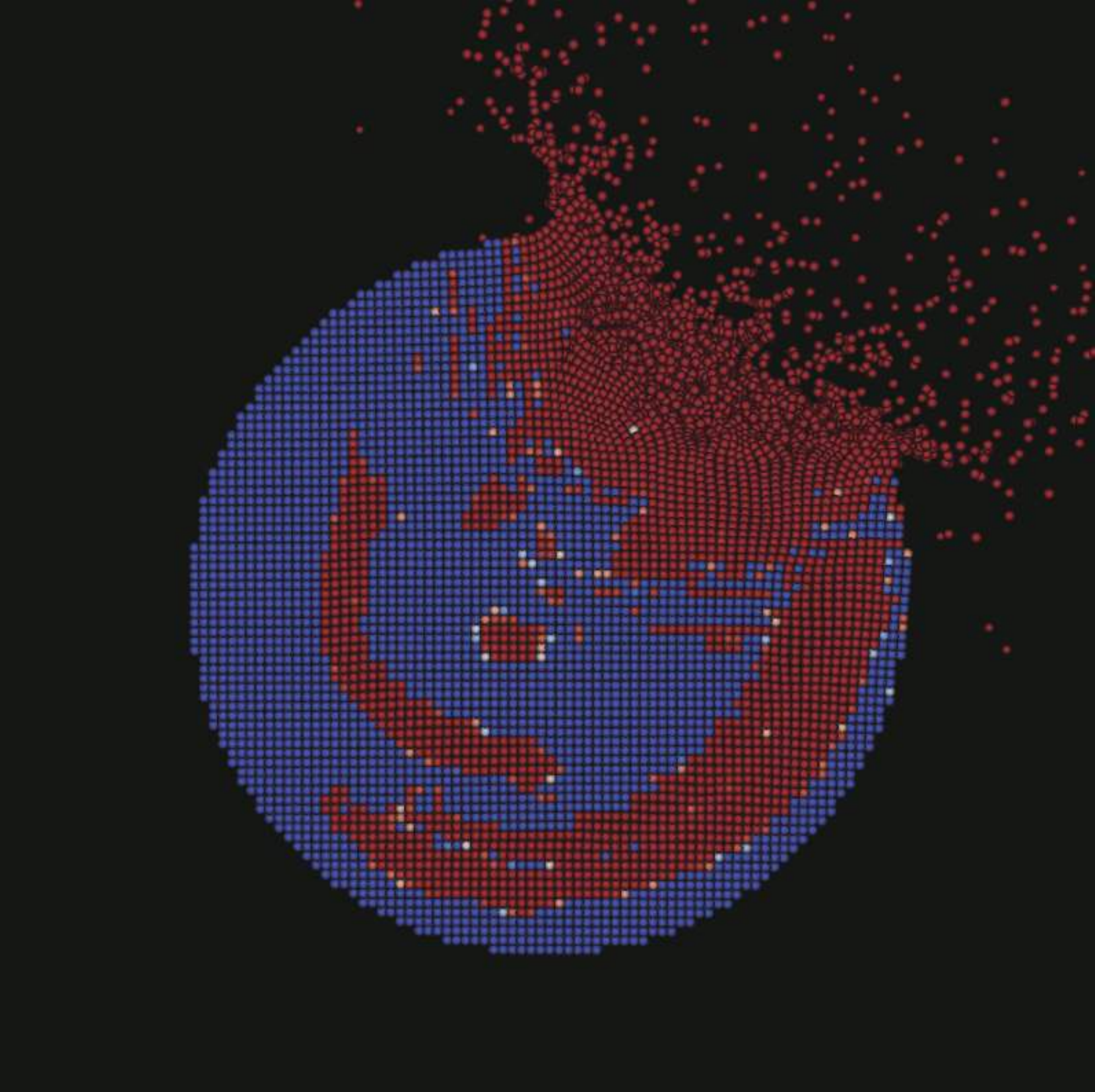}}
\end{minipage}
&
\begin{minipage}[c]{0.1\columnwidth}
    \centering
    {\includegraphics[width=\linewidth]{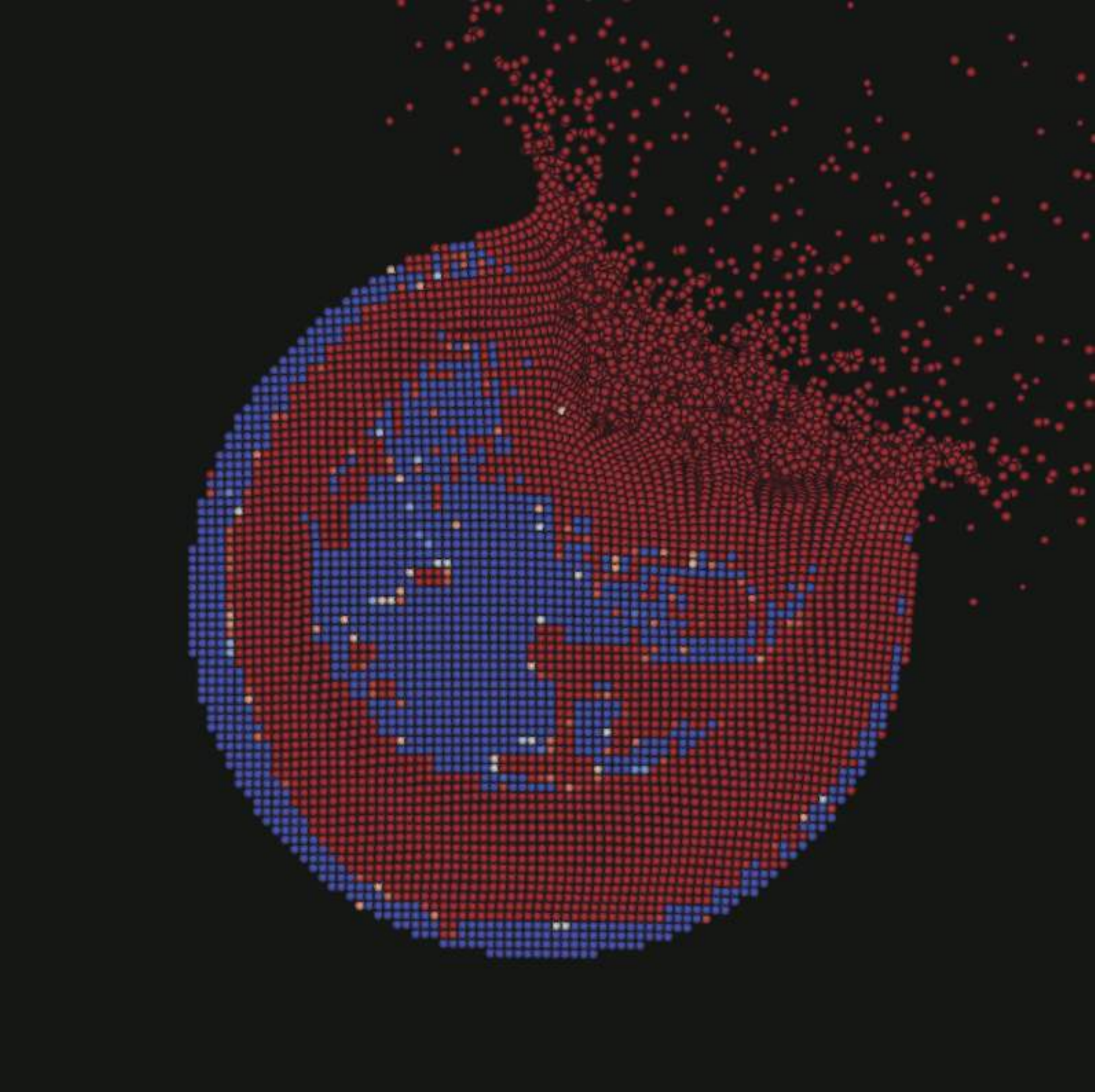}}
\end{minipage}
&
\begin{minipage}[c]{0.1\columnwidth}
    \centering
    {\includegraphics[width=\linewidth]{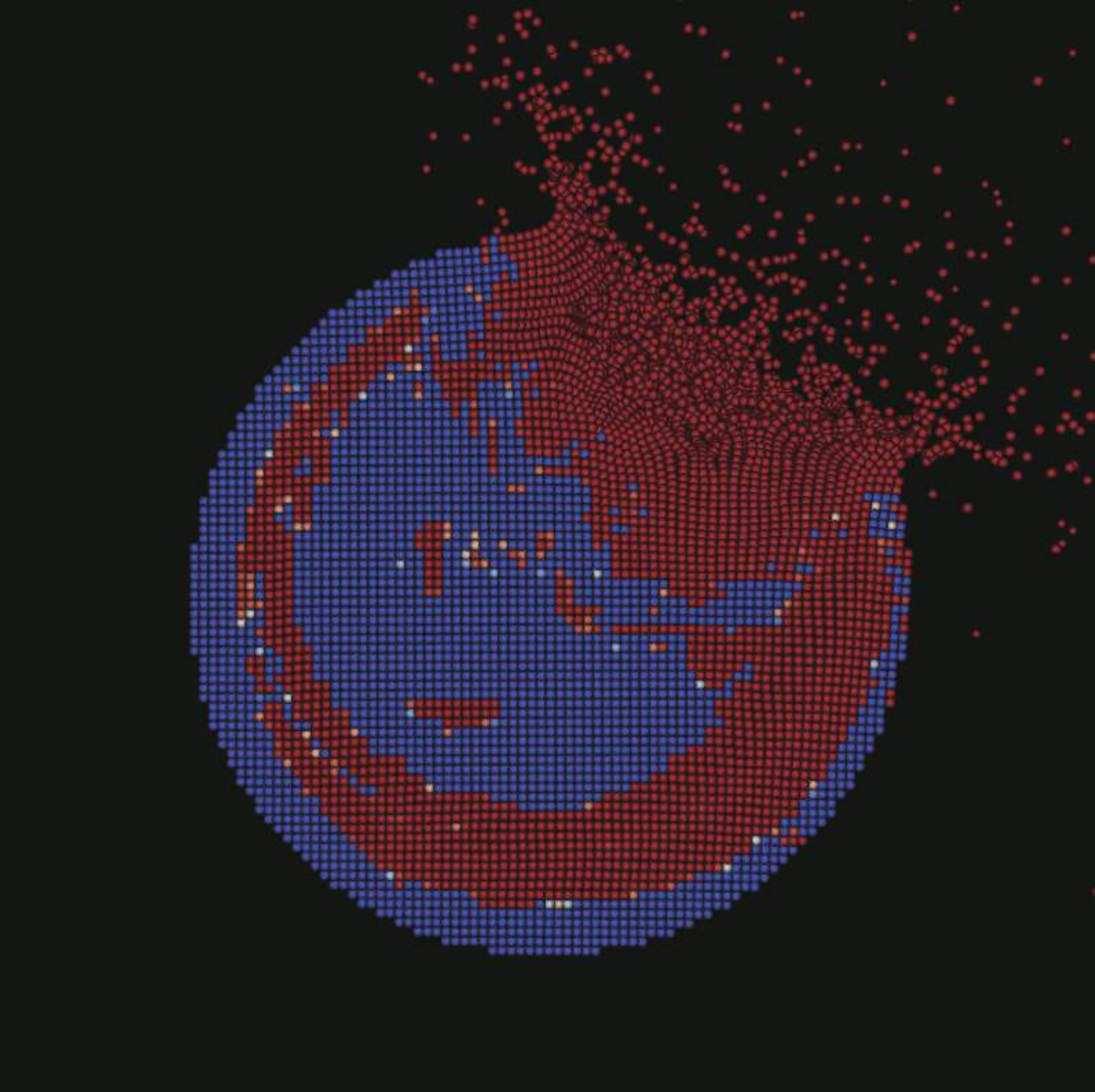}}
\end{minipage}
&
\begin{minipage}[c]{0.1\columnwidth}
    \centering
    {\includegraphics[width=\linewidth]{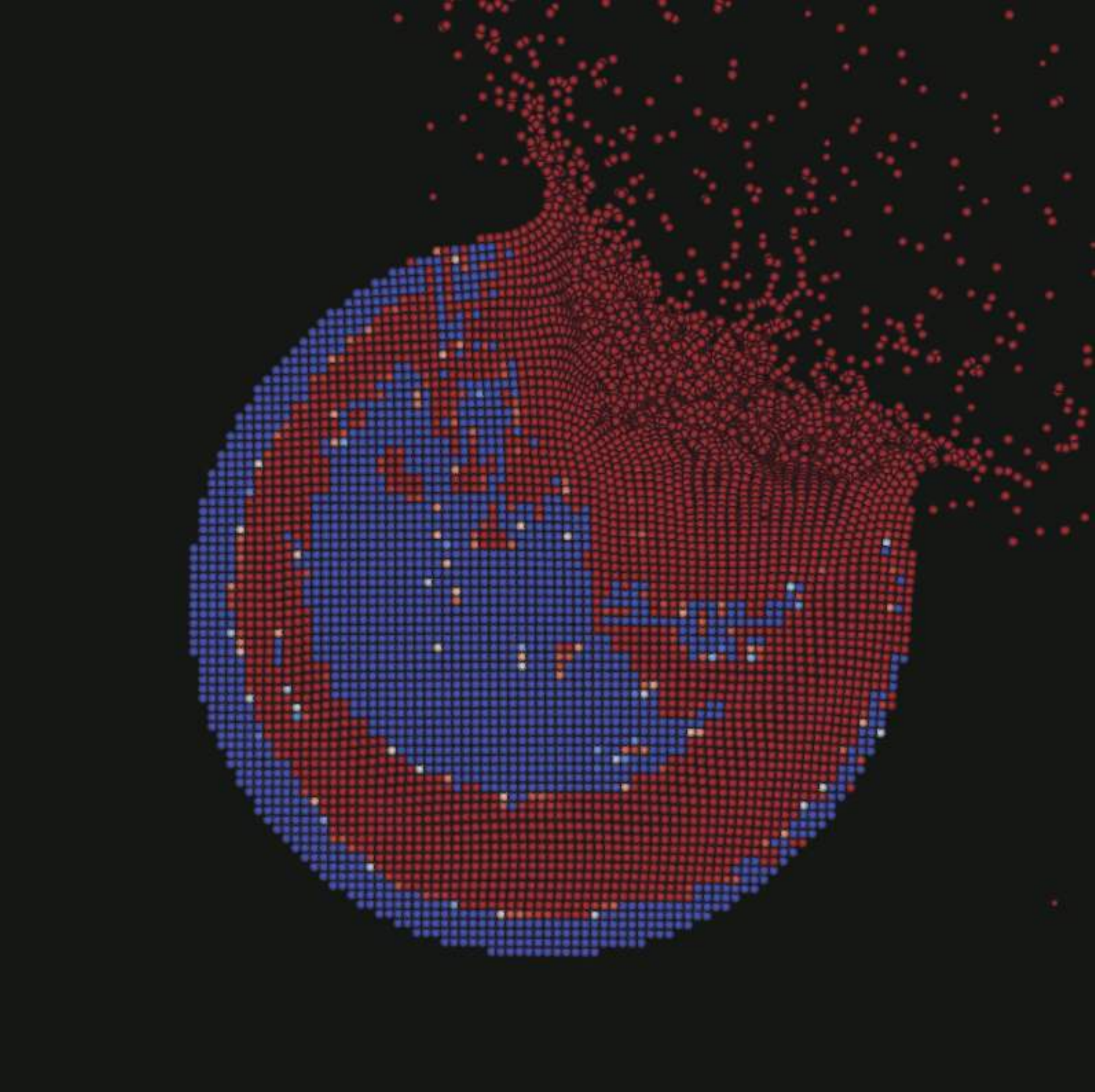}}
\end{minipage}
\\
\arrayrulecolor{white}
\cmidrule(lr){1-8}
Largest fragment$^{\mathrm{c}}$
&
\begin{minipage}[c]{0.1\columnwidth}
    \centering
    {\includegraphics[width=\linewidth]{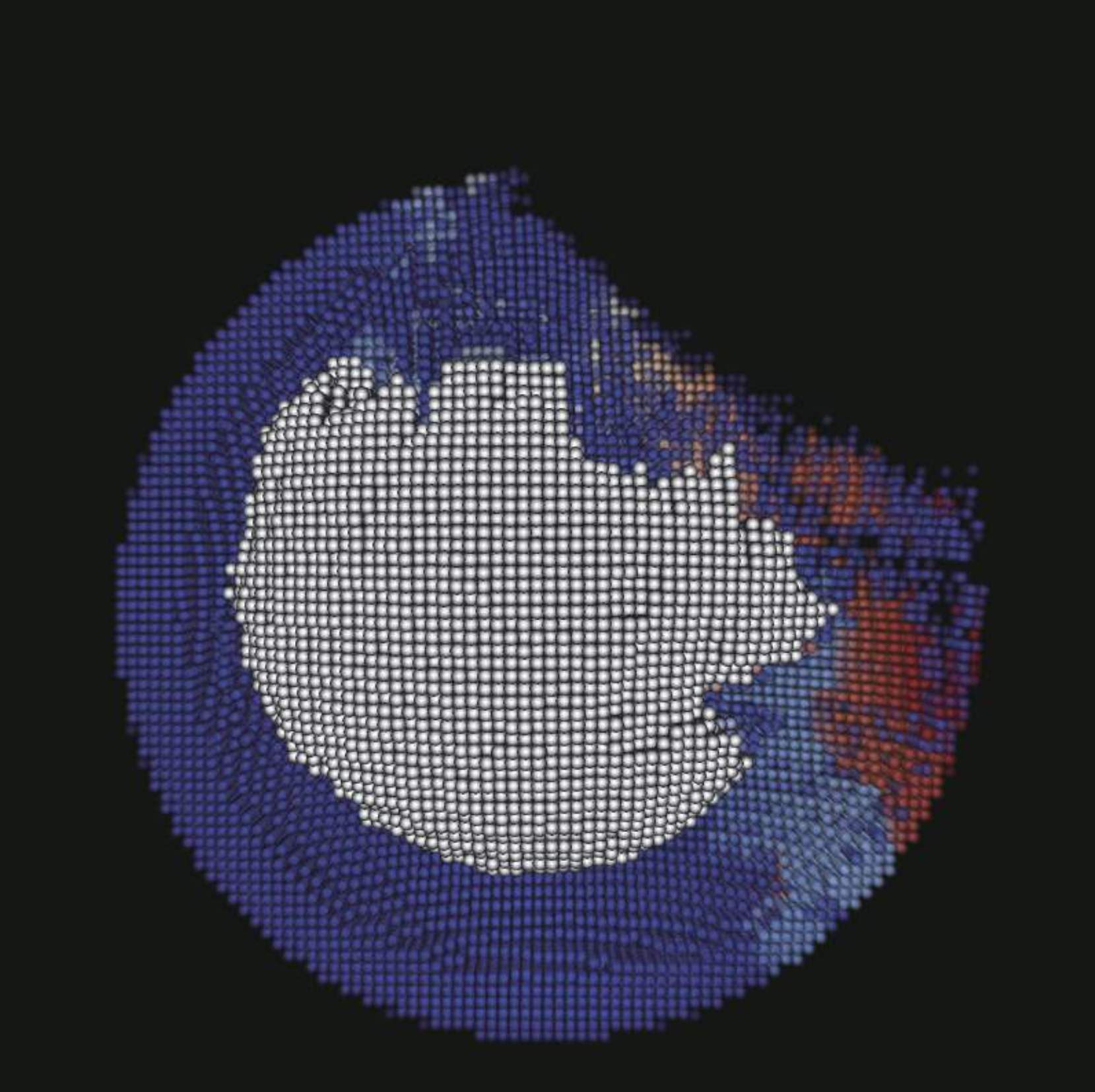}}
\end{minipage}
&
\begin{minipage}[c]{0.1\columnwidth}
    \centering
    {\includegraphics[width=\linewidth]{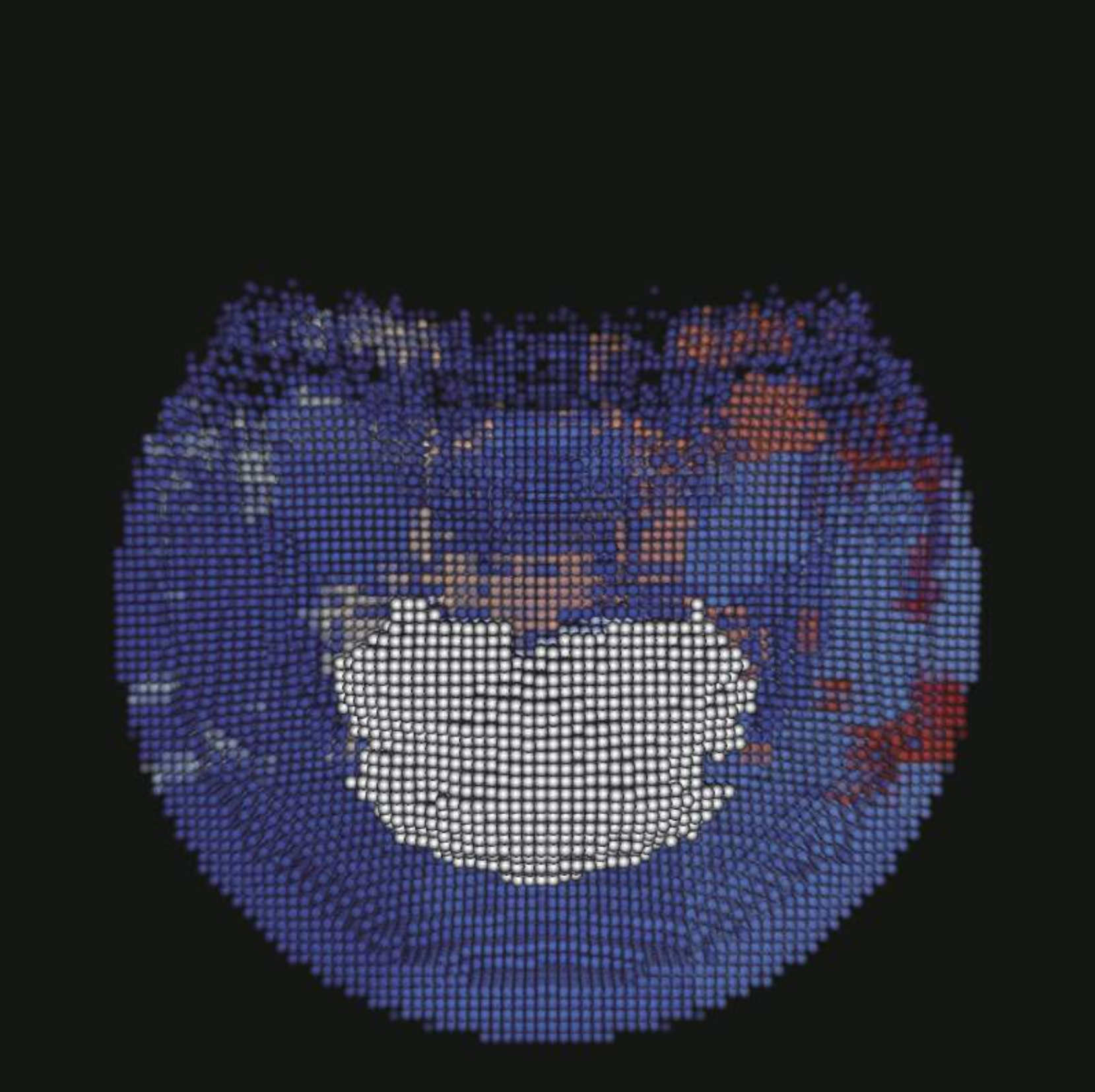}}
\end{minipage}
&
\begin{minipage}[c]{0.1\columnwidth}
    \centering
    {\includegraphics[width=\linewidth]{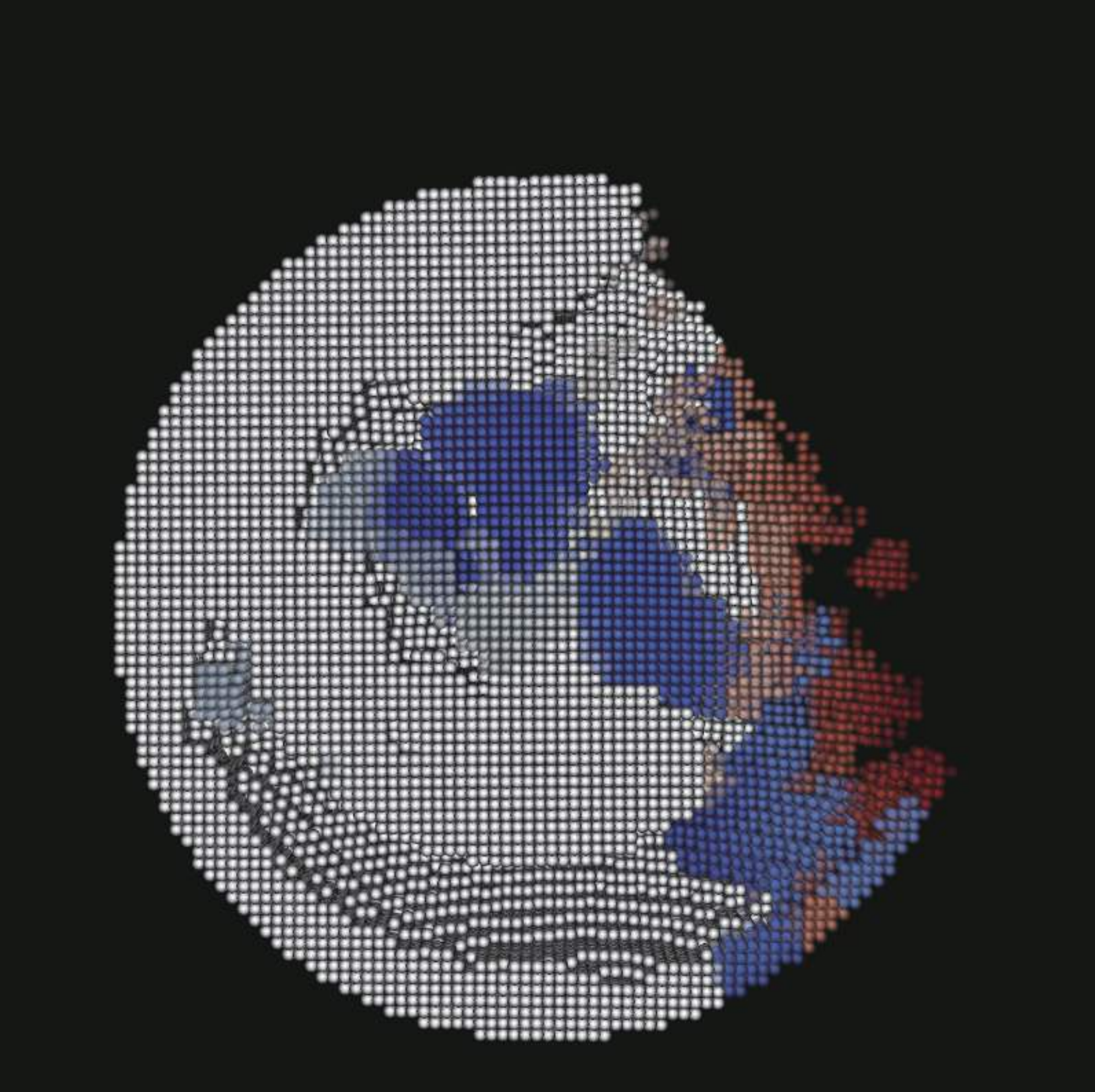}}
\end{minipage}
&
\begin{minipage}[c]{0.1\columnwidth}
    \centering
    {\includegraphics[width=\linewidth]{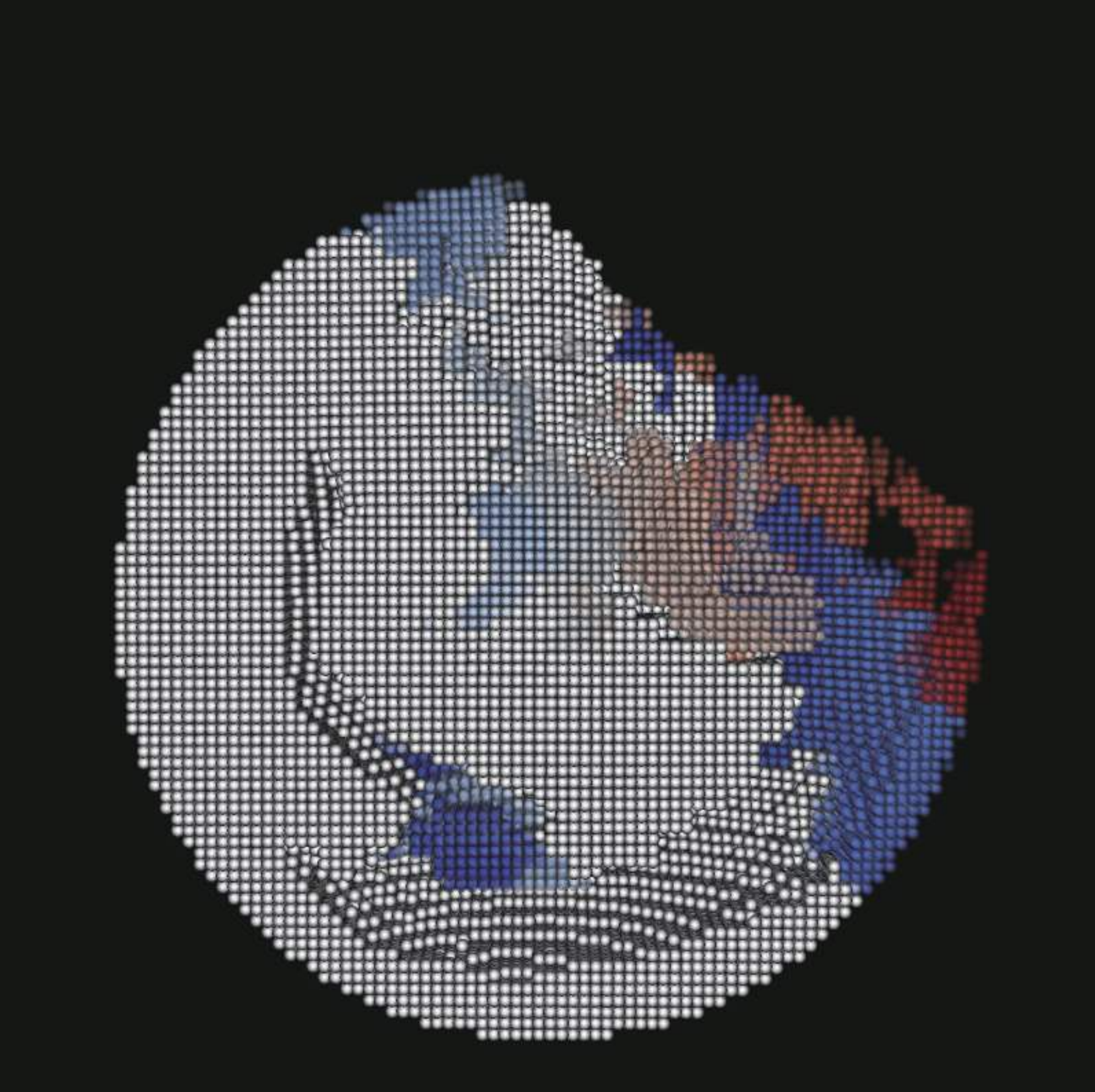}}
\end{minipage}
&
\begin{minipage}[c]{0.1\columnwidth}
    \centering
    {\includegraphics[width=\linewidth]{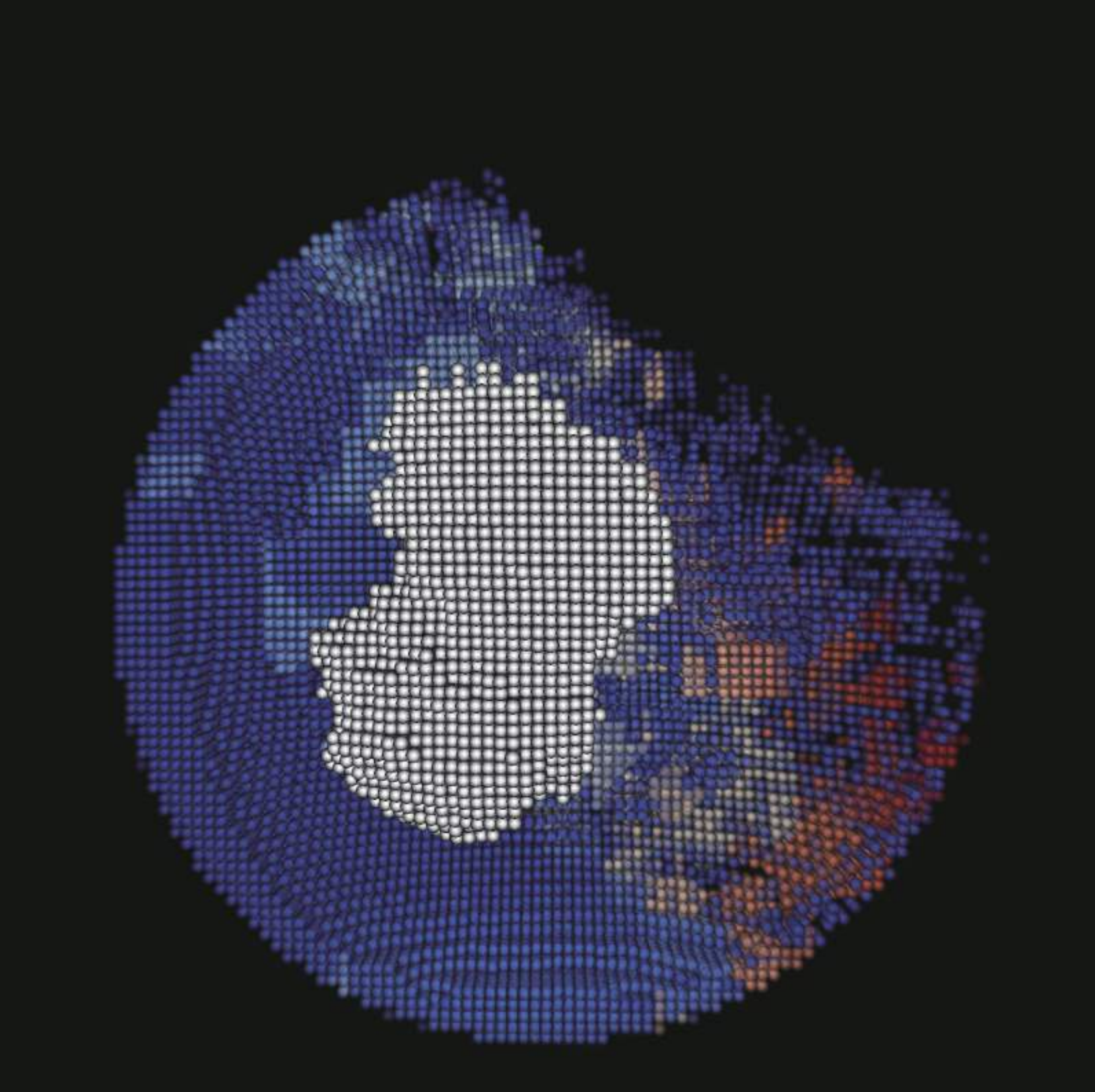}}
\end{minipage}
&
\begin{minipage}[c]{0.1\columnwidth}
    \centering
    {\includegraphics[width=\linewidth]{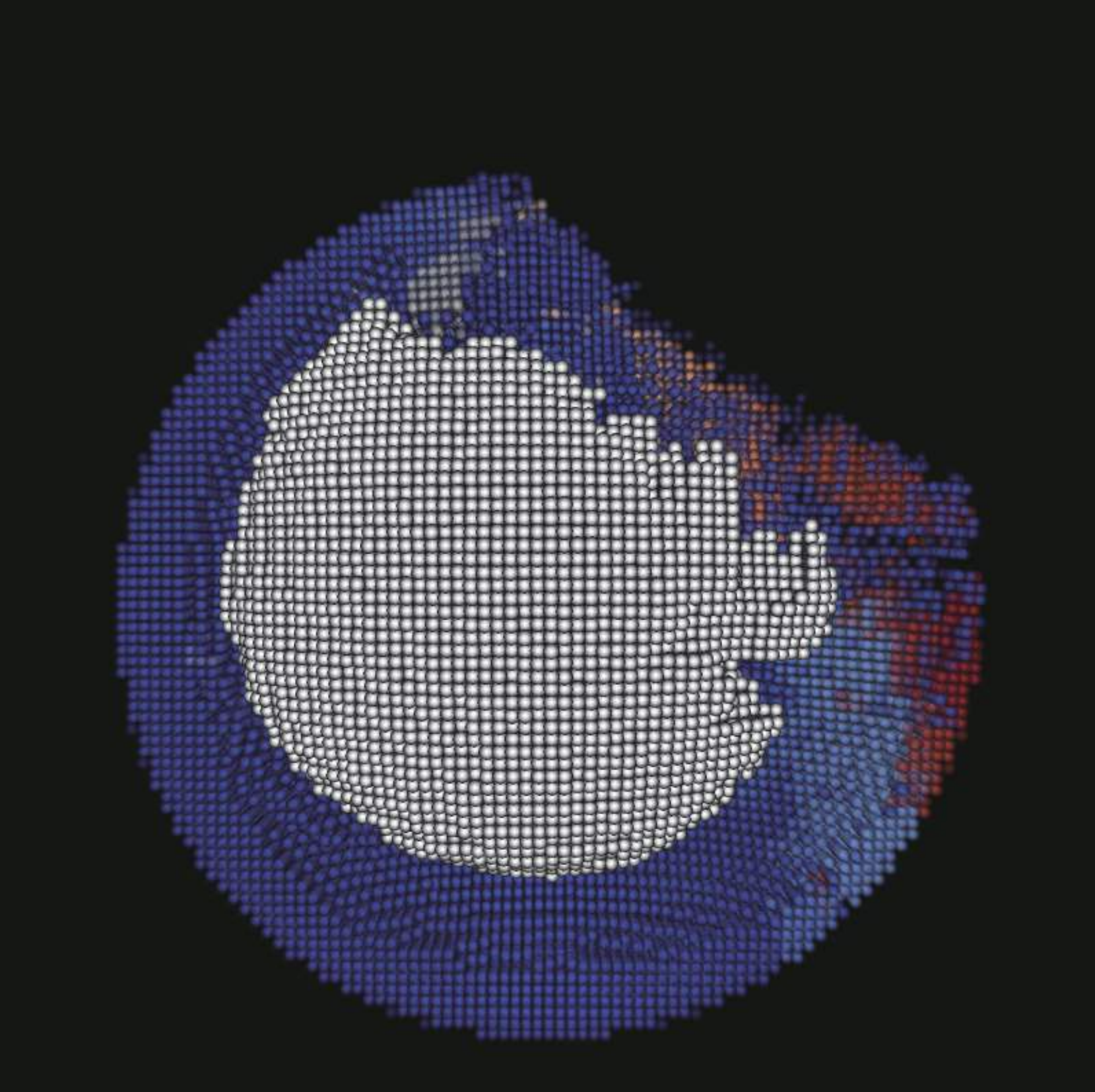}}
\end{minipage}
&
\begin{minipage}[c]{0.1\columnwidth}
    \centering
    {\includegraphics[width=\linewidth]{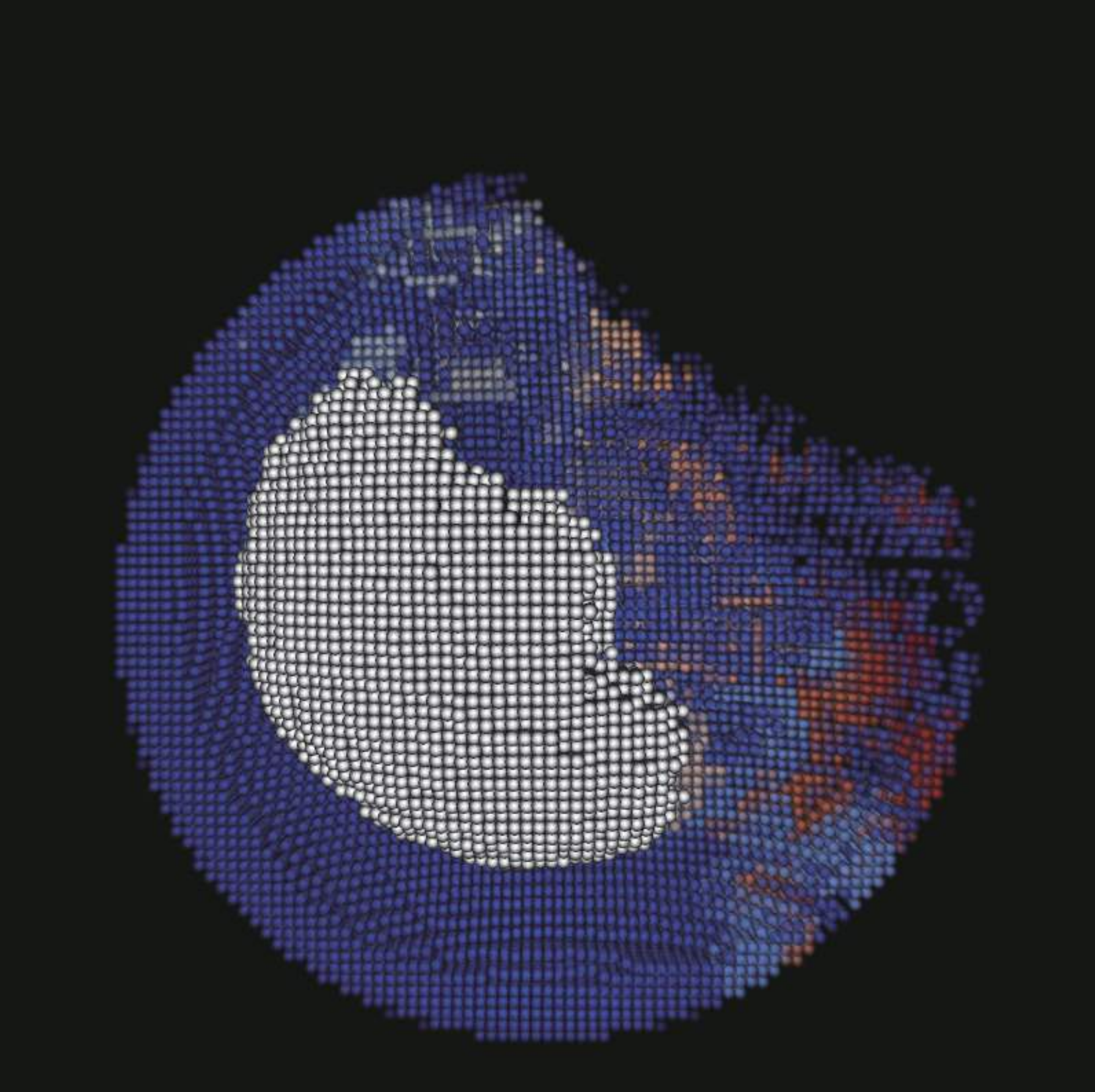}}
\end{minipage}
\\
\arrayrulecolor{black}
\bottomrule
\end{tabular}
\begin{tablenotes}
\item[a] Nominal case: impact angle \SI{30}{\degree}, velocity \SI{3.2}{\kilo\metre\per\second}, \SI{0.2}{\gram} projectile.
\item[b] Cross-sectional views with the same colorbar of damage in Fig.~\ref{fig:precision2}.
\item[c] Shape and position of the largest fragment (white points) and other fragments (color-coded based on their respective identifiers).  For the core-shaped largest fragment, its complete shape is displayed; otherwise, like other fragments, a cross-sectional view along the symmetrical plane is presented.
\end{tablenotes}
\end{threeparttable}
\end{table}

Uncertainty is an inherent aspect of experimental measurements. 
To investigate its influence, this section systematically extends parameters beyond the established nominal case -- specifically, the angle of the impact (both vertical and \ang{45}), the velocity of the impactor (\SI{2.7}{\kilo\metre\per\second} and \SI{3.7}{\kilo\metre\per\second}), and mass of the projectile (altered to \SI{0.175}{\gram} or \SI{0.225}{\gram}, by adjusting material density while maintaining projectile volume constant).
The variables scrutinized include the momentum of the impactor, the likelihood and attributes of core formation, the relative mass and velocity of the core, and the overall distribution of damage and fragments, as elucidated in Table~\ref{tab:initial}.

Comparing the situations that have the same impact angle $\alpha = \ang{30}$ (measured from the surface normal, as illustrated in Fig.~\ref{fig:experiment}), the condition of the core-shaped fragment displays a direct correlation with the initial momentum of the projectile.
An increase in momentum generally yields a smaller core with a higher velocity, reflecting the notion that a more forceful impact intensifies fracture severity and enhances momentum transfer.
Conversely, a lower momentum tends to produce a larger, slower-moving core.
However, the lower momentum of the projectile may also fail to develop a core-shaped fragment, if the stress wave is too weak to complete a shell-like damage pattern on the side opposite to the impact point.
Consequently, in the absence of the shell-like damage pattern to insulate the stress, the velocity of the largest fragment significantly exceeds that of any formed cores.
Besides, for similar changes in momentum, the outcomes demonstrate a greater sensitivity to variations in velocity than to alterations in mass.
This increased sensitivity to velocity may be attributed to its relationship with the time step, a critical factor for integration stability within numerical algorithms discussed in Section~\ref{secA_2}.

In scenarios where the impact angle is varied, while the overall momentum of the projectile remains constant, the distribution of fragments follows the established relationship with the momentum component normal to the impact point.
In the case of a vertical impact, the impact momentum is equivalent to the normal component of a \ang{30} impact with $0.91 \si{\kilogram\metre\per\second}$, resulting in intense fragmentation.
The mass of the core is less than one-quarter of that in the nominal case \citep{Nakamura1991}, providing a closer match to experimental findings \citep{Fujiwara1980}.
At an angle of \ang{45},  the normal component of impact momentum is effectively reduced to that of a \ang{30} impact with $0.52 \si{\kilogram\metre\per\second}$,  insufficient for complete shell-like damage or core formation.
This result is in line with the tendencies observed in our simulations.
In contrast, experimental \citep{Fujiwara1980} and SPH simulation studies \citep{Benz1994} have both observed the formation of a core.
The minor discrepancy in shell-like damage development shown in MPM may originate from a potential deficiency in the smoothness of the shape functions applied within the algorithm, a topic that will be further explored in Section~\ref{sec5}.

The exploration of this sensitivity assesses the robustness of the simulation outcomes under varying conditions and also serves to validate the stability of our MPM algorithm across different impact scenarios.

\subsubsection{Sensitivity on material model}\label{sec4_1_5}

The material models established in Section~\ref{sec3} may behave diversely under a given loading procedure.
In this section, different damage and strength models are tested under the nominal impact condition, to explore the effect of material parameters on the dynamic properties.

\begin{table}
\begin{threeparttable}
\caption{Influence of damage model}
\label{tab:damage}
\setlength{\tabcolsep}{2.5pt}
\begin{tabular}{lcccccccc}
\toprule 
\multirow{3}{*}{Damage model} &
\multicolumn{5}{c}{Weibull distribution cracks} &
\multicolumn{3}{c}{Maximum principal stress} \\
\cmidrule(lr){2-6} \cmidrule(lr){7-9}
  & $m$ & $8.5$ & $8.5^{\mathrm{a}}$ & $8.5$ & $17.2^{\mathrm{b}}$
  & \multirow{2}{*}{\SI{20}{\mega\pascal}} 
  & \multirow{2}{*}{\SI{40}{\mega\pascal}} 
  & \multirow{2}{*}{\SI{80}{\mega\pascal}}\\
  & $k$ (\si{\per\metre\cubed})& \num{1.0e42} & \num{3.0e39} & \num{1.0e38} & \num{4.32e76} & & & \\
\midrule
\multicolumn{2}{l}{\parbox{3.5cm}{Average largest\\
activation stress$^{\mathrm{c}}$\\$\bar{\sigma}_{\mathrm{max}}$ (\si{\mega\pascal})}}
& $20.3$ & $40.2$ & $59.9$ & $40.1$ & $20$ & $40$ & $80$ \\
\arrayrulecolor{white}
\cmidrule(lr){1-9}
\multicolumn{2}{l}{\parbox{3.5cm}{Average smallest\\
activation stress$^{\mathrm{c}}$\\$\bar{\sigma}_{\mathrm{min}}$ (\si{\mega\pascal})}}
& $15.8$ & $31.4$ & $46.8$ & $29.8$ & $20$ & $40$ & $80$ \\
\cmidrule(lr){1-9}
\multicolumn{2}{l}{Core-shaped fragment}
& \Checkmark & \Checkmark & \XSolid & \XSolid & \XSolid & \XSolid & \XSolid \\
\cmidrule(lr){1-9}
\multicolumn{2}{l}{$m_{\mathrm{f,max}} / M_{\mathrm{tar}}$}
& $0.018$ & $0.246$ & $0.644$ & $0.208$ & $0.001$ & $0.120$ & $0.191$ \\
\multicolumn{2}{l}{$v_{\mathrm{f,max}}$ (\si{\metre\per\second})}
&  $1.939$& $2.056$ & $3.722$ & $4.579$ & $7.072$ & $6.403$ & $6.451$ \\
\multicolumn{2}{l}{Damage at \SI{100}{\micro\second}$^{\mathrm{d}}$}
&
\begin{minipage}[c]{0.1\columnwidth}
    \centering
    {\includegraphics[width=\linewidth]{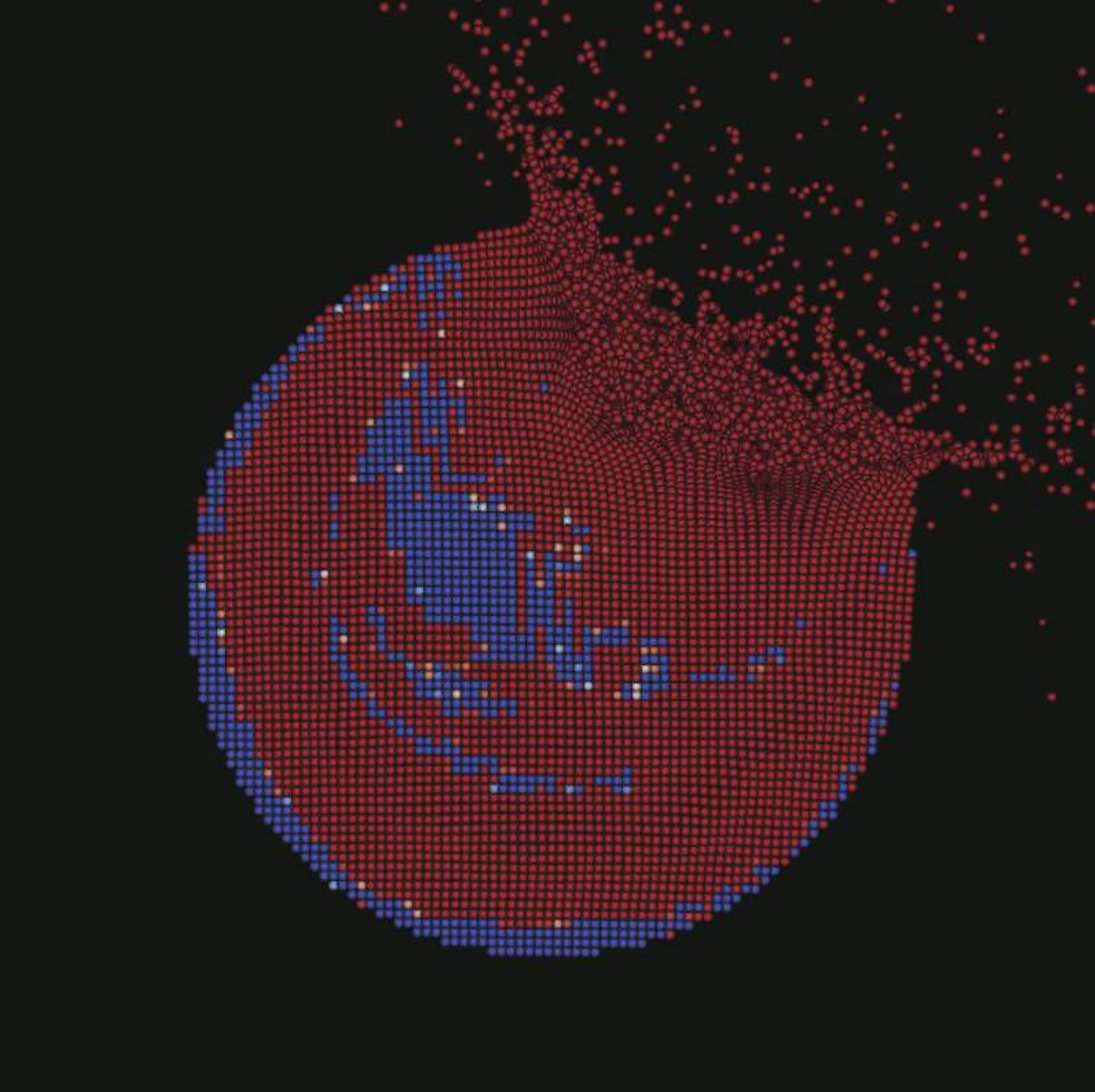}}
\end{minipage}
&
\begin{minipage}[c]{0.1\columnwidth}
    \centering
    {\includegraphics[width=\linewidth]{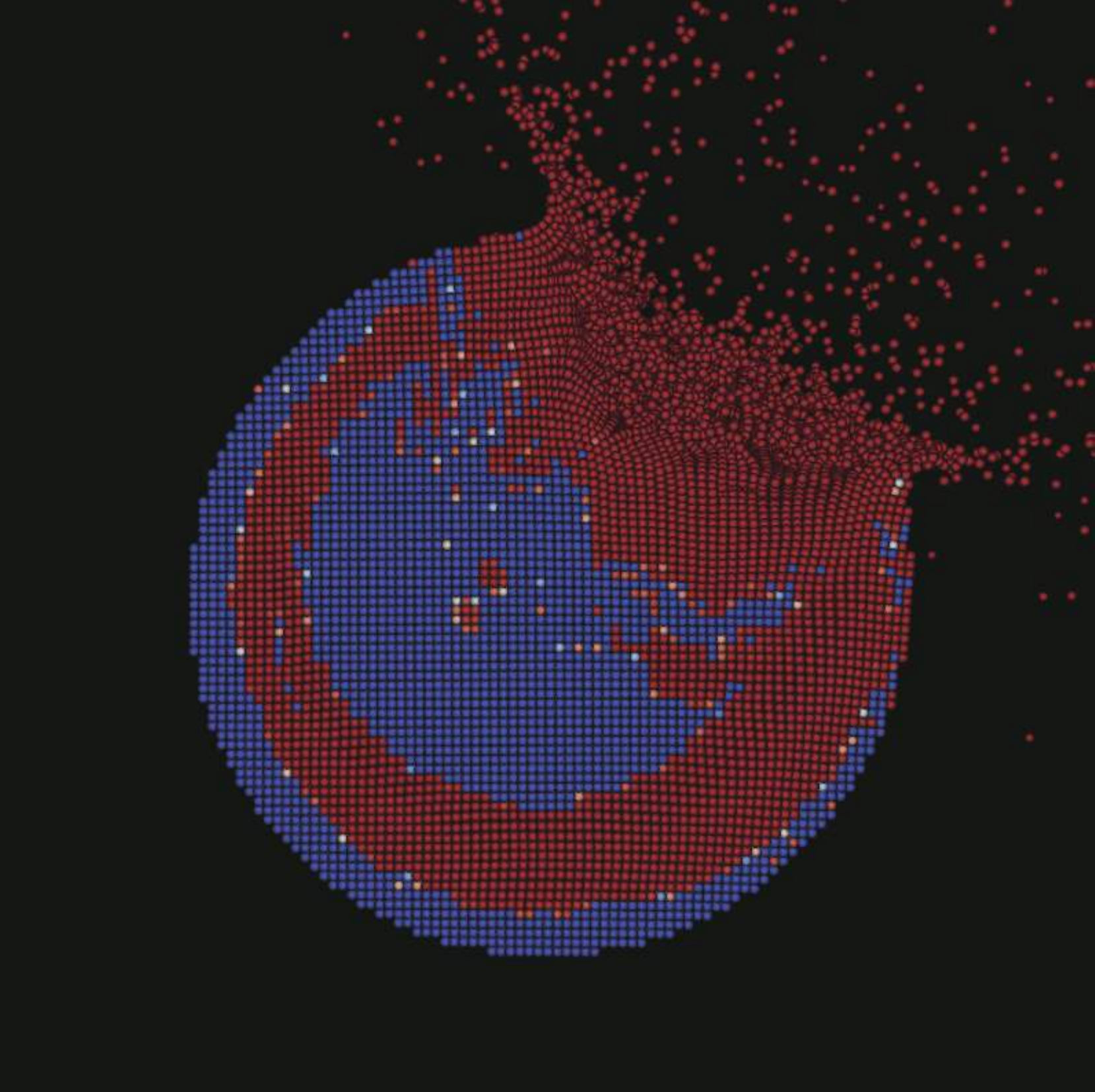}}
\end{minipage}
&
\begin{minipage}[c]{0.1\columnwidth}
    \centering
    {\includegraphics[width=\linewidth]{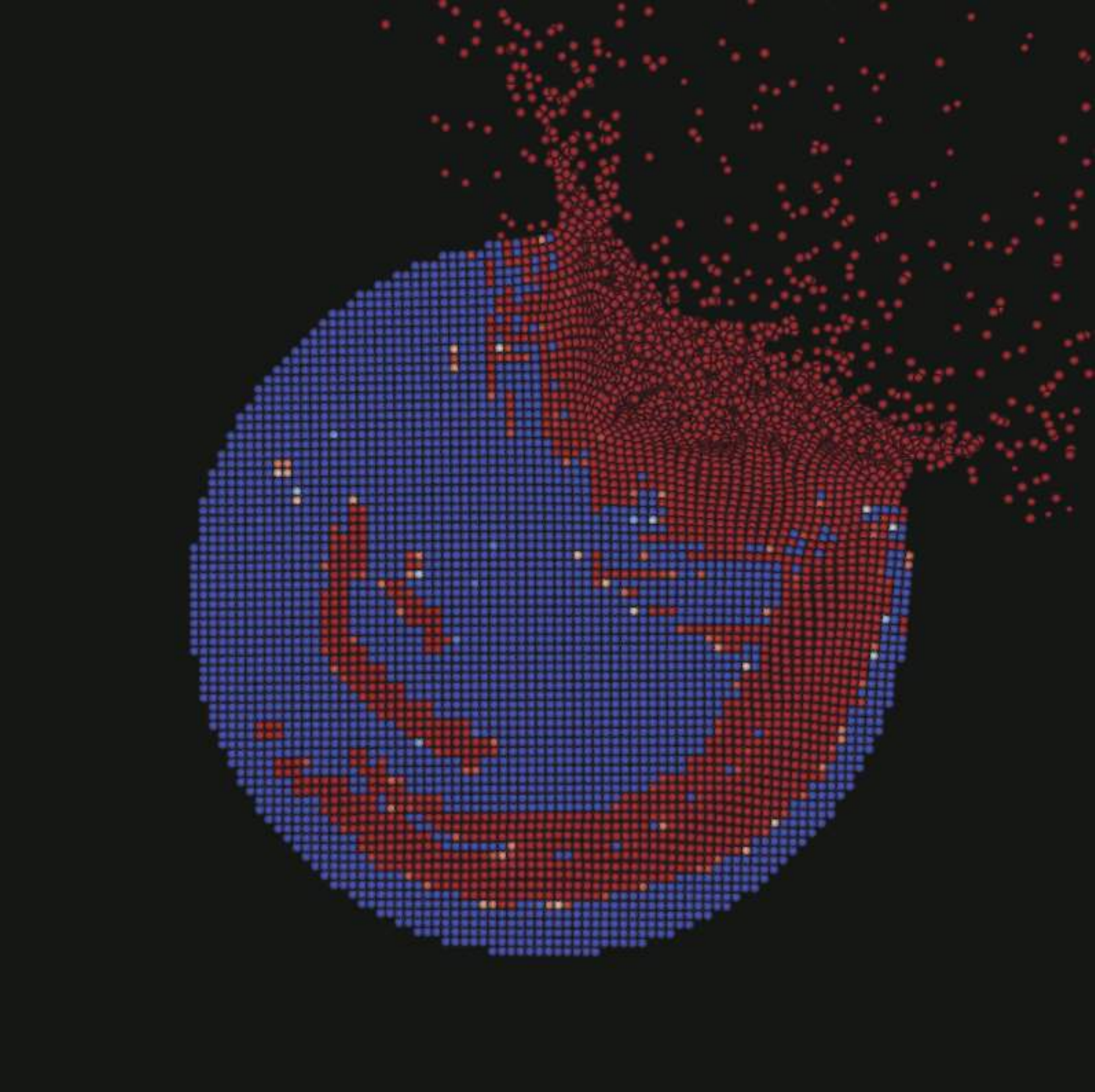}}
\end{minipage}
&
\begin{minipage}[c]{0.1\columnwidth}
    \centering
    {\includegraphics[width=\linewidth]{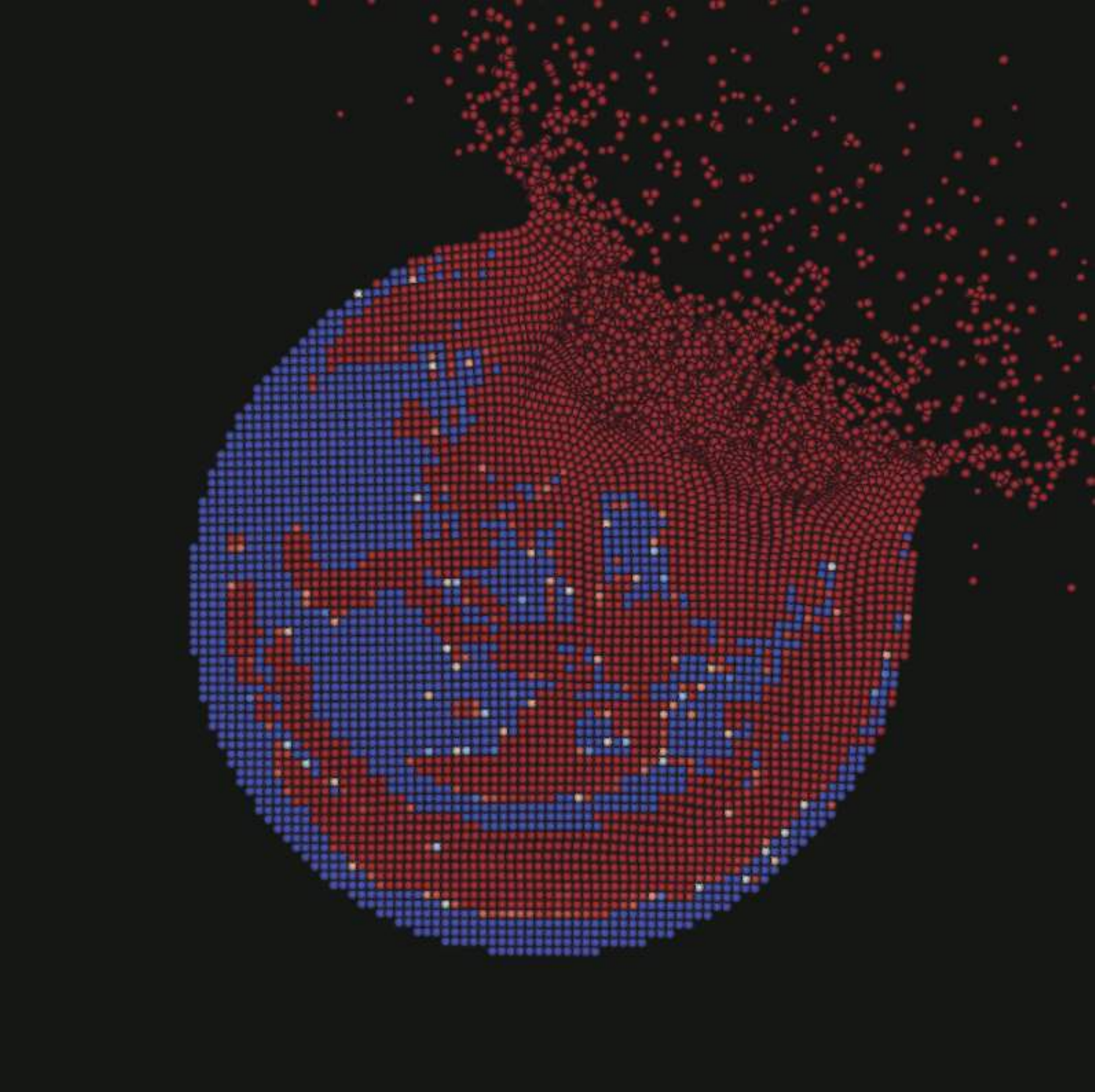}}
\end{minipage}
&
\begin{minipage}[c]{0.1\columnwidth}
    \centering
    {\includegraphics[width=\linewidth]{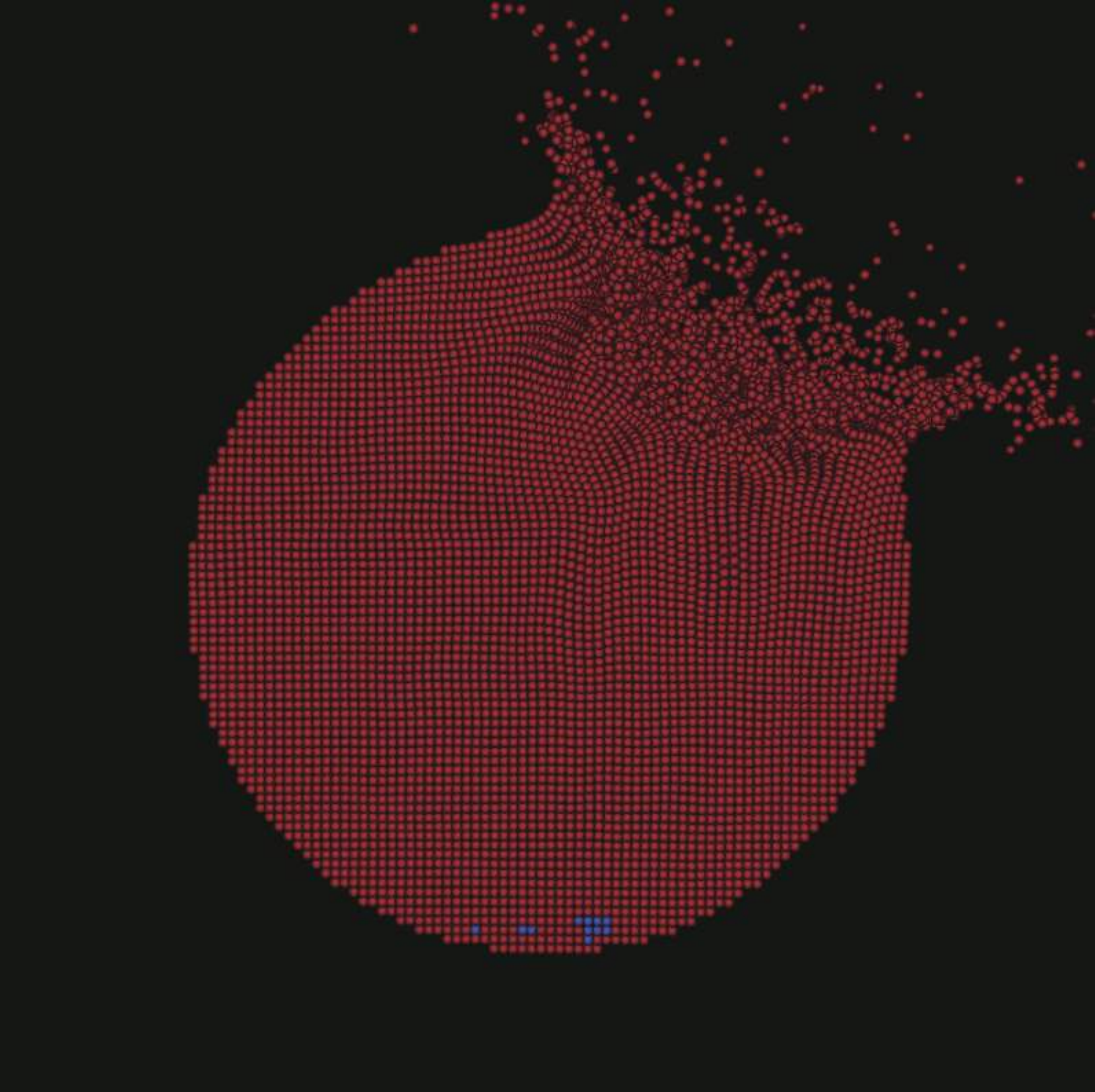}}
\end{minipage}
&
\begin{minipage}[c]{0.1\columnwidth}
    \centering
    {\includegraphics[width=\linewidth]{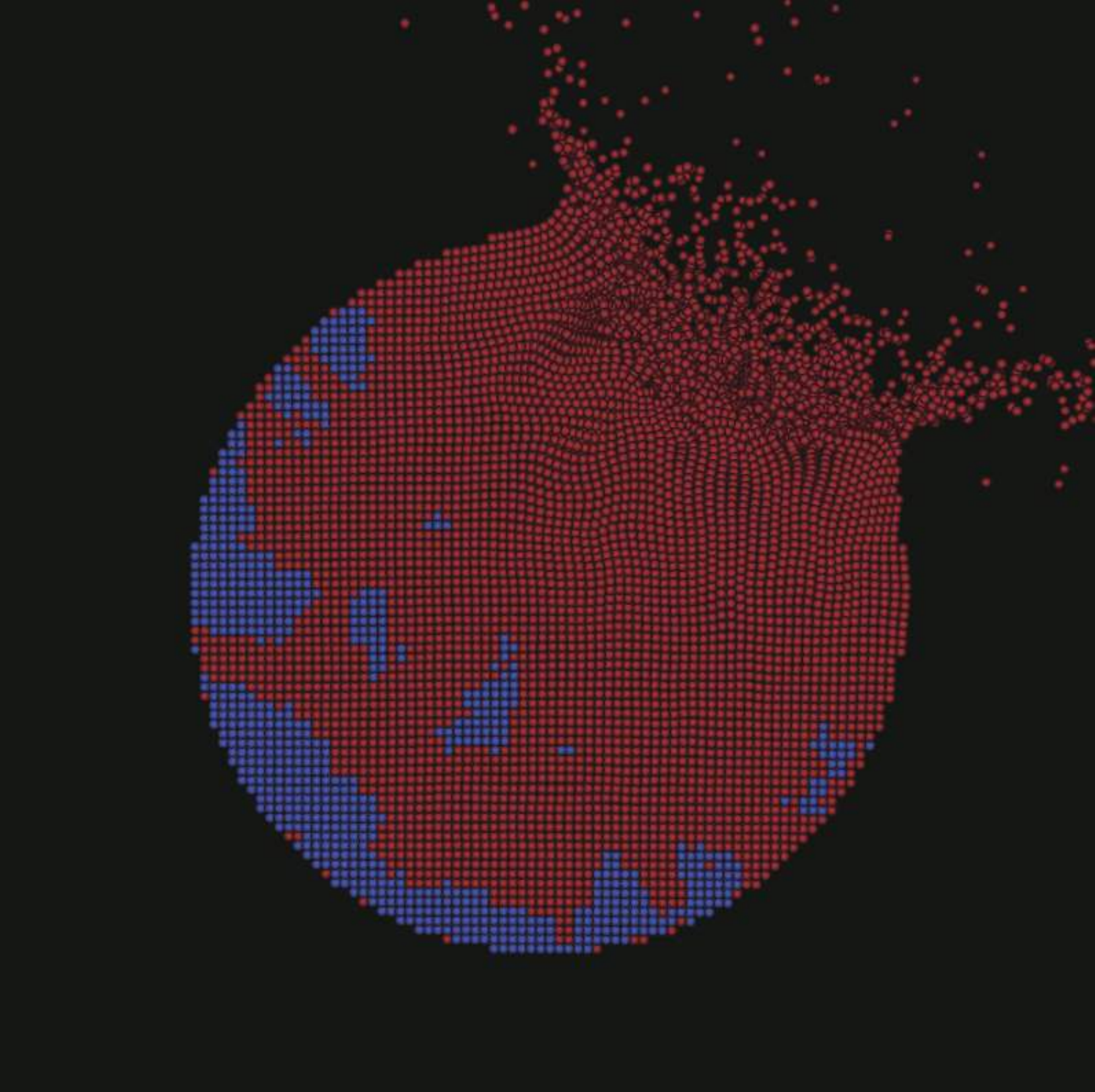}}
\end{minipage}
&
\begin{minipage}[c]{0.1\columnwidth}
    \centering
    {\includegraphics[width=\linewidth]{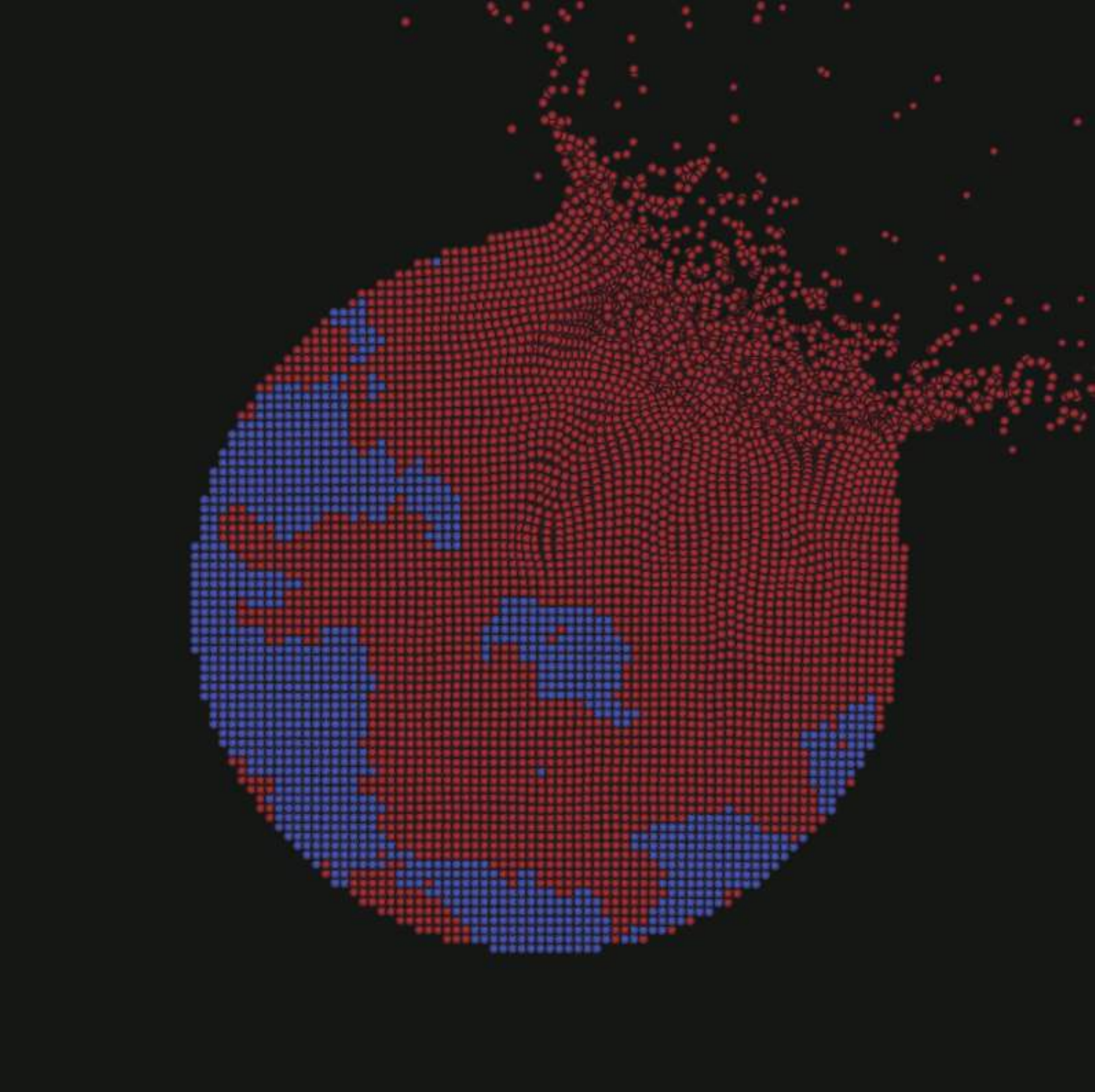}}
\end{minipage}
\\
\cmidrule(lr){1-9}
\multicolumn{2}{l}{Largest fragment$^{\mathrm{e}}$}
&
\begin{minipage}[c]{0.1\columnwidth}
    \centering
    {\includegraphics[width=\linewidth]{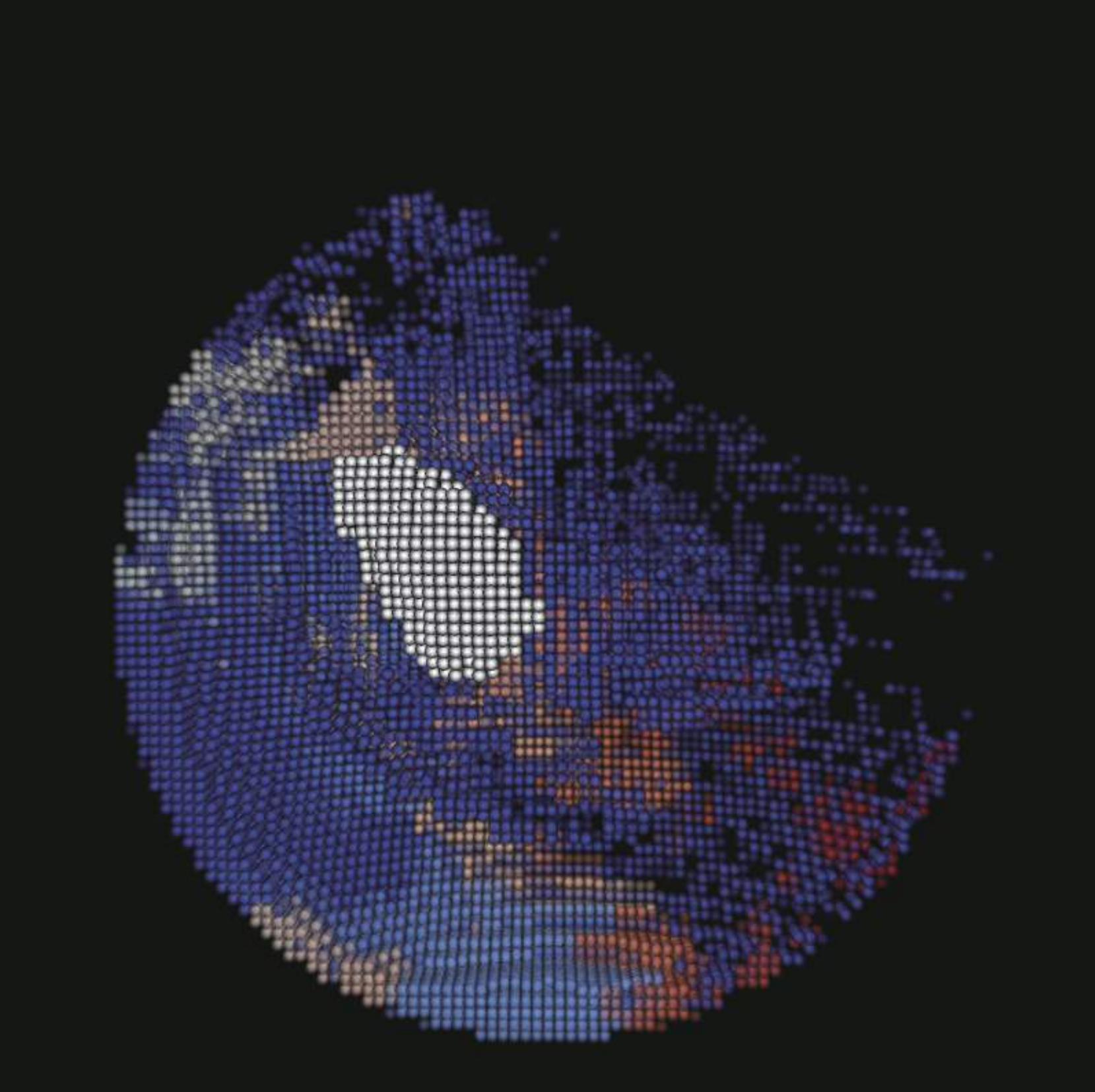}}
\end{minipage}
&
\begin{minipage}[c]{0.1\columnwidth}
    \centering
    {\includegraphics[width=\linewidth]{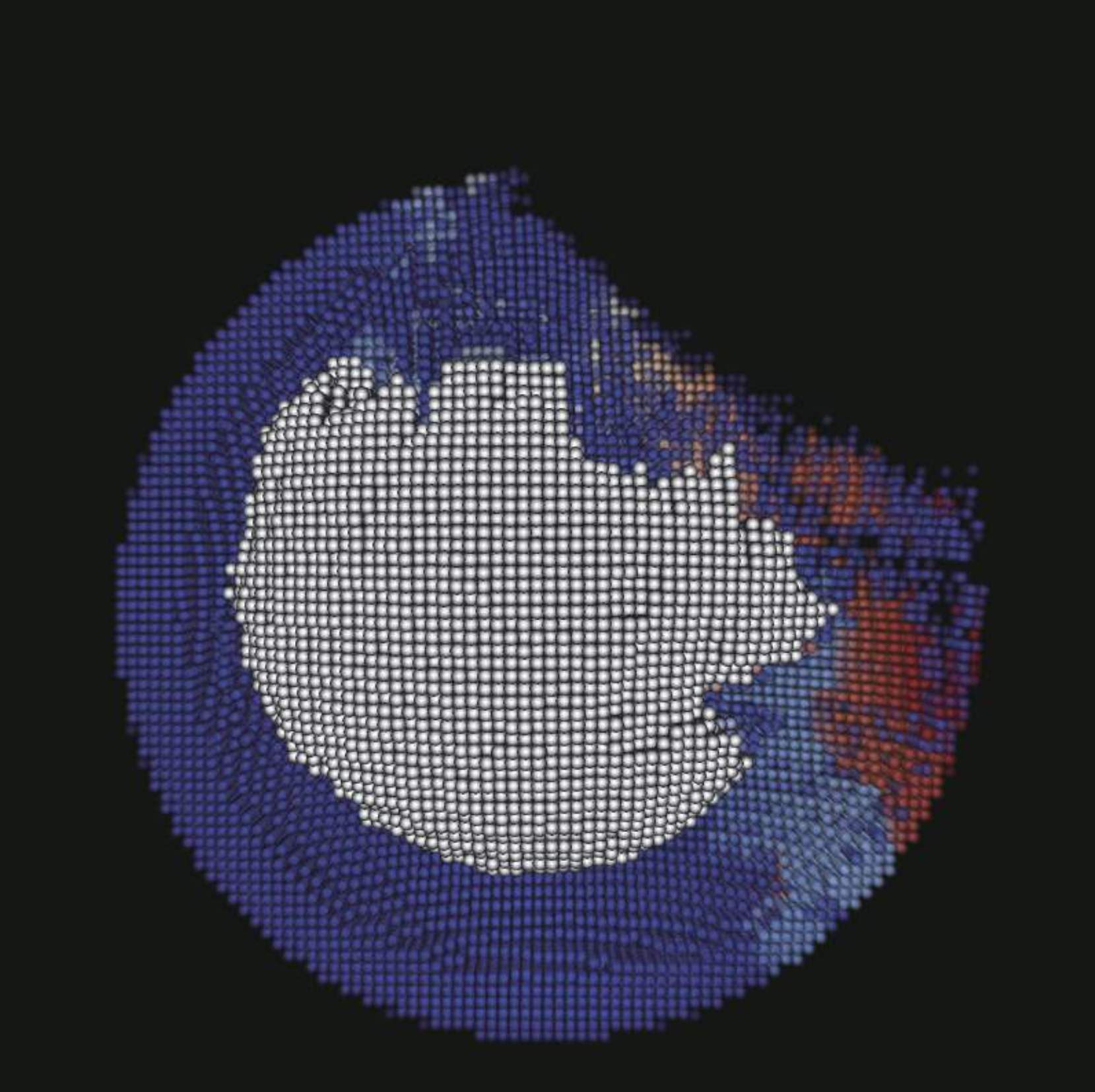}}
\end{minipage}
&
\begin{minipage}[c]{0.1\columnwidth}
    \centering
    {\includegraphics[width=\linewidth]{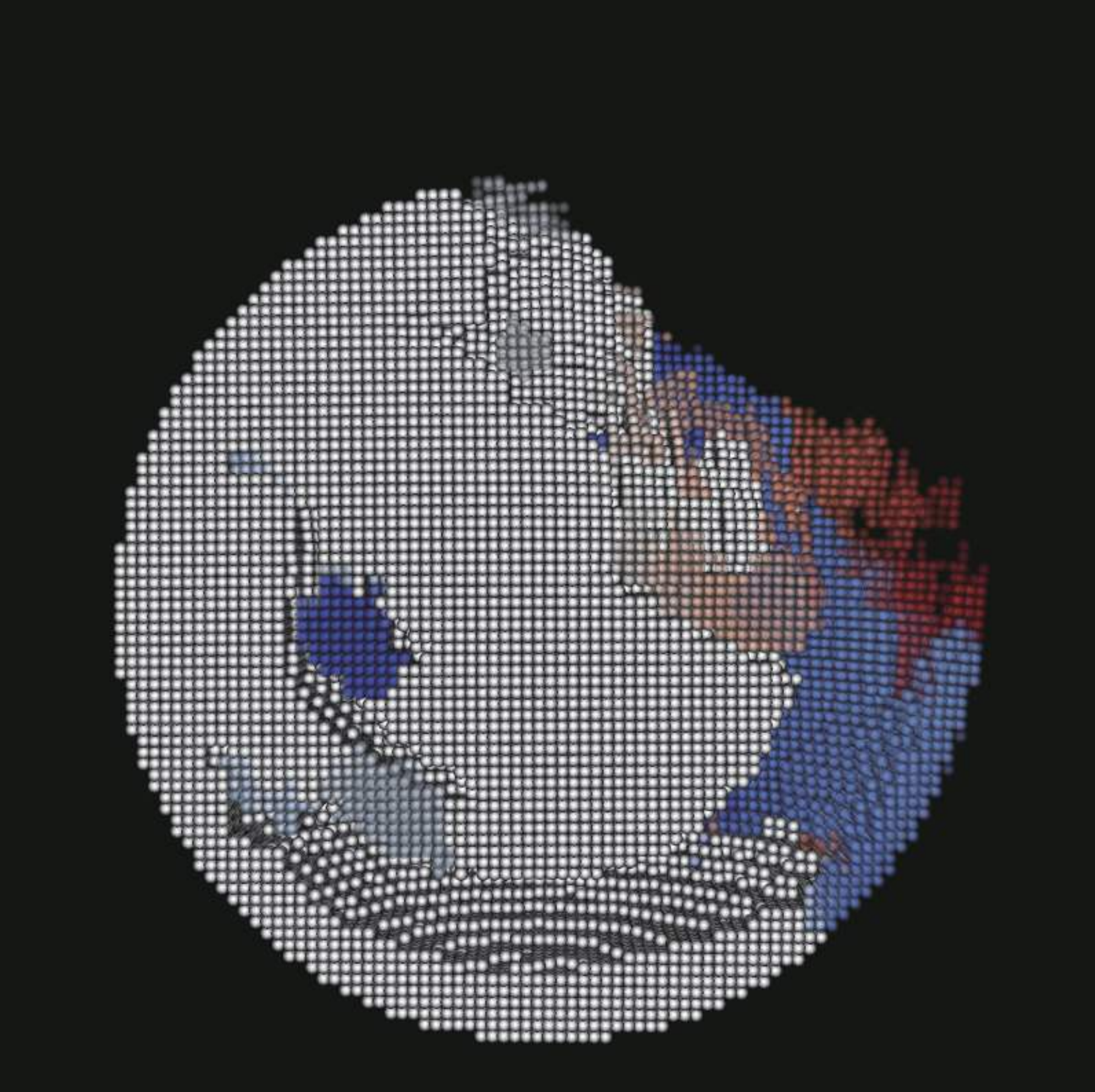}}
\end{minipage}
&
\begin{minipage}[c]{0.1\columnwidth}
    \centering
    {\includegraphics[width=\linewidth]{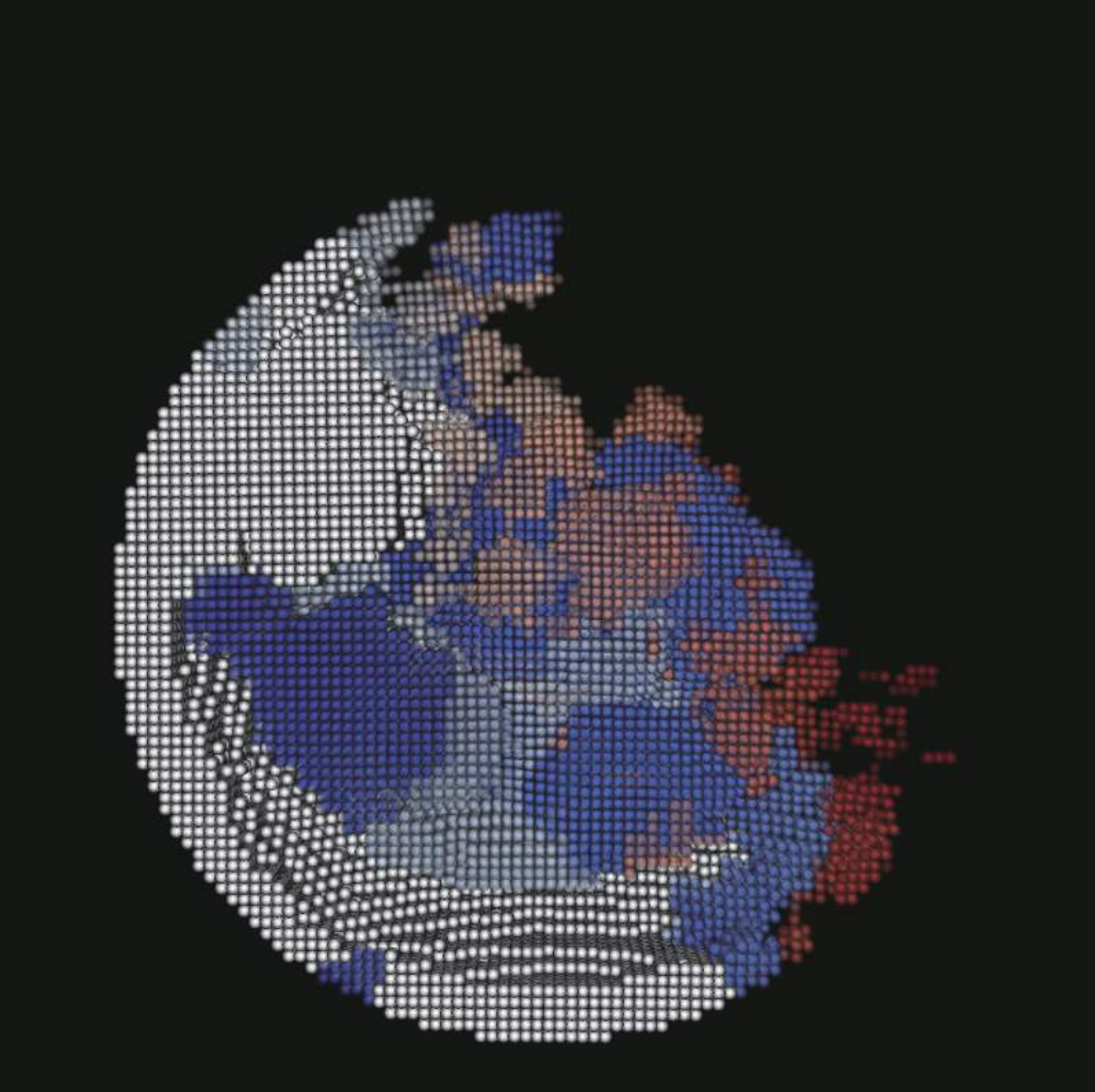}}
\end{minipage}
&
\begin{minipage}[c]{0.1\columnwidth}
    \centering
    {\includegraphics[width=\linewidth]{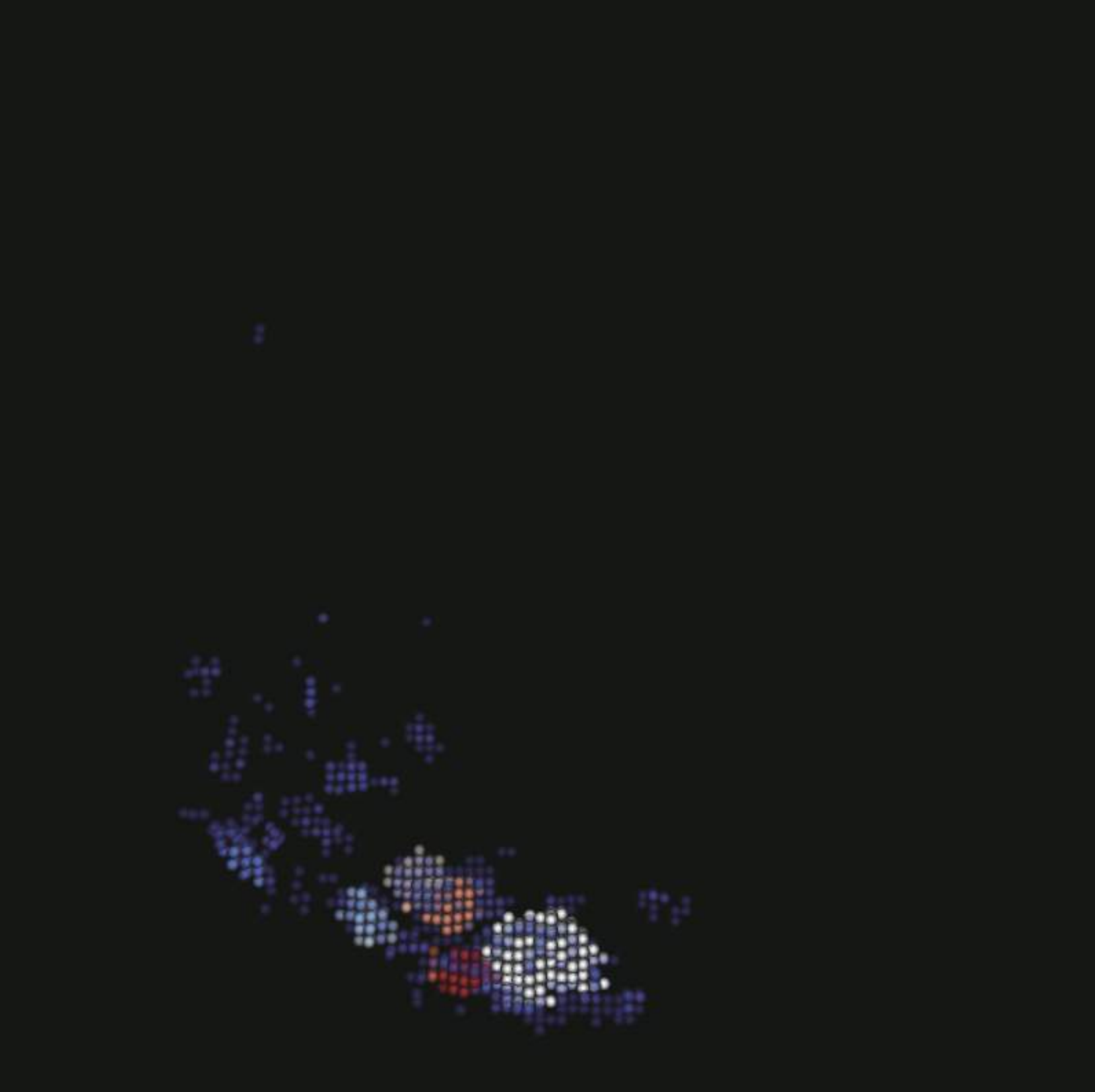}}
\end{minipage}
&
\begin{minipage}[c]{0.1\columnwidth}
    \centering
    {\includegraphics[width=\linewidth]{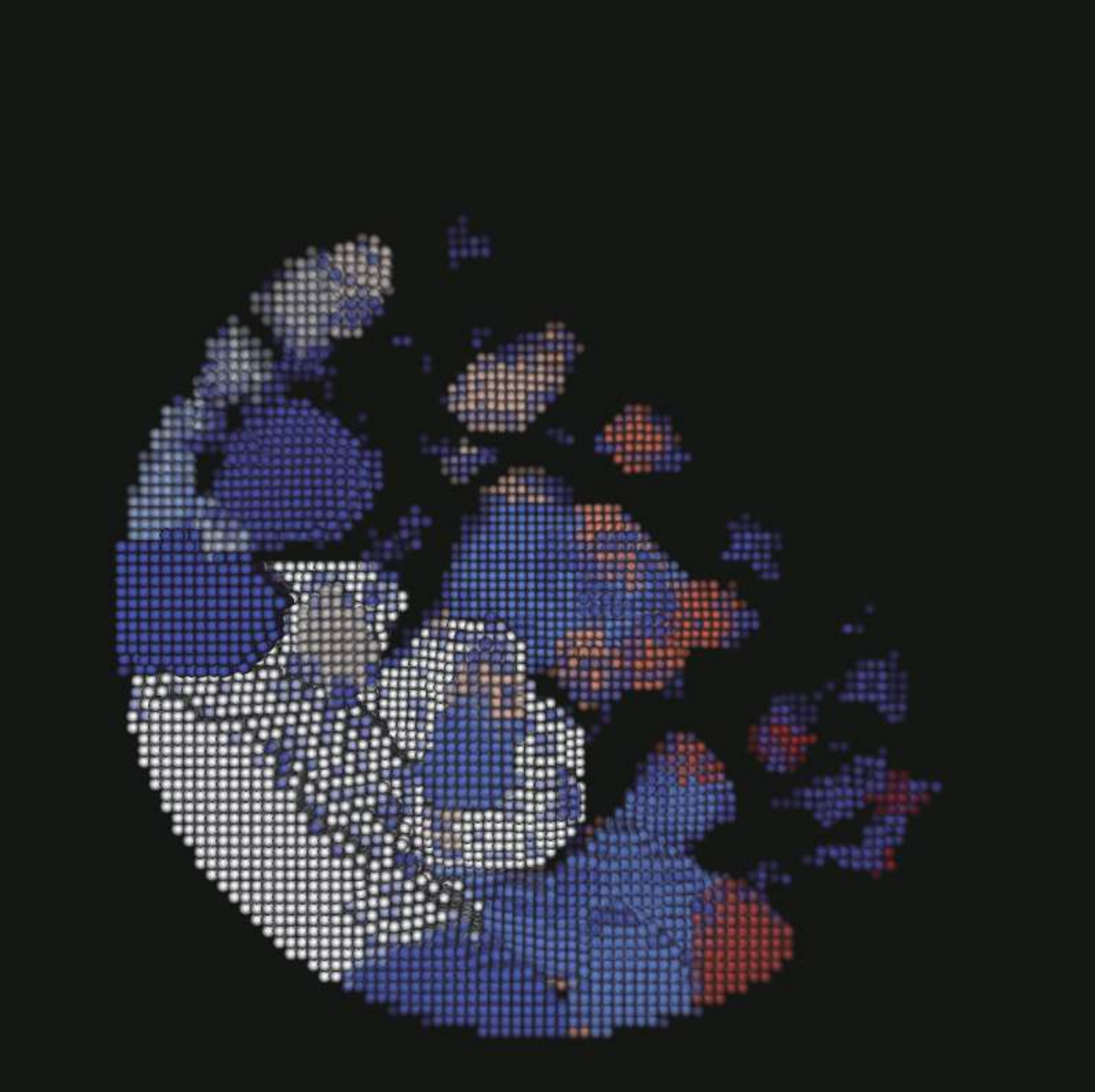}}
\end{minipage}
&
\begin{minipage}[c]{0.1\columnwidth}
    \centering
    {\includegraphics[width=\linewidth]{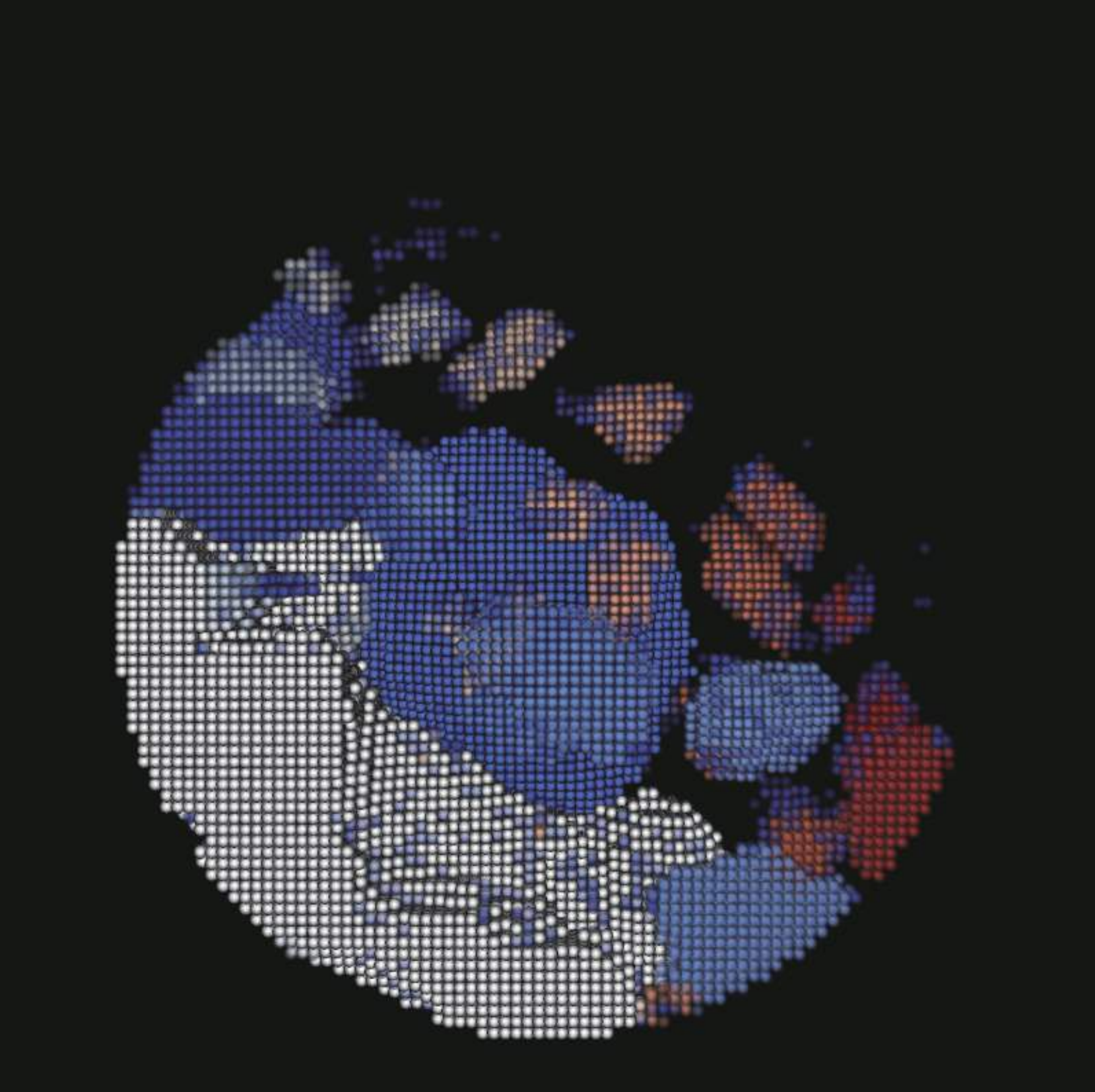}}
\end{minipage}
\\
\arrayrulecolor{black}
\bottomrule
\end{tabular}
\begin{tablenotes}
\item[a] Nominal case.
\item[b] Here the total number of flaws is $n_{\mathrm{tot}} = 1.0\times10^{19}$, to ensure $n_{\mathrm{tot}} > 10^m$.
\item[c] Transforming activation strain to stress with Young's modulus $E_{\mathrm{Y}} = 53100$ \si{\mega\pascal}.
\item[d, e] Same as Table~\ref{tab:initial}.
\end{tablenotes}
\end{threeparttable}
\end{table}

Considering the damage model, the influence of the parameters in two mechanisms, the growth of incipient cracks and the brittle fracture, are evaluated in Table~\ref{tab:damage}.

For the first part, the Weibull parameters $k$ and $m$ are the most pivotal in the Weibull distribution, as they characterize the distribution of flaws.
Consequently, determining these parameters defines the average onset and the extent of damage accumulation within the material, as described in Section~\ref{sec3_3}.
A dimensionless number $\phi = \ln(kV) / m$ can well represent the strength, with a larger $\phi$ indicating a weaker material \citep{Benz1994}.

Using the nominal case with $m = 8.5$ as a baseline, variations in $k$ are implemented to change the average complete failure strength of the target by $\pm \SI{20}{\mega\pascal}$.
This variation does not alter the pattern of damage evolution, but does influence the extent of the damage zone.
A lower strength results in a larger damaged area, leaving a smaller and slower intact core.
Conversely, a higher strength prevents the development of a closed damaged shell, leading to larger fragments moving at higher velocities without forming a core.

If both $k$ and $m$ are altered while maintaining a constant $\phi$, the target's strength theoretically remains unchanged, and the damage development should be similar.
Yet, when we adopt $m = 17.2$ and the corresponding $k$ value based on experimental results \citep{Nakamura2007}, the damage profile differs from the nominal case.
Frame-by-frame analysis of numerical results reveals that these discrepancies stem from the requirements of the Weibull damage algorithm, where $N_{\mathrm{tot}}$ must exceed $10^m$.
When the value of $m$ is elevated, the average number of incipient flaws assigned to each material point increases, and the activation of flaws under the same stress-strain is less frequent, according to Eq.~(\ref{eq:nact}).
This leads to insufficient damage development (again on the side of the target opposite the impact point), causing subsequent stress reflections within the incomplete damage shell and resulting in chaotic damage.
Moreover, the damage strength of the nominal case material (\SI{40}{\mega\pascal}) is higher than both the experimentally determined value (\SI{19.43}{\mega\pascal}) \citep{Nakamura2007} and the strength used in SPH simulations (\SI{13}{\mega\pascal}) \citep{Benz1994}.
This discrepancy may be related to the algorithmic design of the Weibull distribution and calls for further improvement.

In the brittle fracture mode, the maximum principal stress solely determines the material's strength.
However, if the brittle fracture strength matches the equivalent damage strength of Weibull accumulation, the brittle fracture will always cause more severe damage, yielding a greater number of irregularly shaped small fragments.
This is because brittle fracture results in complete failure and instantaneous stress release upon strong impact.
In contrast, in damage accumulation mode, the damage is accumulated at the current time step while the stress is not adjusted until the following time step, which means that the response of stress to damage is delayed.
This phenomenon, akin to a viscous effect, filters the shock wave and prevents the material from damage.

This analysis of the damage model underscores the intricate interplay between damage development and stress wave propagation.
It highlights the necessity to fine-tune material models to better understand the complexity inherent in material behavior under dynamic loading.
Similarly, an exploration into the parameters within strength models is warranted, as listed in Table~\ref{tab:strength}. 

\begin{table}
\begin{threeparttable}
\caption{Influence of strength model}
\label{tab:strength}
\setlength{\tabcolsep}{2.5pt}
\begin{tabular}{lcccccc}
\toprule 
\multirow{2}{*}{Strength model} &
\multicolumn{3}{c}{Linear hardening $\sigma_{\mathrm{Y}}$} &
\multicolumn{3}{c}{Modified Lundborg$^{\mathrm{a}}$ $Y_0$} \\
\cmidrule(lr){2-4} \cmidrule(lr){5-7}
  & \SI{30}{\mega\pascal} & \SI{60}{\mega\pascal} & \SI{3500}{\mega\pascal}$^{\mathrm{b}}$
  & \SI{10}{\mega\pascal} & \SI{60}{\mega\pascal} & \SI{600}{\mega\pascal} \\
\midrule
Core-shaped fragment
& \XSolid & \Checkmark & \Checkmark & \XSolid & \Checkmark & \Checkmark \\
$m_{\mathrm{f,max}} / M_{\mathrm{tar}}$
& $0.221$ & $0.474^{\mathrm{c}}$ & $0.246$ & $0.617$ & $0.301$ & $0.221$ \\ 
$v_{\mathrm{f,max}}$ (\si{\metre\per\second})
& $4.448$ & $4.878$ & $2.056$ & $3.157$ & $1.046$ & $2.087$ \\ 
Damage at \SI{100}{\micro\second}$^{d}$
&
\begin{minipage}[c]{0.1\columnwidth}
    \centering
    {\includegraphics[width=\linewidth]{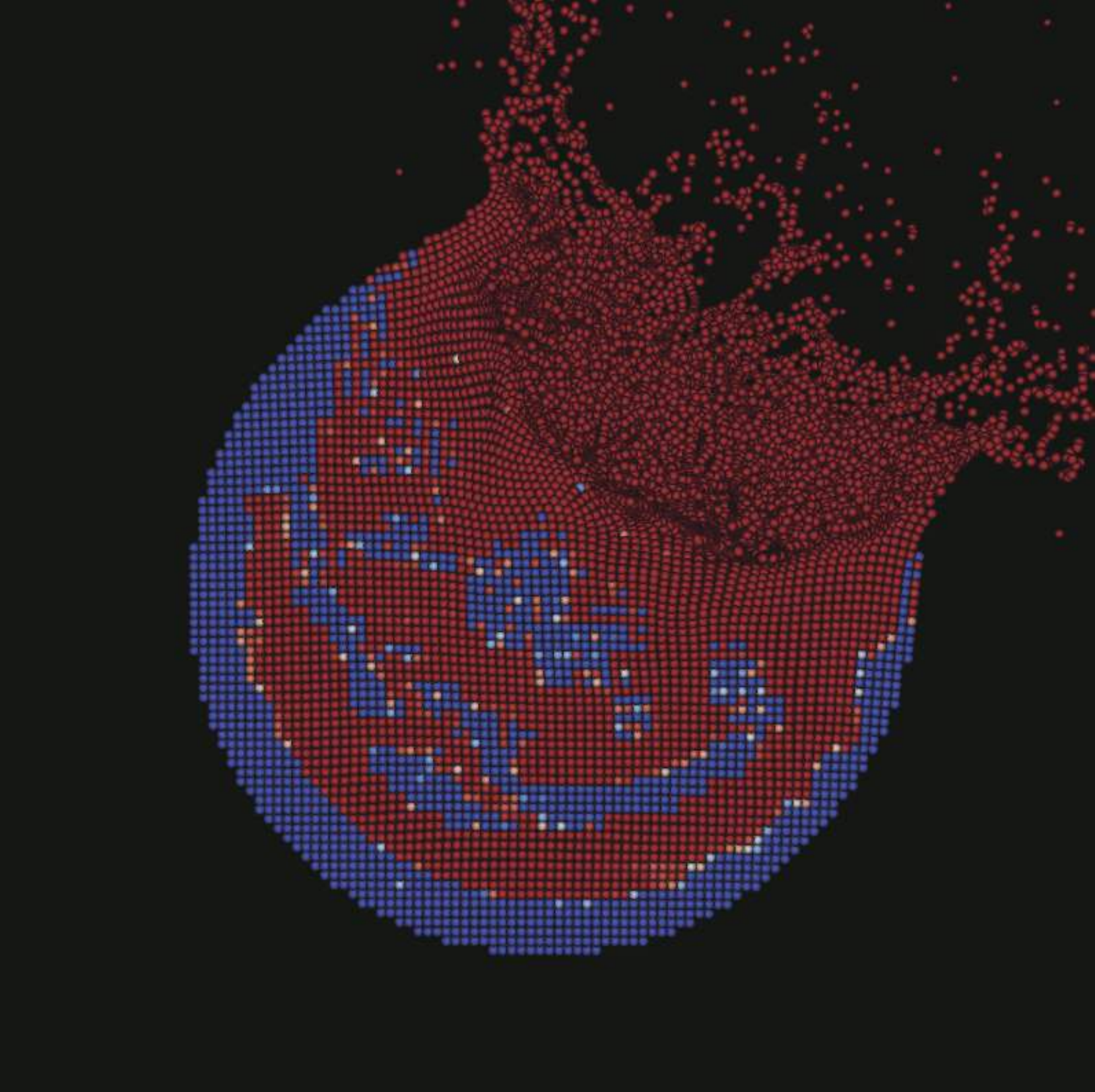}}
\end{minipage}
&
\begin{minipage}[c]{0.1\columnwidth}
    \centering
    {\includegraphics[width=\linewidth]{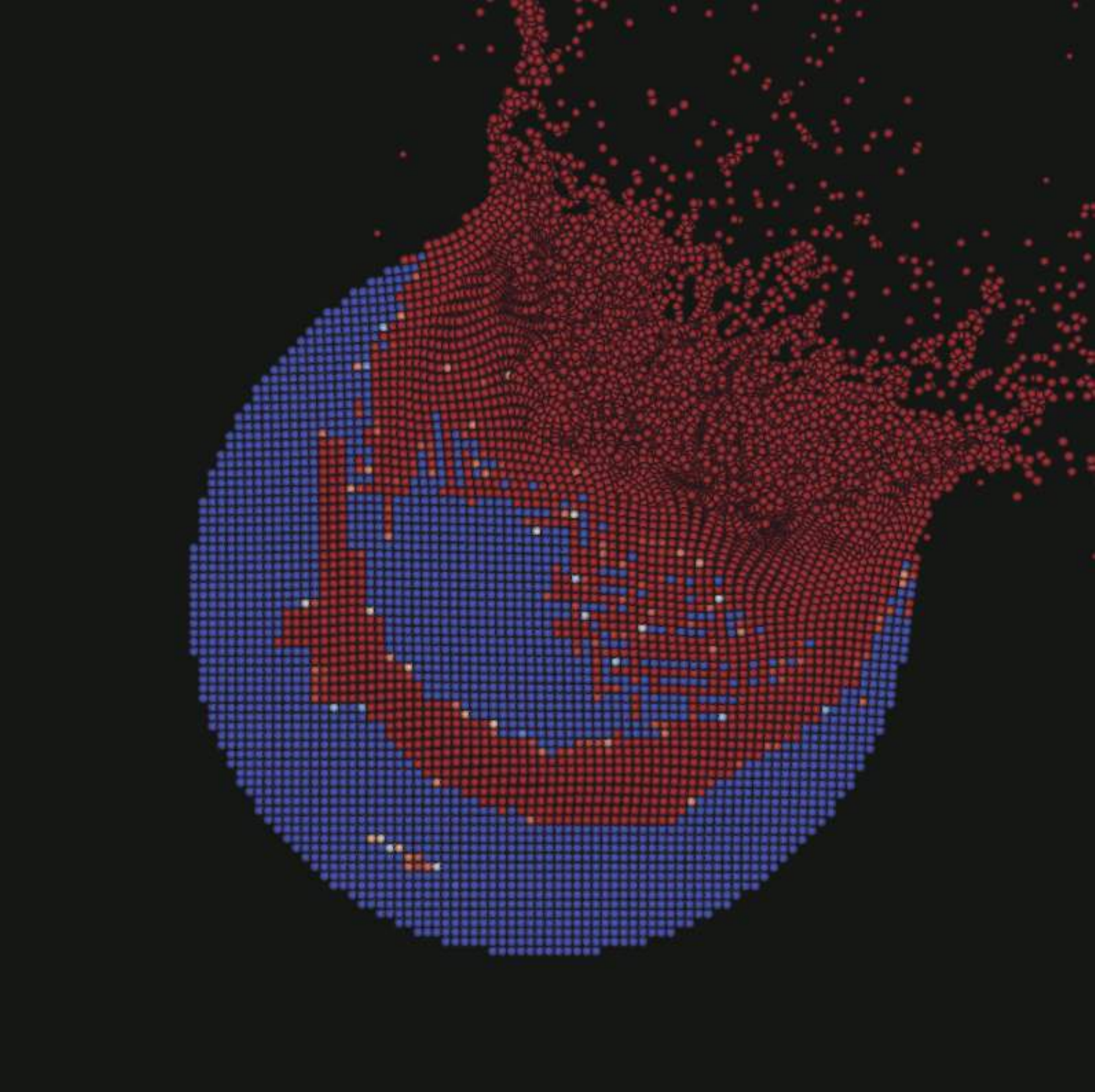}}
\end{minipage}
&
\begin{minipage}[c]{0.1\columnwidth}
    \centering
    {\includegraphics[width=\linewidth]{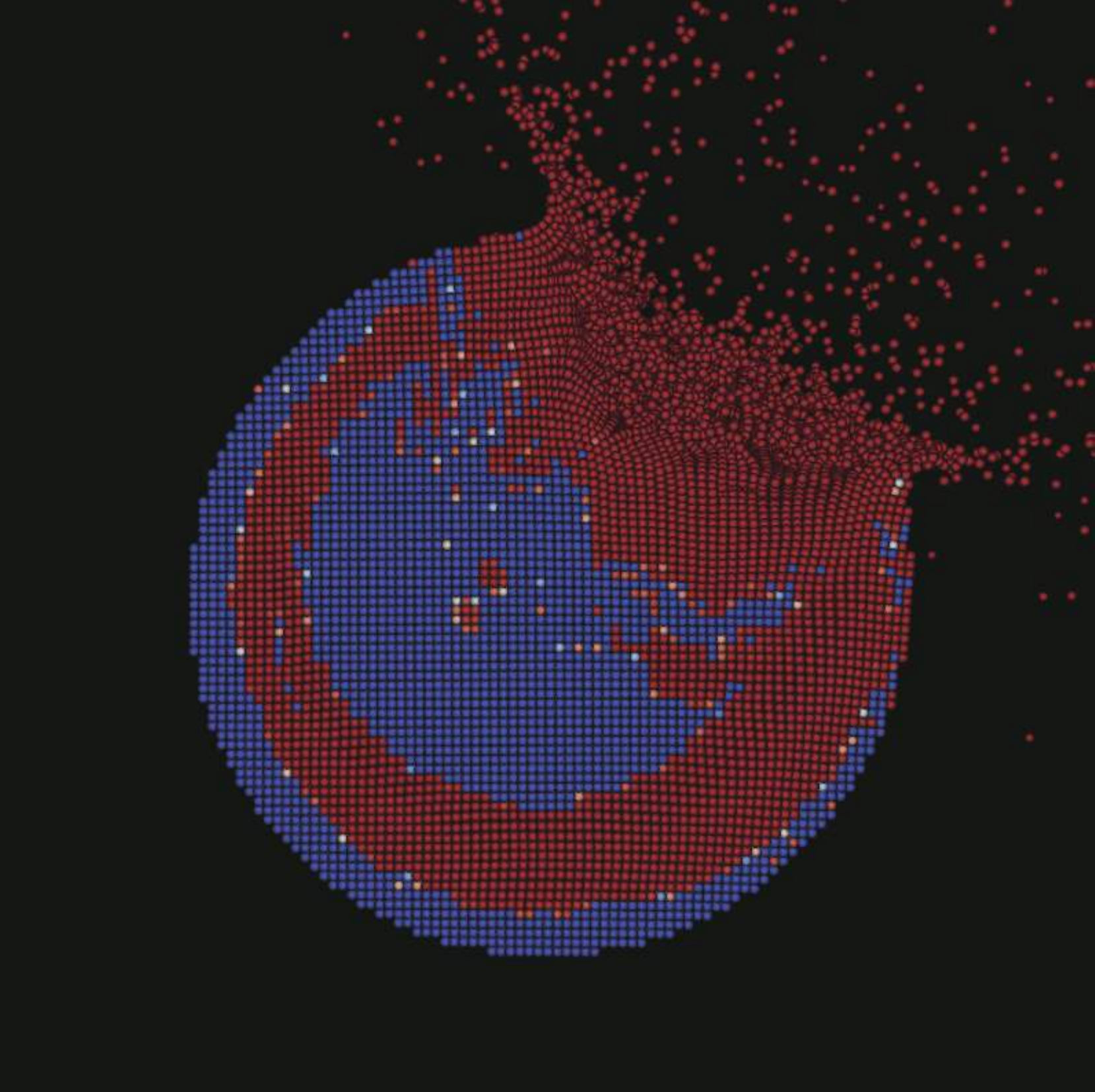}}
\end{minipage}
&
\begin{minipage}[c]{0.1\columnwidth}
    \centering
    {\includegraphics[width=\linewidth]{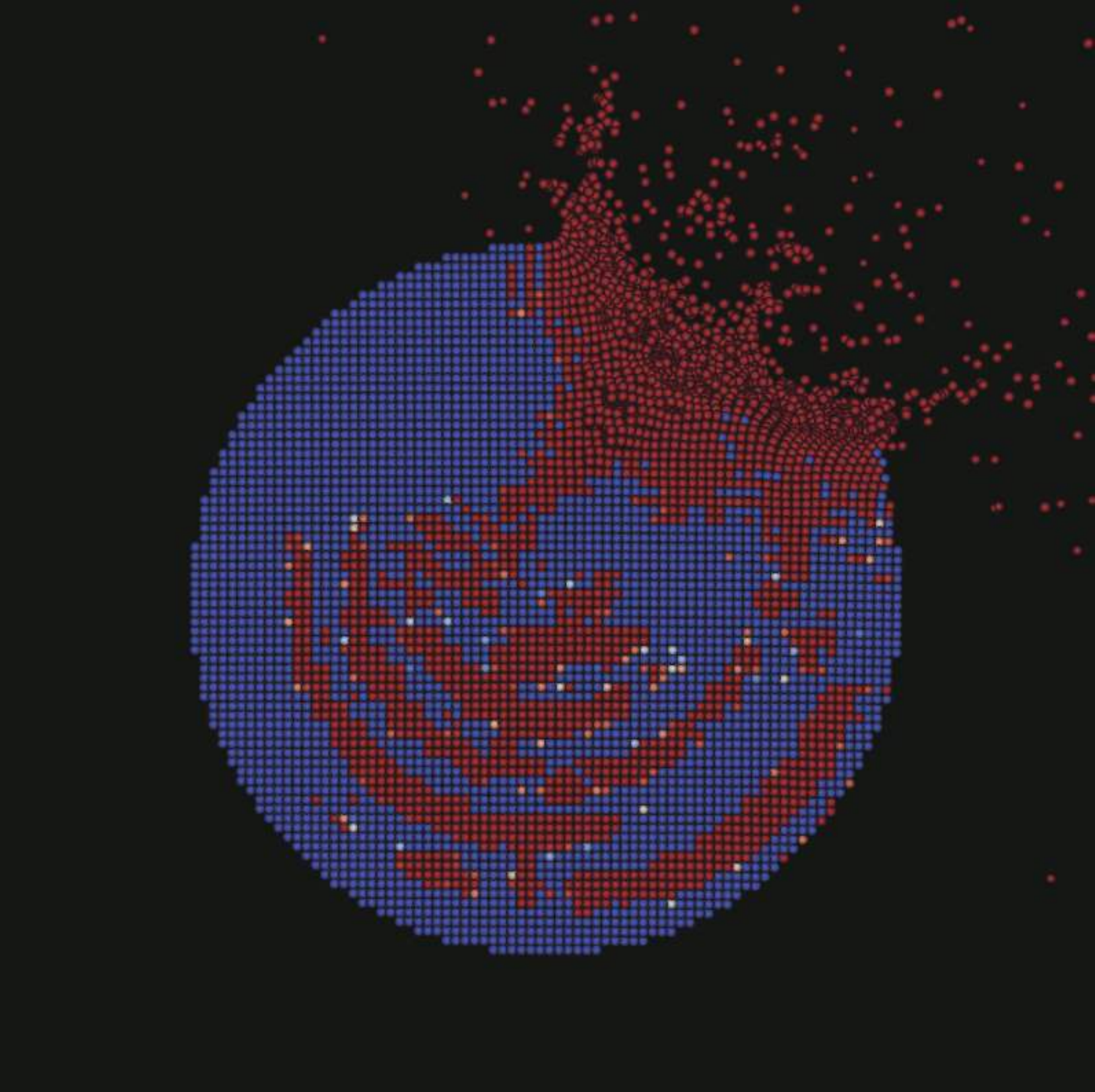}}
\end{minipage}
&
\begin{minipage}[c]{0.1\columnwidth}
    \centering
    {\includegraphics[width=\linewidth]{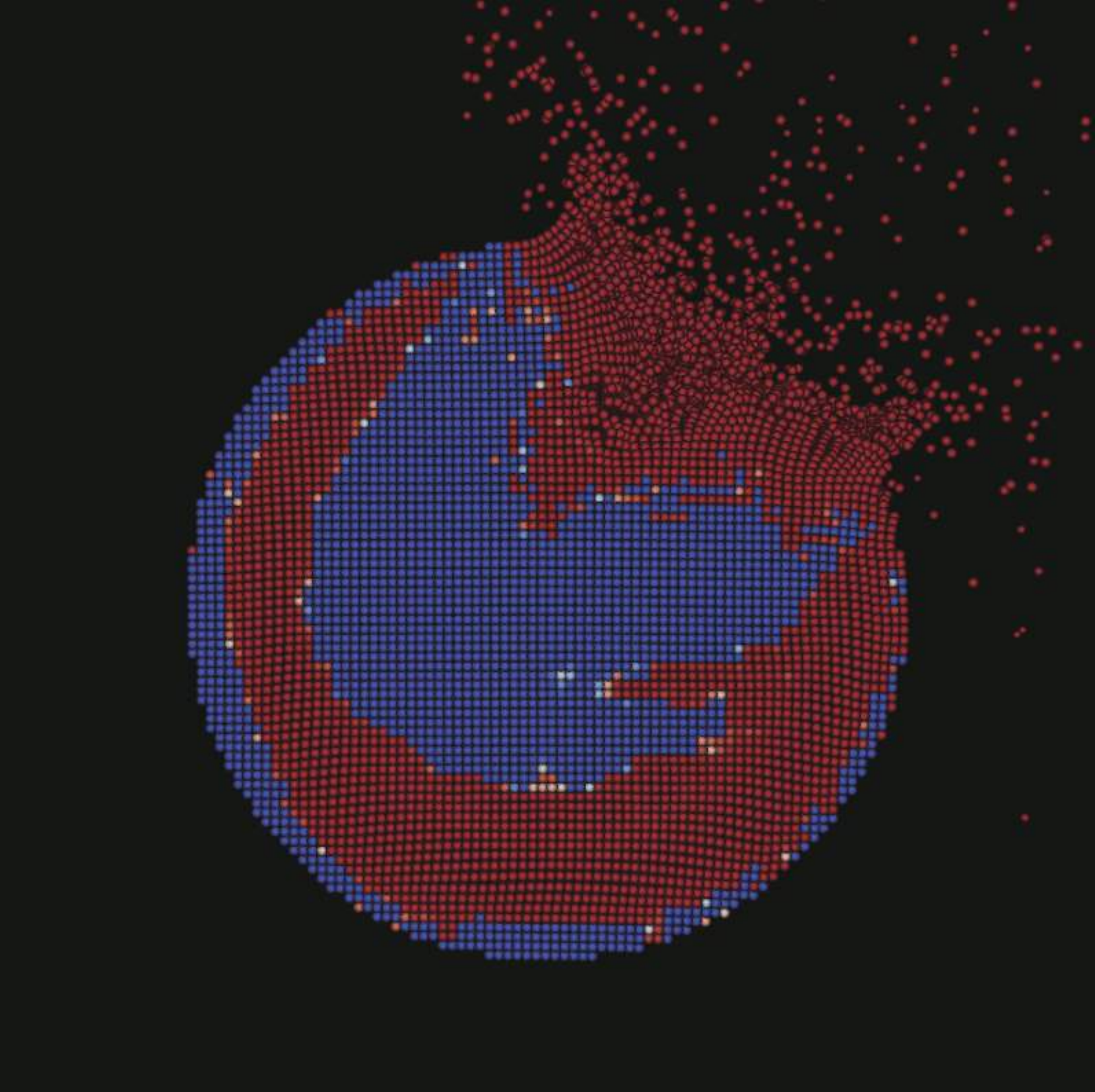}}
\end{minipage}
&
\begin{minipage}[c]{0.1\columnwidth}
    \centering
    {\includegraphics[width=\linewidth]{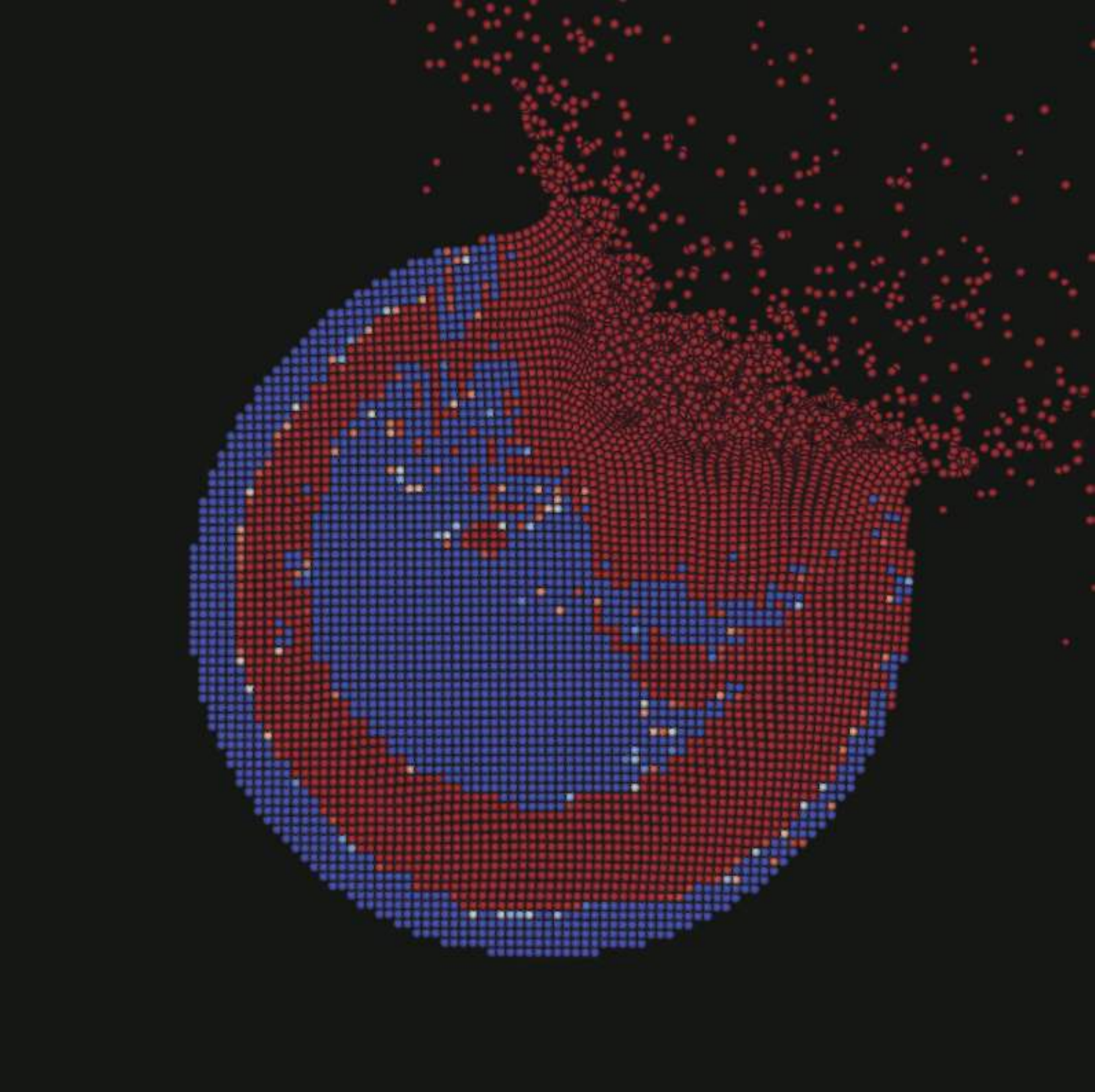}}
\end{minipage}
\\
\arrayrulecolor{white}
\cmidrule(lr){1-7}
Largest fragment$^{e}$
&
\begin{minipage}[c]{0.1\columnwidth}
    \centering
    {\includegraphics[width=\linewidth]{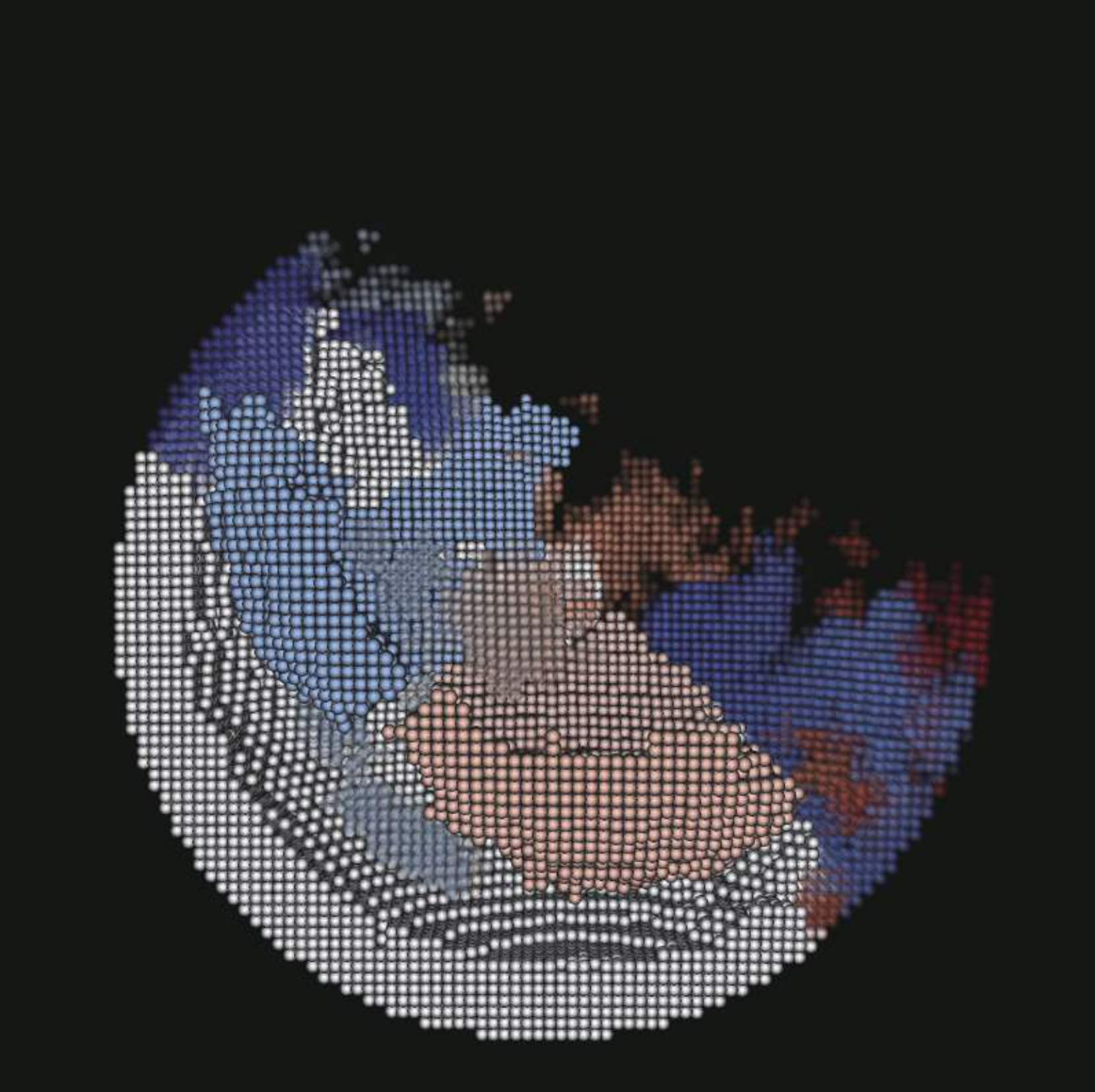}}
\end{minipage}
&
\begin{minipage}[c]{0.1\columnwidth}
    \centering
    {\includegraphics[width=\linewidth]{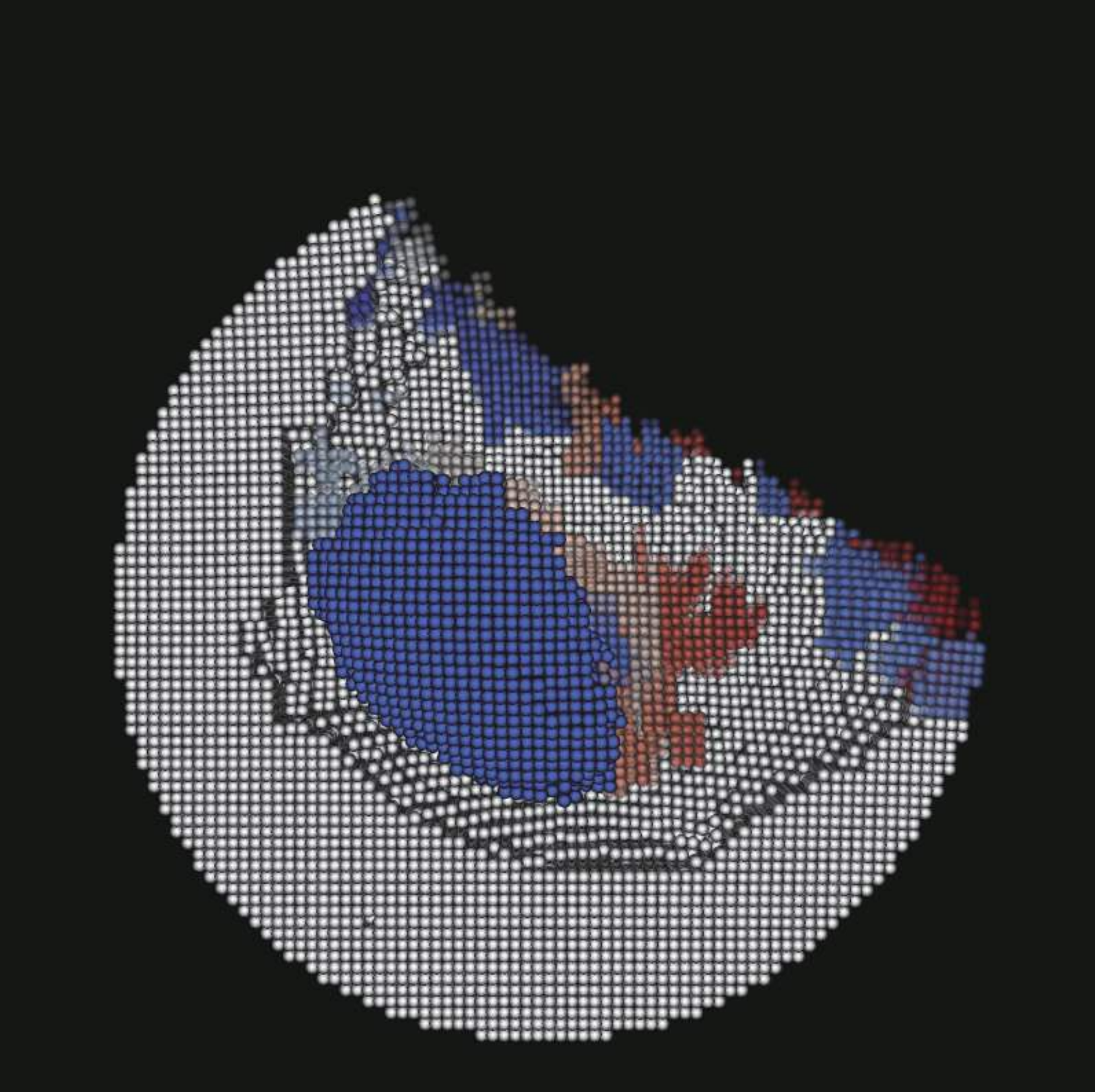}}
\end{minipage}
&
\begin{minipage}[c]{0.1\columnwidth}
    \centering
    {\includegraphics[width=\linewidth]{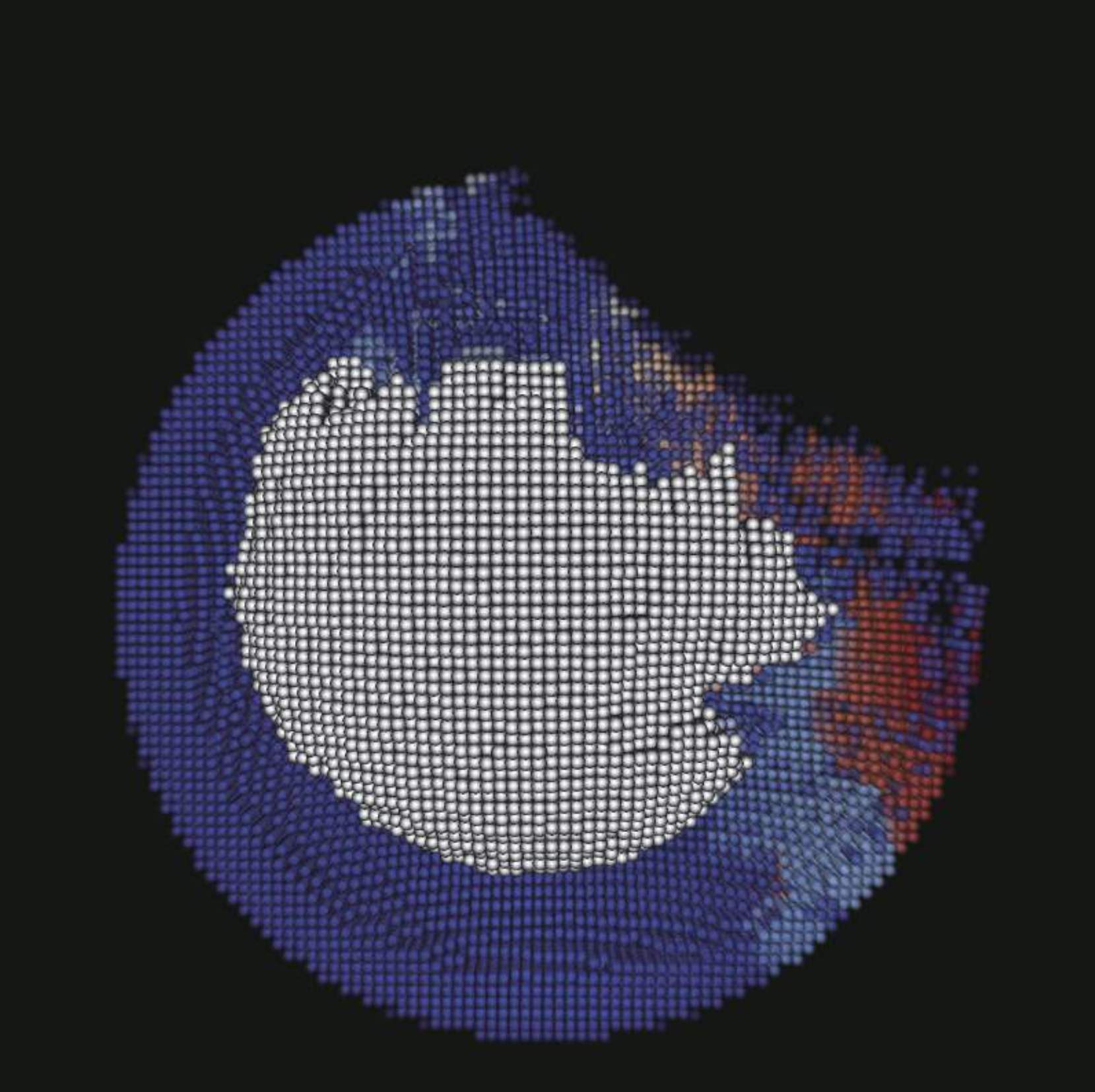}}
\end{minipage}
&
\begin{minipage}[c]{0.1\columnwidth}
    \centering
    {\includegraphics[width=\linewidth]{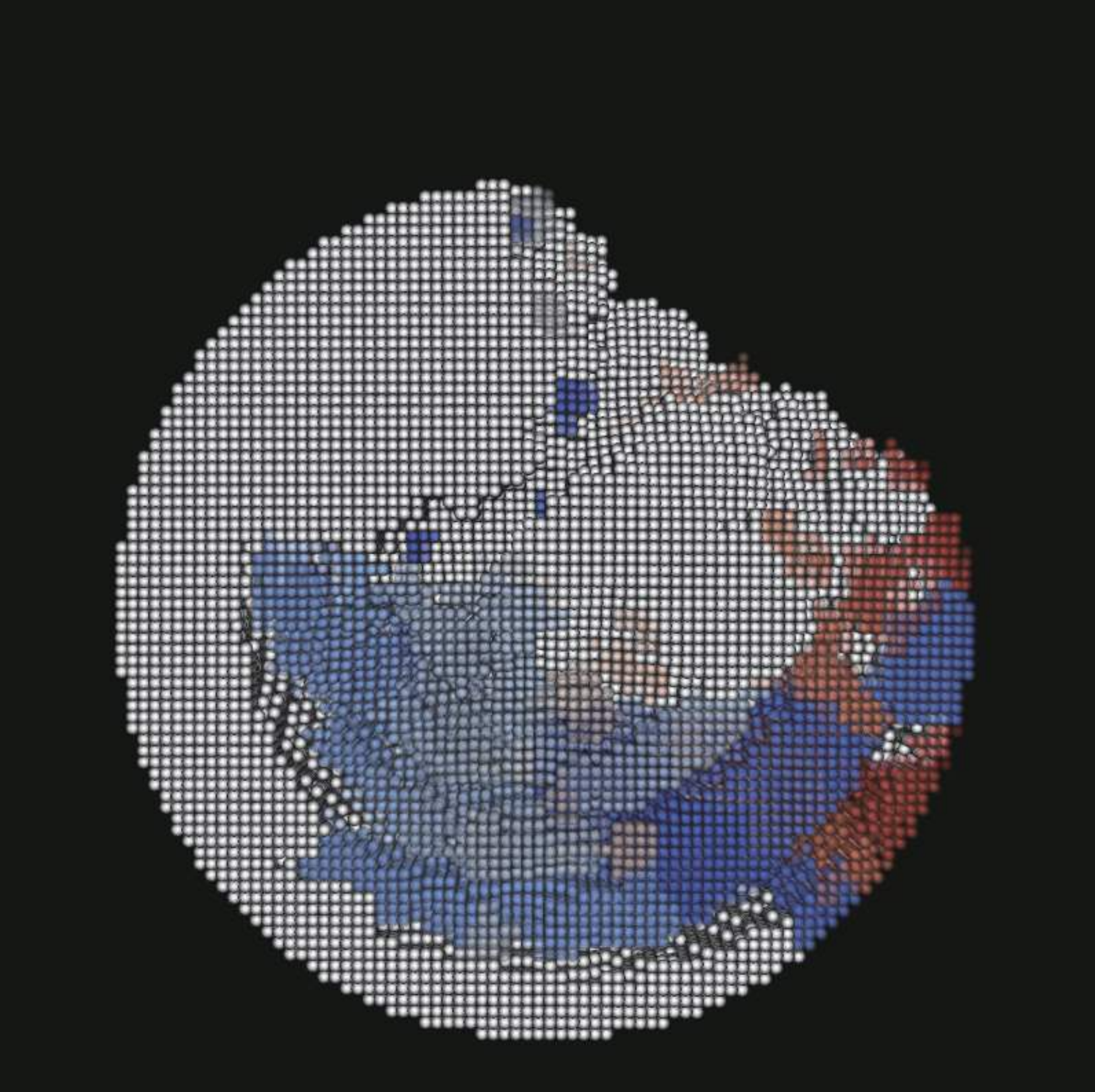}}
\end{minipage}
&
\begin{minipage}[c]{0.1\columnwidth}
    \centering
    {\includegraphics[width=\linewidth]{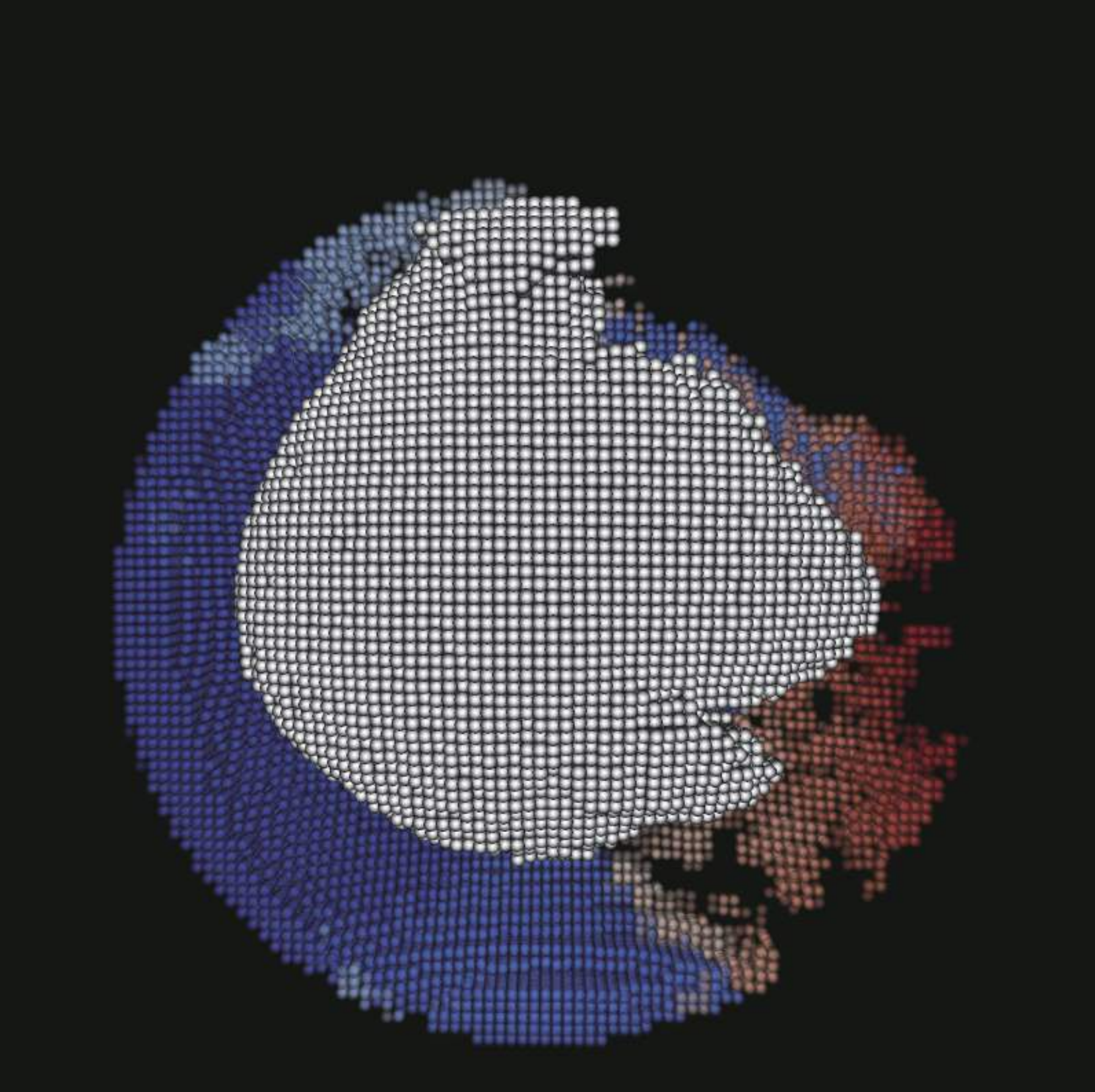}}
\end{minipage}
&
\begin{minipage}[c]{0.1\columnwidth}
    \centering
    {\includegraphics[width=\linewidth]{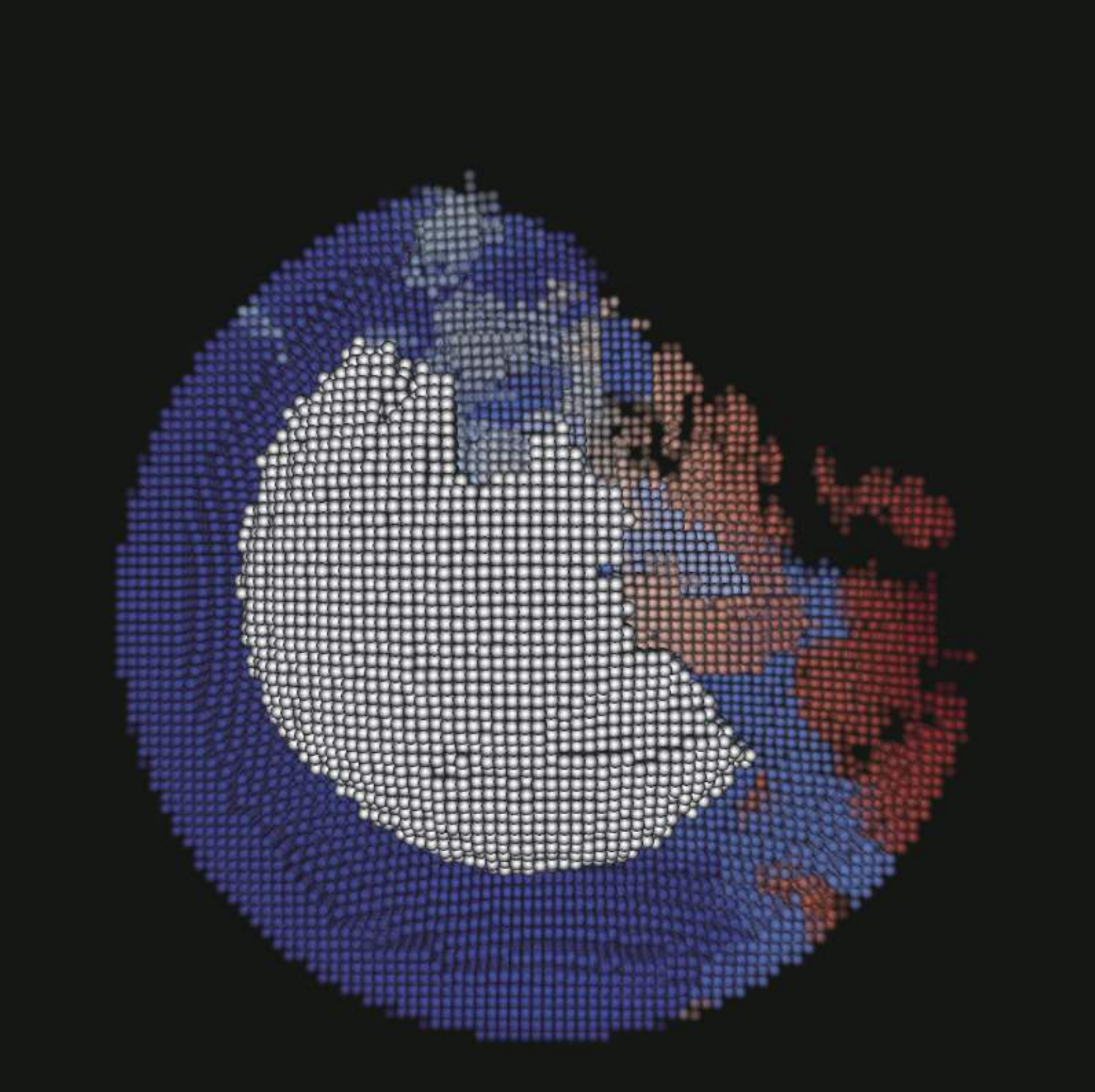}}
\end{minipage}
\\
\arrayrulecolor{black}
\bottomrule
\end{tabular}
\begin{tablenotes}
\item[a] Other parameters includes $\mu_{\mathrm{i}} = 1.5$, $\mu_{\mathrm{d}} = 0.8$, $Y_{\mathrm{m}} = \SI{3500}{\mega\pascal}$ and $q_{\psi} = 0.256$. 
\item[b] Nominal case.
\item[c] The core-shaped fragment is the second largest, with $m_{\mathrm{f}} / M_{\mathrm{tar}} = 0.057$ and velocity \SI{6.043}{\metre\per\second}.
\item[d, e] Same as Table~\ref{tab:initial}.
\end{tablenotes}
\end{threeparttable}
\end{table}

The nominal case employs the simple linear hardening strength model with a yield strength set at \SI{3500}{\mega\pascal}.
This yield strength is significantly higher than the damage strength, resulting in the material rarely yielding during simulation.
To explore the influence of the yield surface and plastic correction, the yield strength is adjusted to \SI{30}{\mega\pascal} and \SI{60}{\mega\pascal}, values slightly below and above the damage strength, respectively.
Spallation still occurred under these conditions, but the location shifted, and a double-layered damage shell even emerged at the lower yield strength.

Further investigation is conducted using the modified Lundborg model proposed in this study, and the effects of cohesive shear strength ($Y_0$) are examined.
At a $Y_0$ of \SI{10}{\mega\pascal}, plasticity was pronounced, leading to disordered damage patterns.
As $Y_0$ surpasses the initial damage threshold, a sealed shell-like damage zone and a core-shaped largest fragment are formed, mirroring the experimental findings in terms of the distribution of core mass and velocity.

The divergent outcomes between the linear hardening and modified Lundborg model arise from their differential approaches to stress constraints.
As illustrated in Fig.~\ref{fig:yield}(\textit{b}), the path that corrects the trial elastic stress back to the yield surface is indicative of the models' differing methodologies.
Linear hardening, evolving from the $J_2$ flow theory, applies plastic corrections only to shear stress by stress scaling.
This leads to a discrepancy in the velocities of shear and compression waves and thus affects the damage development.
Conversely, the modified Lundborg model, with its plastic flow adjustments addressing both shear and tensile stresses, produces a convergence in the damage zones across different material parameters.

In conclusion, the material models developed and tested in this study have passed scrutiny, confirming the stability of the algorithm.
It is evident that the material models are capable of capturing distinct dynamic response characteristics.
The development of material models enhances the universality of the algorithm across different materials.
Furthermore, by coupling experimental data with numerical analysis, the parameters of the material models can be reverse-engineered to determine the physical properties of the materials, such as strength.
This synthesis of empirical data and computational modeling is instrumental in advancing our understanding of material behavior under impact conditions.

\subsection{Hypervelocity impact between asteroids}\label{sec4_2}

The collisional origin of asteroid families was, for the first time, entirely and successfully reproduced via numerical simulation by \citet{Michel2001}.
The fragmentation and the gravitational reaccumulation phases are sequentially executed to complete the simulation of the collisional process, where the fragmentation phase, calculated using a shock physics code, provides detailed outputs on the size and velocity field of the impact remnants \citep{Michel2015}. 
These outputs serve as initial conditions for the gravitational reaccumulation phase.
By employing various material models or different compaction structures of the parent body, simulations of large asteroid disruptions ultimately result in the generation of diverse types of asteroid families \citep{Michel2001, Michel2002, Michel2003, Michel2004}, asteroids of specific shapes \citep{Michel2013, Michel2020}, and even comets \citep{Schwartz2018}. 

\begin{figure}[pos=h]
\centering
  \includegraphics[width = 0.17\textwidth]{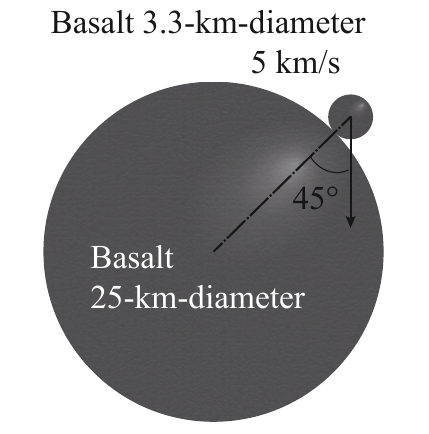}
  \caption{Schematic of a pseudo catastrophic collision between S-type asteroids.
  The simulation involves a $3.3$-\si{\kilo\metre}-diameter basalt spheroidal asteroid colliding with another $25$-\si{\kilo\metre}-diameter basalt spheroidal target, at a velocity of \SI{5}{\kilo\metre\per\second} and a \ang{45} impact angle.}
  \label{fig:asteroid}
\end{figure} 

To validate the applicability of the Material Point Method for simulating the fragmentation phase of impacts on the scale of small celestial bodies, we envisioned a pseudo-catastrophic collision between S-type asteroids, drawing inspiration from the setup described by \citet{Michel2013} (see Fig.~\ref{fig:asteroid}).
Our simulation considers a spheroidal asteroid with a diameter of $3.3$-\si{\kilo\metre} impacting a $25$-\si{\kilo\metre}-diameter target body at an impact velocity of \SI{5}{\kilo\metre\per\second} and an impact angle of \ang{45}.
While real asteroids of this size often exhibit irregular shapes and potential internal differentiation, we adopted an idealized spherical and homogeneous geometry.
This choice serves to represent a generic, primordial parent body and provides a controlled baseline.
It allows us to isolate the specific effects of the implemented material models and directly benchmark our MPM outcomes against established SPH studies (e.g., \citet{Michel2013}) without the compounding variables introduced by complex topography.
Both asteroids are modeled as S-type bodies, composed primarily of basalt, with general material parameters detailed in Table~\ref{tab:astromat}. 

Consistent with the approach, we also compared the outcomes of simulations using different strength models or damage models.
The strength models included a linear hardening model and a modified Lundborg model, as seen in Table~\ref{tab:astrostr}, which in the presentation of results are labeled `\textit{a}' and `\textit{b}', respectively.
For the damage model, we adopted Weibull-distributed cracks with either the same Weibull parameters as the nominal case (\SI{0.4}{\mega\pascal} low damage strength) or the same damage strength (\SI{40}{\mega\pascal} high damage strength), detailed in Table~\ref{tab:astrodmg}, and these variations are denoted as `1' and `2' in the display of results.

\begin{table}[h]
\begin{threeparttable}
\caption{General characteristics for modeling asteroids}
\label{tab:astromat}
\begin{tabular}{lrr}
\toprule 
Description & Target & Projectile \\
\midrule
Material & Basalt & Basalt \\
Radius (\si{\kilo\metre}) & $12.5$ & $1.65$ \\
Density (\si{\kilogram\per\cubic\metre}) & $2700.2$ & $2692.0$ \\
Mass (\si{\kilogram}) & \num{2.2091e16} & \num{5.0653e13} \\
\bottomrule
\end{tabular}
\end{threeparttable}
\end{table}

\begin{table}
\begin{threeparttable}
\caption{Strength model parameters}
\label{tab:astrostr}
\begin{tabular}{ccccccccc}
\toprule 
\multirow{2}{*}{Young's modulus $E_{\mathrm{Y}}$} & \multirow{2}{*}{Poisson's ratio $\nu$} &
\multicolumn{2}{c}{Linear hardening} &
\multicolumn{5}{c}{Modified Lundborg}\\
\cmidrule(lr){3-4} \cmidrule(lr){5-9}
& & Yield stress $\sigma_{\mathrm{Y}}$ & Plastic modulus $E^{\mathrm{p}}$ & 
$\mu_{\mathrm{i}}$ & $\mu_{\mathrm{d}}$ & $Y_0$ & $Y_{\mathrm{m}}$ & $q_{\psi}$ \\
\midrule
\SI{53100}{\mega\pascal} & $0.152$ & \SI{3500}{\mega\pascal} & \SI{5.31}{\mega\pascal} &
$1.5$ & $0.8$ & \SI{90}{\mega\pascal} & \SI{1500}{\mega\pascal} & $0.256$\\
\bottomrule
\end{tabular}
\end{threeparttable}
\end{table}

\begin{table}
\begin{threeparttable}
\caption{Weibull damage parameters}
\label{tab:astrodmg}
\begin{tabular}{ccccccccc}
\toprule 
\multirow{2}{*}{case} & \multicolumn{4}{c}{Target} & \multicolumn{4}{c}{Projectile}\\
\cmidrule(lr){2-5} \cmidrule(lr){6-9}
& $m$ & $k$ (\si{\per\metre\cubed}) & $\bar{\sigma}_{\mathrm{min}}$ (\si{\mega\pascal}) & $\bar{\sigma}_{\mathrm{max}}$ (\si{\mega\pascal}) &
$m$ & $k$ (\si{\per\metre\cubed}) & $\bar{\sigma}_{\mathrm{min}}$ (\si{\mega\pascal}) & $\bar{\sigma}_{\mathrm{max}}$ (\si{\mega\pascal}) \\
\midrule
1 & $8.5$ & \num{3.0e39} & $0.326$ & $0.417$ & $8.5$ & \num{3.0e39} & $0.666$ & $0.853$ \\
2 & $8.5$ & \num{5.0e22} & $30.703$ & $39.292$ & $8.5$ & \num{1.5e25} & $32.075$ & $41.049$ \\
\bottomrule
\end{tabular}
\end{threeparttable}
\end{table}

\begin{figure}[pos=htb!]
    \centering
    \begin{subfigure}{0.18\textwidth}
        \centering
        \includegraphics[width=\textwidth]{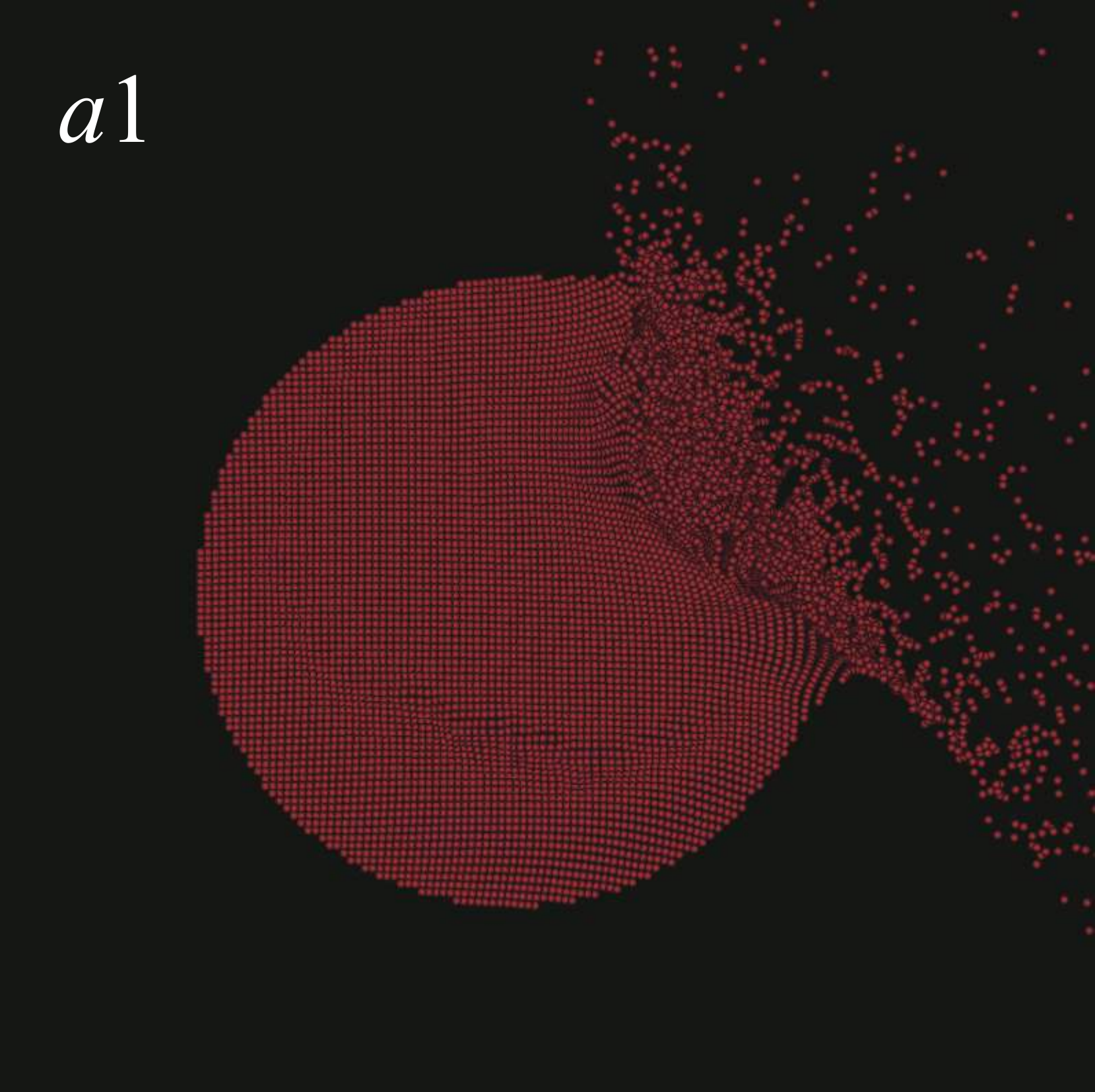}
    \end{subfigure}
    \begin{subfigure}{0.18\textwidth}
        \centering
        \includegraphics[width=\textwidth]{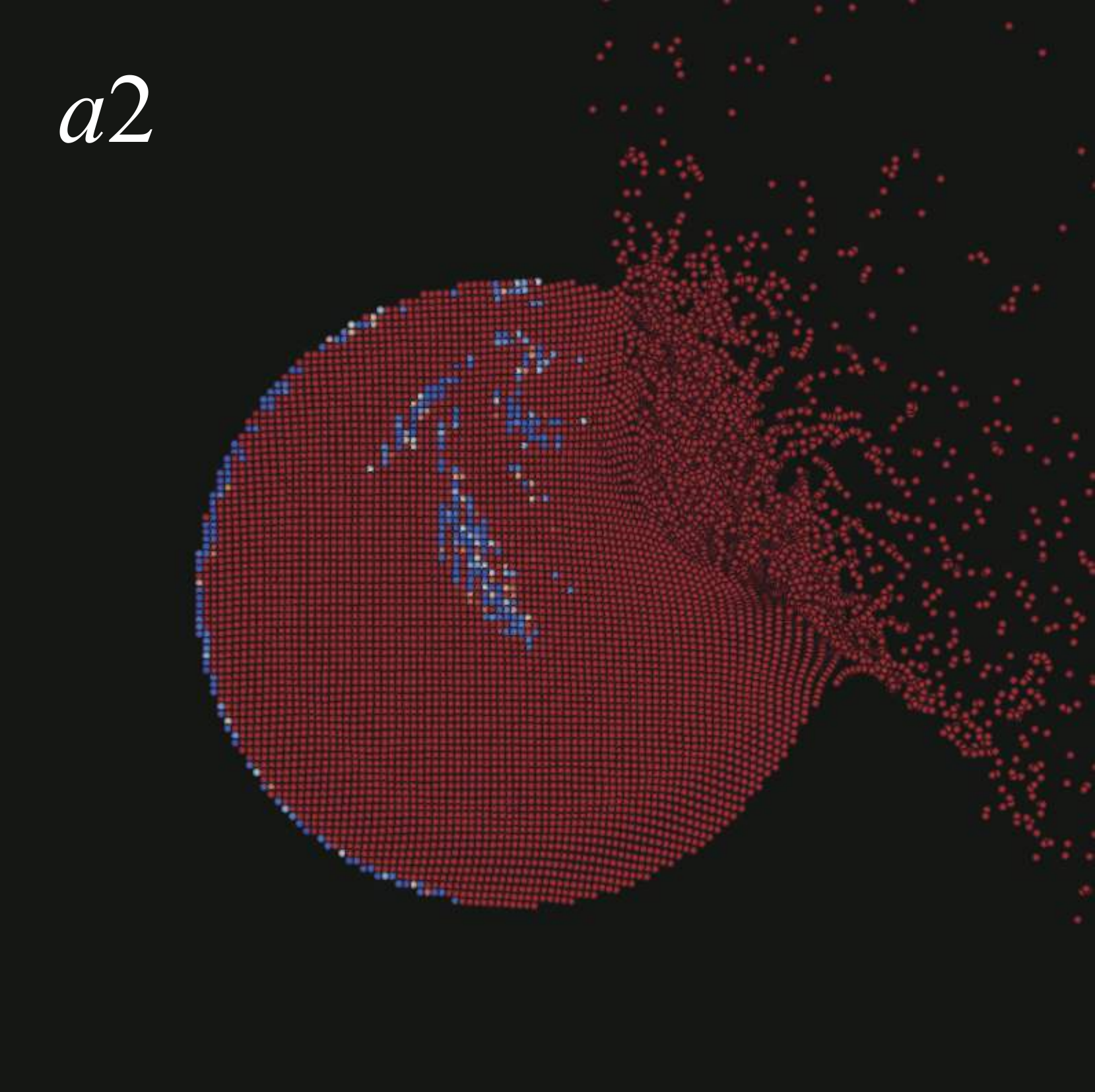}
    \end{subfigure}

    \begin{subfigure}{0.18\textwidth}
        \centering
        \includegraphics[width=\textwidth]{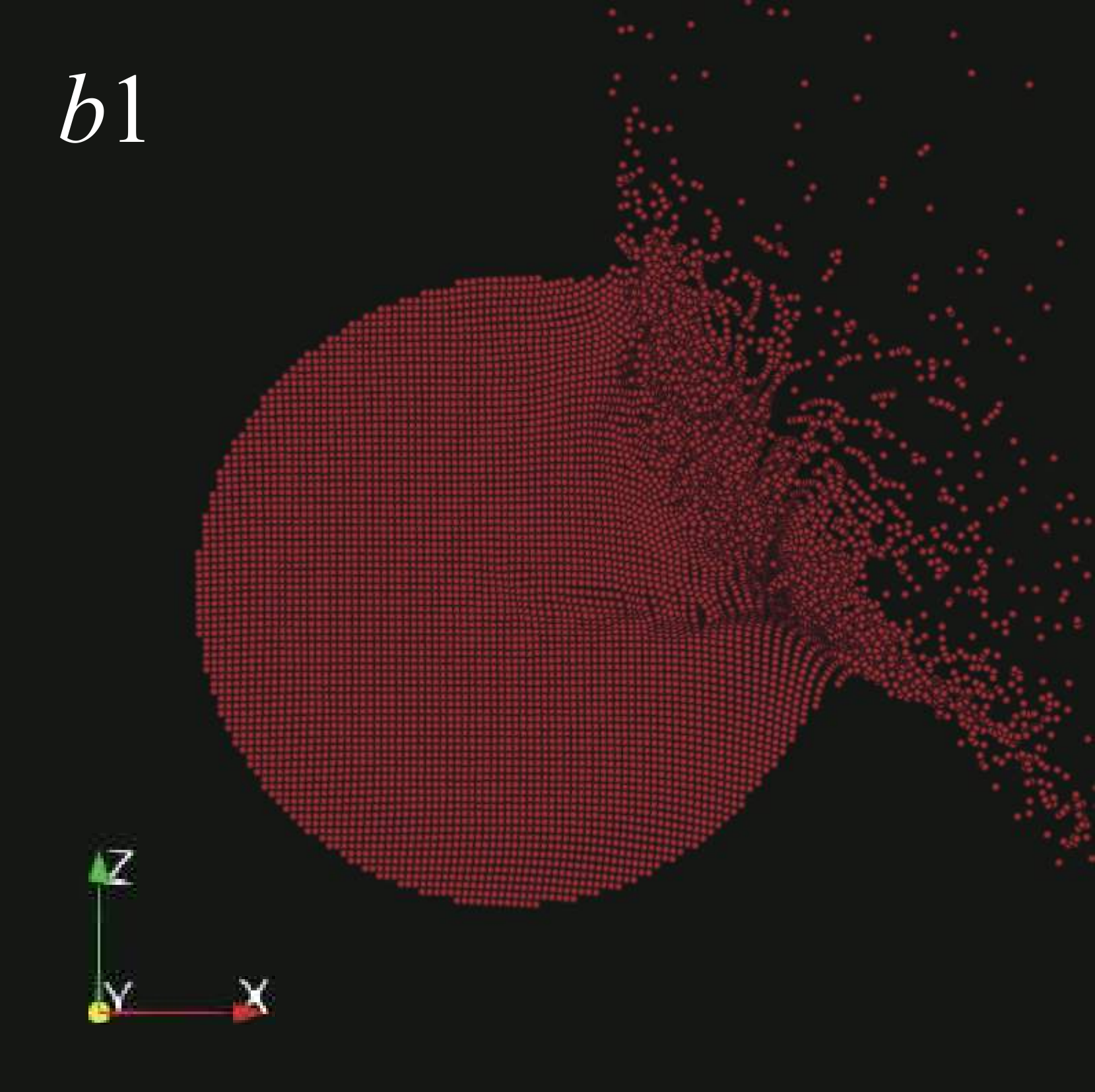}
    \end{subfigure}
    \begin{subfigure}{0.18\textwidth}
        \centering
        \includegraphics[width=\textwidth]{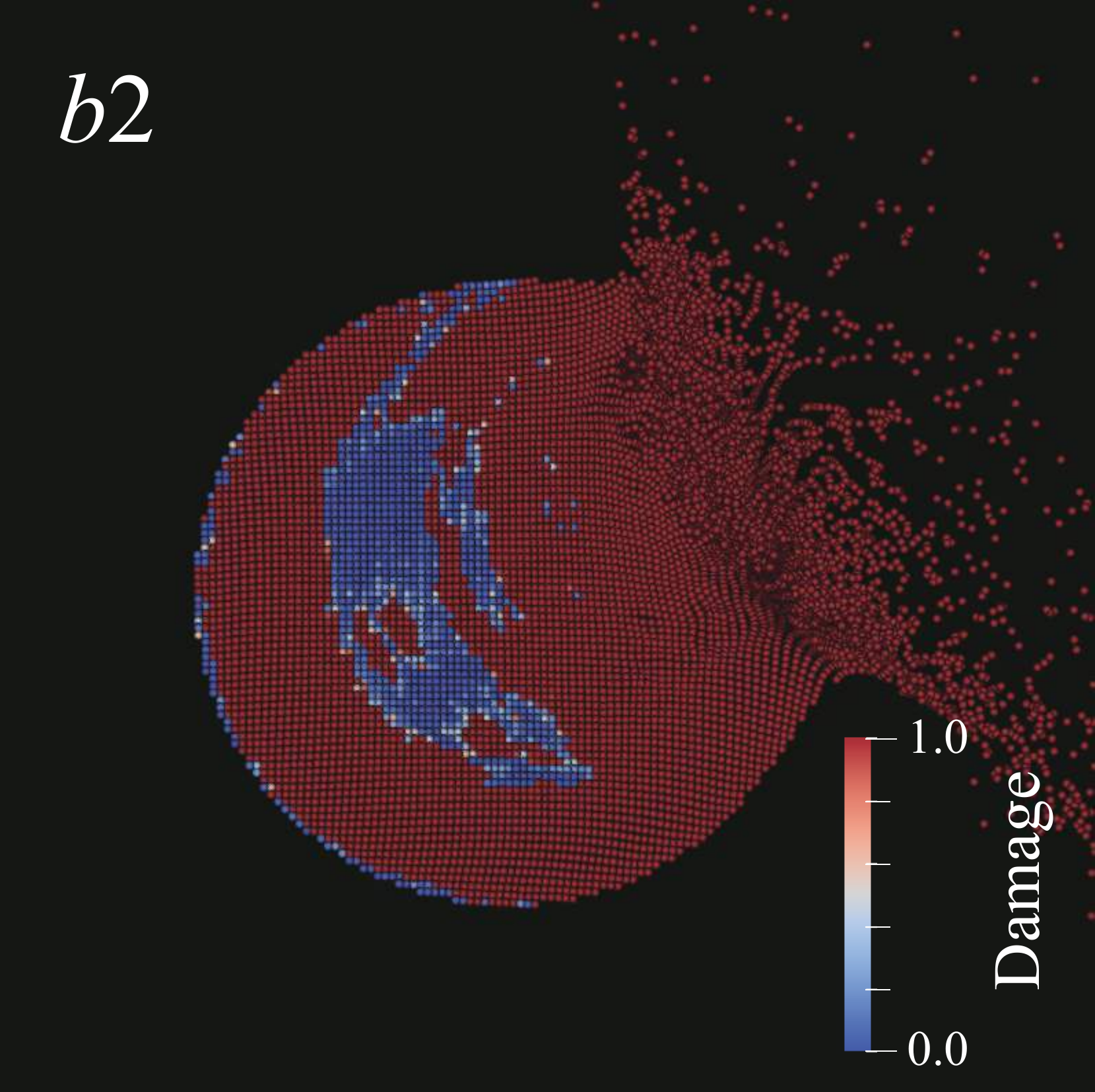}
    \end{subfigure}
    \caption{The degree of damage at \SI{15}{\second} after impact, presented in a central symmetrical cross-section.
    Linear hardening strength model is used in the '\textit{a}' series, and the modified Lundborg model in the '\textit{b}' series.
    The '$1$' series employs Weibull parameters consistent with the nominal case of laboratory impact simulation, while the '$2$' series sets the same average largest activation stress.}
    \label{fig:astrodmg}
\end{figure}

\begin{figure}[pos=hbt!]
    \centering
    \begin{subfigure}{0.18\textwidth}
        \centering
        \includegraphics[width=\textwidth]{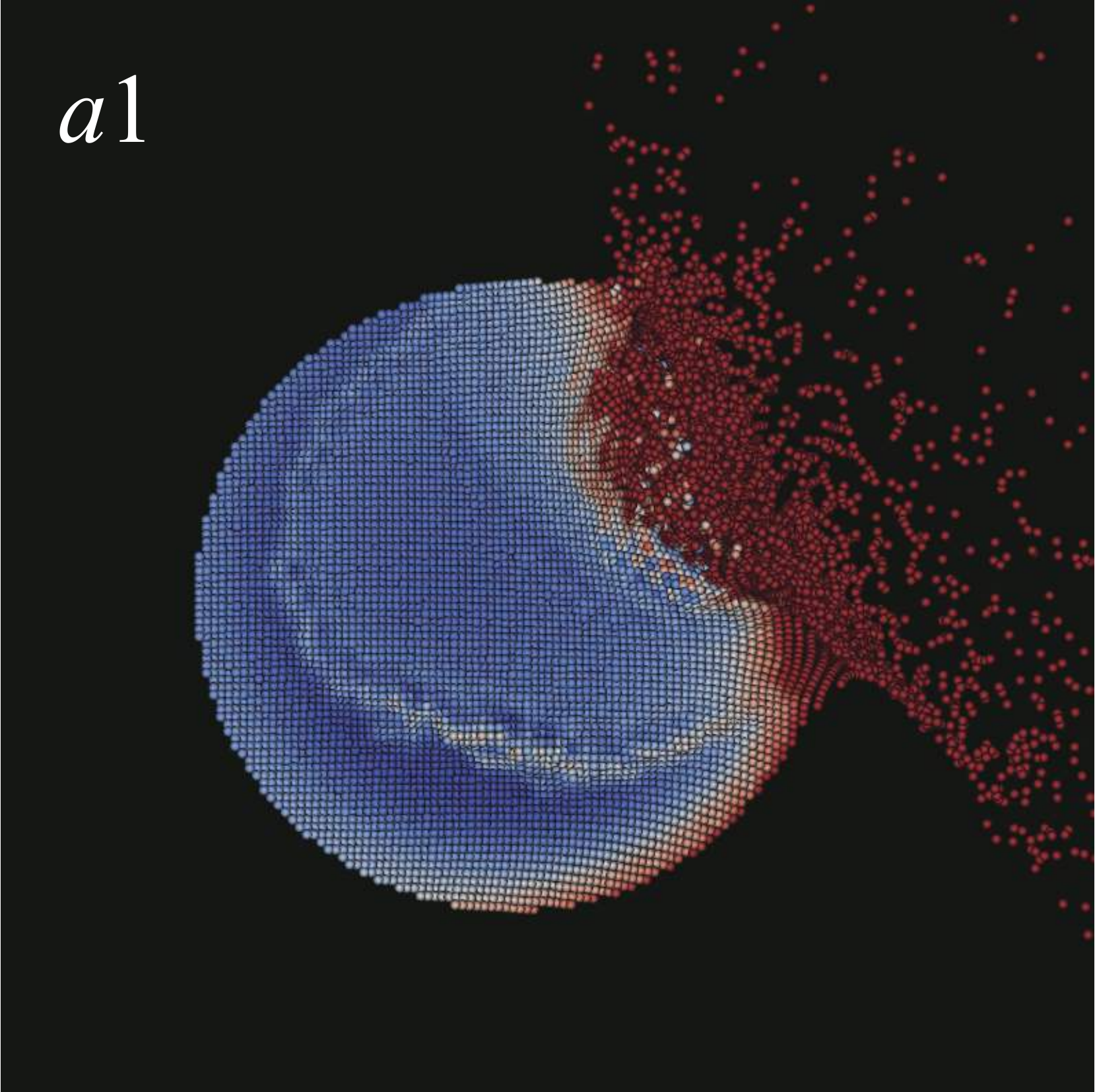}
    \end{subfigure}
    \begin{subfigure}{0.18\textwidth}
        \centering
        \includegraphics[width=\textwidth]{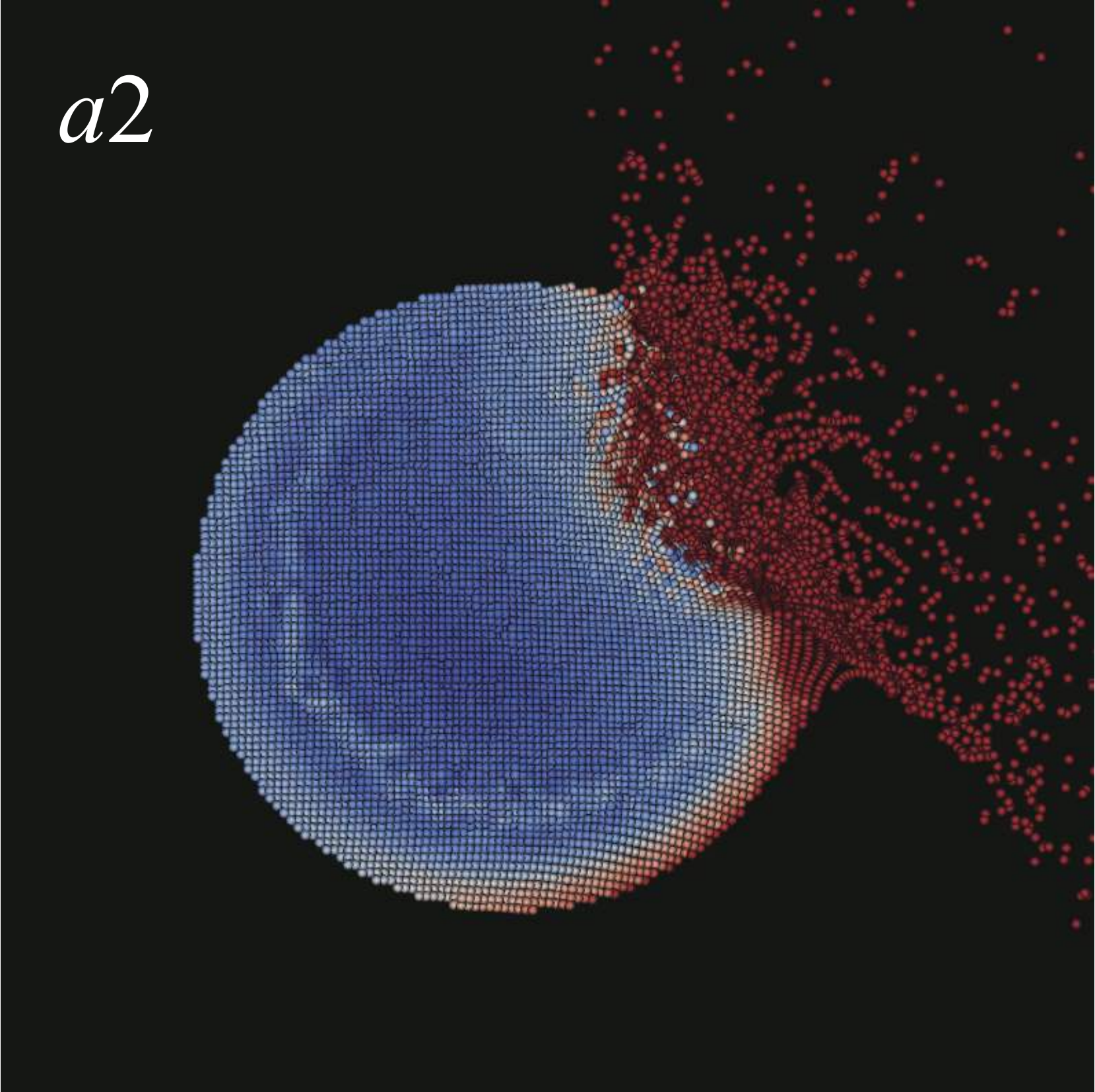}
    \end{subfigure}

    \begin{subfigure}{0.18\textwidth}
        \centering
        \includegraphics[width=\textwidth]{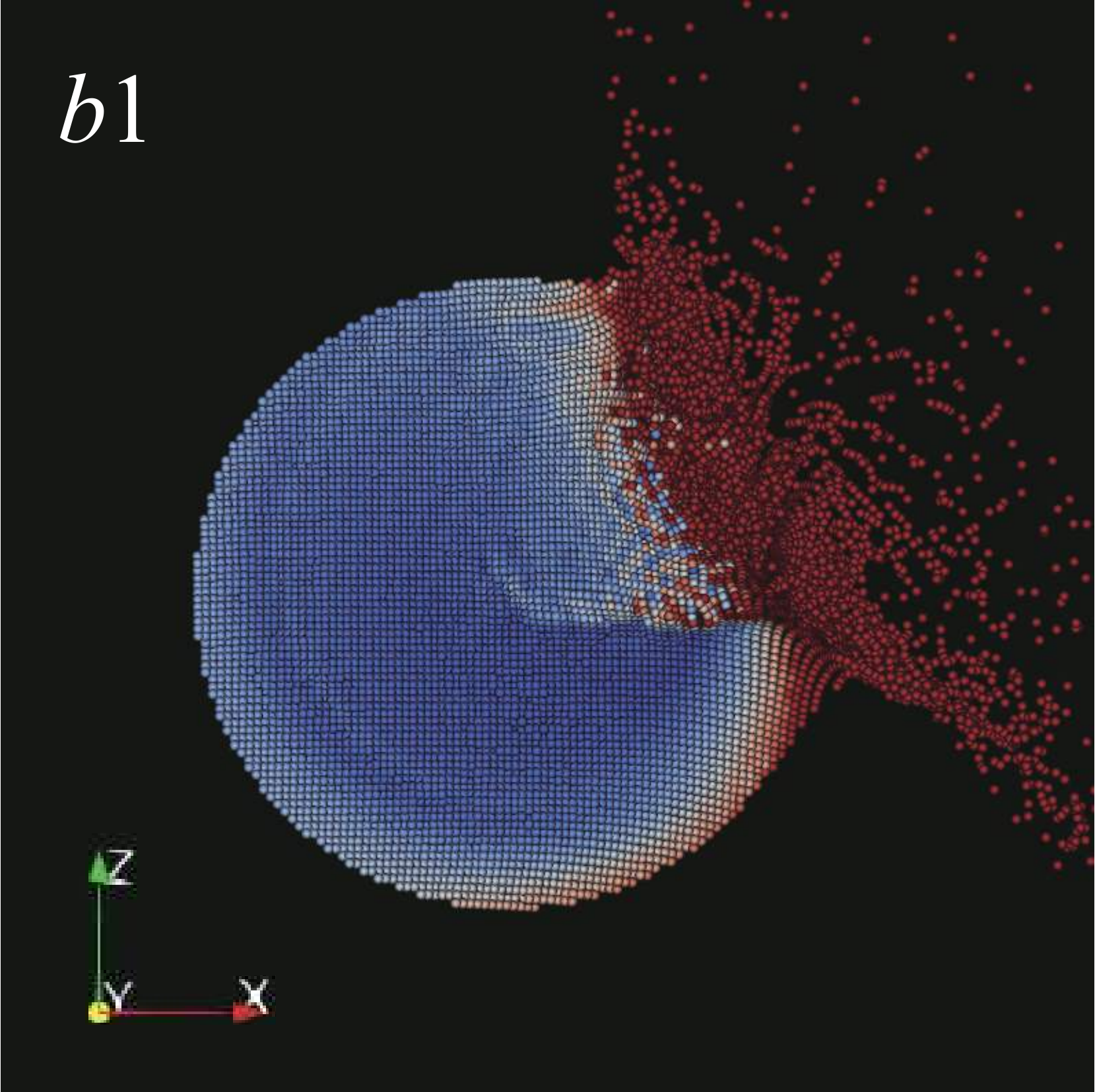}
    \end{subfigure}
    \begin{subfigure}{0.18\textwidth}
        \centering
        \includegraphics[width=\textwidth]{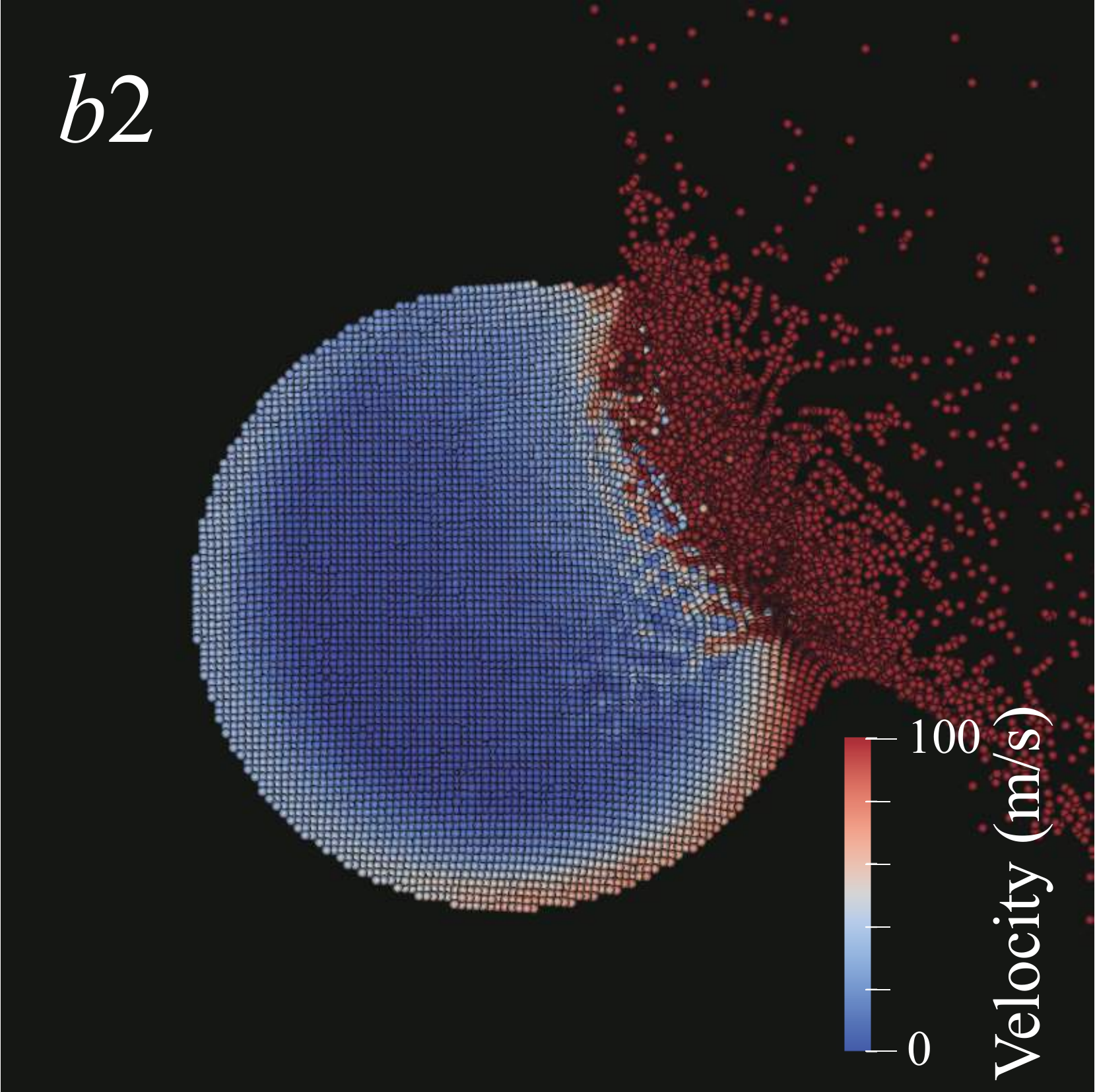}
    \end{subfigure}
    \caption{Velocity profiles at \SI{15}{\second} post-impact, visualized on a central symmetrical cross-section.
    The same grouping methodology as previously described in Fig.~\ref{fig:astrodmg}. }
    \label{fig:astrov}
\end{figure}

The simulation chronicles the post-impact evolution over a span of \SI{100}{\second}.
In reality, the attenuation of the shock wave is exceedingly rapid, with the later stages dominated by the displacement of fragments, which, unfortunately, can obscure the direct fragmentation effects caused by the impact.
Given that the stress wave speed is \SI{4.435}{\kilo\metre\per\second}, it takes approximately \SI{5.638}{\second} for a wave to traverse the diameter of the target once.
It is assumed that after two reflections, the stress waves no longer cause significant widespread damage accumulation.
Therefore, we focus on the results at \SI{15}{\second} after impact as a representative snapshot.
The damage distribution at this juncture is illustrated in Fig.~\ref{fig:astrodmg}, and the velocity distribution of the material points is depicted in Fig.~\ref{fig:astrov}.

The choice of strength model exhibits a relatively minor effect on the final damage patterns.
However, subtle differences emerge in the velocity fields, where the linear hardening model produces more pronounced shear bands.
This disparity, stemming from distinct plastic correction approaches, could influence the initial ejection and subsequent evolution of fragments.

In contrast, the initial damage strength of the target body dramatically influences the collision outcome.
A low damage strength of approximately \SI{0.4}{\mega\pascal} (case 1 with \textit{a} or \textit{b}) is insufficient to withstand the hypervelocity impact, leading to the complete failure of both bodies.
This result is consistent with findings from many previous studies using the SPH method, which often report pervasive fragmentation and subsequent gravitational reaccumulation to form rubble-pile bodies \citep{Michel2015, Walsh2018}.
This low-strength scenario likely represents the collision of second-generation asteroids, which have already undergone at least one catastrophic disruption.

\begin{figure}[pos=h]
\centering
  \includegraphics[width = 0.4\textwidth]{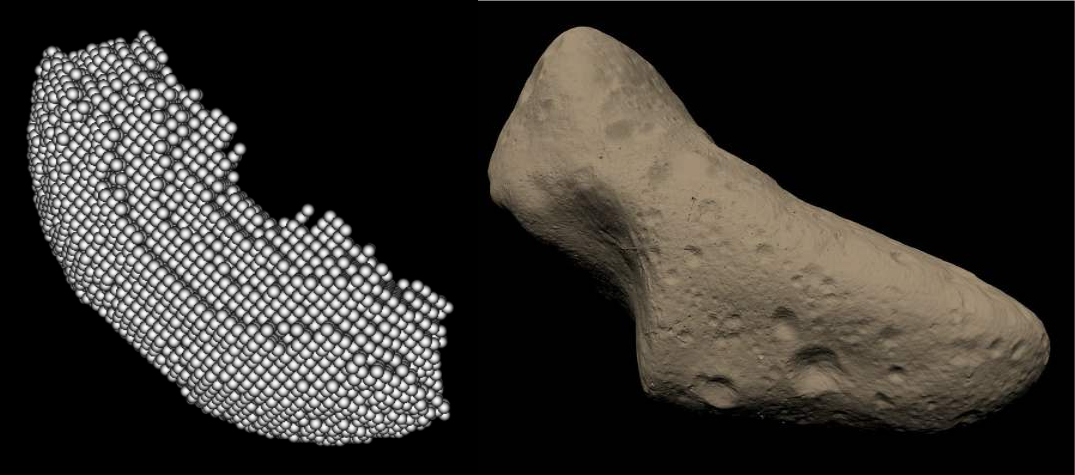}
  \caption{The largest fragment (left) of asteroid collision in Fig.~\ref{fig:astrodmg} \textit{b}$2$, comparing with the model of asteroid ($433$) Eros (right) that NASA reconstructed from NEAR-spacecraft observations.\protect\footnotemark[2]}
  \label{fig:asteroidfrag}
\end{figure}

Conversely, a high damage strength of approximately \SI{40}{\mega\pascal} (case 2 with \textit{a} or \textit{b}) does not lead to complete disintegration.
Simulations with both strength models reveal the presence of undamaged large fragments, a novel finding in the context of these studies.
Furthermore, our simulations with the modified Lundborg model reveal the survival of a large, coherent fragment.
The largest remnant produced in our \textit{b}2 simulation is a prolate object with a maximum dimension of \SI{16}{\kilo\metre} (Fig.~\ref{fig:asteroidfrag}).
Its size and elongated shape bear a striking resemblance to the S-type near-Earth asteroid (433) Eros, which has an effective diameter of \SI{16.84}{\kilo\metre} and a triaxial shape of $34.4 \times 11.2 \times 11.2$ \si{\kilo\metre} \citep{Yeomans2000, Veverka2000}.
While the actual simulated fragment possesses a complex 3D topology with significant concavities, the projection shown in Fig.~\ref{fig:asteroidfrag} optimally highlights its maximum dimensions and overall prolate profile, although the detailed 3D features would likely be subject to further modification during subsequent long-term evolution.

\footnotetext[2]{https://science.nasa.gov/resource/eros-3d-mode}

The origin and evolution of Eros, as well as its internal structure, have long been a subject of debate.
One view holds that it is a "fractured monolith", a large shard from a parent body that remains structurally intact despite extensive fracturing \citep{Richardson2005, Thomas2005}.
This hypothesis is supported by simulations from \citet{Tonge2016a}, who demonstrated that a strong internal flaw distribution is required to reproduce Eros's observed surface features from major impacts.
However, a recent study by \citet{Ballouz2025} analyzing seismic data suggested that Eros's interior properties are more consistent with those of a rubble pile.
They argue that if Eros were a monolith, it must have been shattered by a Solar System age worth of collisions to mimic the seismic properties of a rubble pile.

Our MPM simulations provide a critical, missing piece to this puzzle.
Hydrocode simulations of parent-body disruptions using traditional SPH methods have struggled to produce large, monolithic fragments like Eros.
They typically result in smaller debris that reaccumulates into rubble piles.
Our \textit{b}2 simulation, for the first time, numerically demonstrates a physical pathway where a single, catastrophic impact on a strong parent body can directly produce an Eros-sized and shaped shattered monolith.
This result reconciles the need for a strong primordial body, as suggested by \citet{Tonge2016a}, with the existence of a large, coherent fragment.
A heavily shattered but not disrupted body generated from Himeros-forming impact could also exhibit seismic properties that appear similar to a rubble pile \citep{Tonge2016a}, potentially bridging the gap with the observations of \citet{Ballouz2025}.
This finding suggests that some SPH models may overestimate damage in asteroid disruption events.
While not definitively settling the debate, our work provides the first strong numerical evidence supporting the hypothesis that Eros is a giant shard from a primordial parent, offering a new perspective on the formation and evolution of asteroids in its size range.

\section{Discussion}\label{sec5}
Our results not only validate the Material Point Method framework as a robust tool for simulating hypervelocity impacts on asteroids but also offer deeper insights into two critical aspects.
First, it highlights the intricate relationship between constitutive models and material behavior.
Second, it demonstrates the unique capabilities of MPM to address long-standing challenges in planetary science.
This section explores these two themes, highlighting how our work paves the way for new avenues of research.

\subsection{The interplay between material models and dynamic behavior}\label{sec5_1}
\begin{figure}[pos=htbp]
\centering
  \includegraphics[width = 0.5\textwidth]{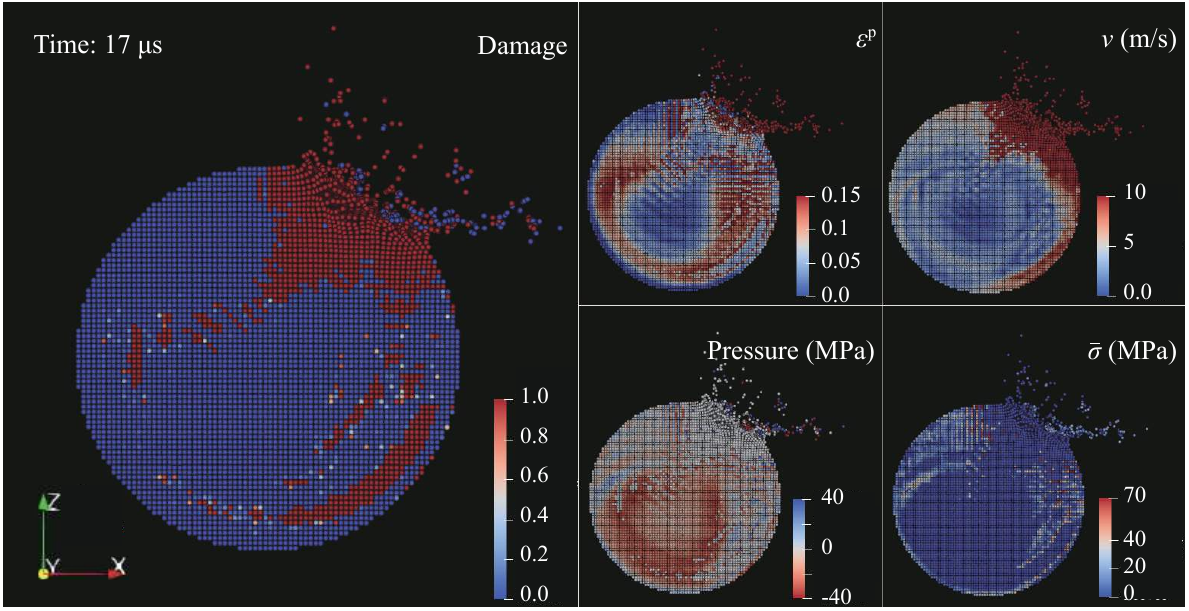}
  \caption{Snapshot at \SI{17}{\micro\second} of the simulation case using modified Lundborg strength model with $Y_0 = \SI{10}{\mega\pascal}$ in Table~\ref{tab:strength}. Outputs include damage,  accumulated plastic strain, velocity, pressure, and von Mises effective stress, which serves as an illustration of MPM's capacity to carry extensive physical state information for in-depth analysis.}
  \label{fig:process}
\end{figure}

A key contribution of this work is the detailed exploration of how constitutive models govern the simulated physical behavior of impacted bodies.
Our MPM framework, capable of tracking a rich set of state variables (Fig.~\ref{fig:process}), allows us to deconstruct this relationship at multiple levels.

At the highest level, the choice of model paradigm dictates the failure mode.
As demonstrated in Table~\ref{tab:damage}, employing a maximum-principal-stress criterion leads to brittle fracture, whereas a damage accumulation model based on crack growth naturally reproduces spallation.
This underscores the necessity of selecting a model paradigm that reflects the known physics of the target material.

At a finer level, model parameters control the material's strength and response. 
This effect is generally consistent and continuous, barring other influencing factors, as shown in Table~\ref{tab:damage}, Table~\ref{tab:strength}, and Section~\ref{sec4_2}.
Crucially, we found that even with different parameters, a physically sound model preserves the fundamental failure patterns.
For instance, both our laboratory-scale and asteroid-scale impacts exhibited spallation and an intact core, with the primary difference being the global extent of damage due to scaling of impact energy.

At the most fundamental level, the mathematical formulation of a model determines its physical fidelity.
This is often overlooked but is critical for avoiding unphysical artifacts.
For example, in situations of extensive yielding, as seen in Table~\ref{tab:strength}, the $J_2$ flow theory causes shear and compression wave velocities to decouple, leading to less realistic simulations compared to those using the modified Lundborg model.
Furthermore, in exploring factors causing shell-like damage, we experimented with using stress instead of the effective local tensile strain to calculate the number of activated cracks, i.e., removing the damage correction in the denominator of Eq.~(\ref{eq:dmgsoft}).
With the same damage parameters, this stress-based model significantly increased material strength (far beyond reality, so we didn't show these results).
In practice, crack propagation reduces the effective cross-sectional area of the material matrix that bears the stress, and the damage correction in the denominator of Eq.~(\ref{eq:dmgsoft}) depicts this relationship, proving its necessity.

These numerical experiments affirm that grasping the interplay between models and their physical manifestations is paramount for predictive simulations.
Our work not only provides a benchmarked library of models for basalt but also establishes a methodology for their validation.
The modular design of our code facilitates the future inclusion of models for other materials, such as metals or ices, enabling a broader investigation into the diverse collisional histories of small bodies.

\subsection{Features of MPM: A new tool for unresolved problems in asteroid science}\label{sec5_2}
Beyond material modeling, our simulations highlight the unique advantages of MPM's hybrid Eulerian-Lagrangian formulation, positioning it to tackle scientific questions that have been challenging for traditional methods like SPH.

The Lagrangian nature of the material points provides a natural and powerful way to track material history and interfaces.
Unlike grid-based methods that struggle with large deformations, or SPH where fragment boundaries require post-processing algorithms (e.g., $\alpha$-shape), MPM explicitly tracks fragments and their complex morphologies (Fig.~\ref{fig:frag}).
This capability is not merely a technical convenience.
It is scientifically transformative.
While other coupled frameworks like FDEM or SPH-DEM already exist, MPM provides a uniquely seamless and natural alternative for direct coupling with DEM codes for long-term gravitational evolution studies and enables detailed analysis of fragment shape statistics, which are crucial for understanding the collisional grinding process and the dynamics of irregular particle assemblies \citep{Ferrari2020}.

Furthermore, MPM's ability to handle discontinuities opens the door to studying the complex internal structures of asteroids.
The explicit grid allows for a natural and robust treatment of contact between distinct bodies or internal layers \citep{Bardenhagen2001}.
While not the focus of this paper (based on findings by \citet{Liu2015} that the friction coefficient is negligible in hypervelocity impacts and MPM's inherent non-slip constraint effectively addresses contact issues), our ongoing work leverages this to model contact forces between boulders in a rubble pile or between regolith and bedrock.
This is a significant advantage over SPH, where contact is often difficult to define without particle interpenetration.
Consequently, MPM is an ideal tool to investigate fundamental, unresolved questions: How do shock waves propagate through a rubble-pile asteroid versus a fractured monolith?
How does the presence of a deep regolith layer or large subsurface voids alter the outcome of an impact?
These are questions that MPM is uniquely suited to answer.

Eulerian representation in MPM is manifested at the beginning of each time step with the resetting of a regular background grid.
The use of a regular background grid allows for the implementation of various boundary conditions beyond natural ones.
For instance, the simulation with symmetric boundary condition, as shown in Fig.~\ref{fig:precision1} and Fig.~\ref{fig:precision2}, holds the same spatial discretization precision as the nominal case but half the total number of particles.
Using symmetry reduces computational load, shortening computation time by approximately \SI{25}{\percent} without significant deviations in results.
Additionally, take the situation that targets significantly larger than the impact crater as an example, transmitting boundary conditions can be set to avoid allocating excessive computational resources to areas with minimal stress impact \citep{Shen2005}.
These features add flexibility to MPM applications in diverse scenarios.

The hybrid particle-grid approach is a key design feature of MPM.
The particles bridge the Eulerian background grid in each time step, avoiding the advection issues common in Eulerian methods, and the mesh tangling problem common in Lagrangian methods.
The presence of the grid also accelerates the process of locating the spatial positions of particles, reducing the computational burden of searching for neighboring particles.
This design not only ensures numerical stability but also enhances the efficiency of the MPM method.
All simulations in this study were conducted on a personal desktop with an Intel Core i7-7700 CPU, and each of them was completed within a few hours without parallelization.

Naturally, MPM is not without its own numerical challenges.
Cell-crossing noise, which occurs when particles traverse grid boundaries, can introduce numerical artifacts, particularly in high-speed, large-deformation impacts.
We mitigate this using the Generalized Interpolation Material Point (GIMP) method \citep{Bardenhagen2004}, which, while effective, slightly increases computational time.
Moreover, in the nominal cases, the fragment locating the antipode of the impact site was larger than expected, possibly due to the over-suppression of cell-crossing noise of GIMP.
Future improvements could involve exploring alternative schemes like the Staggered Grid Material Point (SGMP) method for enhanced stability \citep{Liang2019}, combining MPM with DEM and then simulating more complex and complete processes of hypervelocity impacts on small celestial bodies, and integrating MPM with other shock physics codes that have been benchmarked and validated to simulate collision problems collectively.
This collaborative approach in simulations can provide a more comprehensive and profound understanding of collision dynamics.

In conclusion, MPM is more than just another numerical method.
It bridges the gap between continuum mechanics and discrete systems, offering a robust framework to simulate the entire collisional process, from initial impact and fragmentation to the interaction of resulting fragments.
By providing a clearer view of phenomena like contact mechanics and interface tracking, MPM empowers us to move beyond idealized targets and begin exploring the rich, complex, and discontinuous nature of real-world asteroids.

\section{Summary}\label{sec6}
Understanding the collisional evolution of asteroids requires numerical tools that can accurately capture fracture, fragmentation, and the behavior of complex geological structures.
This paper introduces a comprehensive 3D Material Point Method (MPM) framework tailored for this purpose, demonstrating its power to bridge critical gaps left by traditional simulation techniques.

Our framework combines Lagrangian material points and Eulerian background grid description, with variable step size leapfrog integration and a Modified Update-Stress-Last (MUSL) scheme.
It incorporates significant enhancements to material constitutive models, including a modified $C^1$ continuous Lundborg yield criterion and a plastic corrector corresponding to the plastic flow law, a Tillotson equation of state with sound speed modification, and a resolution-independent Grady-Kipp fragmentation model for consistent fragmentation analysis.
These models were rigorously benchmarked against laboratory impact experiments and existing SPH simulations, demonstrating high fidelity in predicting fragment mass and velocity distributions.
The primary conclusions of this work are:

\begin{itemize}

\item The MPM framework was validated by simulating the experimental hypervelocity impact case, and benchmarked with SPH simulation.
By identifying fragments and comparing the cumulative mass distribution and the mass-velocity distribution with previous research, reasonable agreement was achieved, with results consistent within $3 \mathrm{\sigma}$.
It also demonstrates exceptional robustness and computational efficiency, making complex 3D impact simulations accessible on standard desktop hardware.

\item The analysis elucidates the critical link between constitutive models and physical realism.
The strength model with a smoothed Lundborg yield surface and plastic correction reproduces the shear and tensile stress waves well without decoupling them, and the fragmentation model evaluates the damage pattern and the maximum stress each material can afford.
Elastic-plastic behavior and damage softening are essential for capturing realistic phenomena like spallation and avoiding unphysical simulation artifacts.

\item Simulations of asteroid-scale collisions reveal that large, coherent remnants, analogous to ($433$) Eros, can survive catastrophic disruption.
The size of the largest remnant is shown to be highly sensitive to the statistical strength (Weibull parameters) of the parent body, linking an observable property to its internal structure.

\item The hybrid Lagrangian-Eulerian nature enables the explicit tracking of complex fragment morphologies, managing boundary conditions and evolving interfaces.
This overcomes a key challenge in impact simulations and allows for direct analysis of debris fields, a significant advantage for coupling with long-term gravitational codes.
\end{itemize} 

In summary, this work establishes MPM not merely as an alternative to existing methods, but as a powerful new paradigm for planetary impact science.
Its inherent ability to handle discontinuities and contact mechanics paves the way for future investigations into previously intractable problems, such as shock propagation in rubble-pile asteroids, the influence of subsurface layering, and the dynamics of binary asteroid formation.
By providing a more holistic tool to unravel the collisional history of our Solar System, this research also contributes to a deeper understanding necessary for future planetary defense initiatives.

\appendix
\section{Essentials of Material Point Method}\label{secA}
\subsection{Governing equations and MPM discretization}\label{secA_1}
To solve dynamic problems, a closed set of governing equations, consisting of the conservation equations of mass, momentum, and energy, constitutive equation, kinematic relation, boundary condition, and initial data, is essential. 
The relevant variables include position $\bm{x}$ (or displacement $\bm{u}$), strain $\bm{\varepsilon}$, stress $\bm{\sigma}$, as well as other state variables that can be deduced such as the density $\rho$, specific internal energy $E$, etc.

In MPM, the updated Lagrangian approach is used to derive these equations. The kinematic condition links the movement and strain, which is given by the rate of deformation tensor
\begin{equation}
    \bm{D} = \frac{1}{2}(\bm{L} + \bm{L}^{\mathrm{T}}) \, \text{,}
    \label{eq:D}
\end{equation}
where $\bm{L} = \partial \bm{v} / \partial \bm{x}$ is the gradient of velocity.
$\bm{D}$ could describe both the large deformation and the rigid body motion, and is equal to the rate of true strain $\dot{\bm{\varepsilon}}$, or the rate of Cauchy strain with respect to the current configuration in other words. 
The constitutive equation relates the strain and stress, and a detailed description is provided in Section~\ref{sec3}. 
It is crucial to recognize that the material response must exhibit material objectivity, ensuring independence from the chosen frame of reference.
This requirement dictates that the stress rates employed in constitutive equations must also adhere to objectivity.
However, the Cauchy stress tensor $\bm{\sigma}$, which characterizes the actual stress state and is integral to the conservation equations, does not maintain this objectivity when differentiated with respect to time.
The co-rotational rate of the Cauchy stress, called the Jaumann rate and denoted by $\bm{\sigma}^{\triangledown}$, satisfies the objectivity requirement and is used in MPM. 
The relation between the Jaumann rate and the Cauchy stress tensor can be expressed as
\begin{equation}
    \bm{\sigma}^{\triangledown} = \dot{\bm{\sigma}} - \bm{\mathit{\Omega}} \cdot \bm{\sigma} - \bm{\sigma} \cdot \bm{\mathit{\Omega}}^{\mathrm{T}}  \, \text{,}
    \label{eq:stress}
\end{equation}
where
\begin{equation}
    \bm{\mathit{\Omega}} = \tfrac{1}{2} (\bm{L} - \bm{L}^{\mathrm{T}})
    \label{eq:Omega}
\end{equation}
is a skew-symmetric tensor, referred to as the spin tensor. 
Up to this point, only two questions remain to be further discussed: the formulation of the conservation equations, and the relationship between motion and stress.

Considering that the material domain is represented by the material points, and the mass of the points never changes, the conservation of mass is automatically satisfied. 
Besides, the specific internal energy $E$, taking the form of $\rho \dot{E} = \dot{\bm{\varepsilon}} \colon \bm{\sigma}$ if neglecting heat transfer, is naturally evaluated with the variables stored at material points (or iterated with pressure by substituting the equation of state into the expression), which validates the conservation property of the framework but does not contribute to the closed-form nature of the governing equations. 
On top of that, it is the conservation of momentum that connects the movement and the stress, and embodies the foundational and essential concept of MPM.

The weak form of the updated Lagrangian formulation equivalent to the momentum equation and the traction boundary condition could be given by
\begin{equation}
    \int_{\mathit{\Omega}} \rho \ddot{\bm{u}} \cdot \delta \bm{u} \mathrm{d}V
    + \int_{\mathit{\Omega}} \rho \bm{\sigma}^{\mathrm{s}} \colon \delta (\nabla \bm{u}) \mathrm{d}V
    - \int_{\mathit{\Omega}} \rho \bm{b} \cdot \delta \bm{u} \mathrm{d}V
    - \int_{\mathit{\Gamma}_{\mathrm{t}}} \rho \overline{\bm{t}}^{\mathrm{s}} \cdot \delta \bm{u} \mathrm{d}\mathit{\Gamma}
    = 0  \, \text{.}
    \label{eq:weakform}
\end{equation}
if taking the virtual displacements $\delta \bm{u}$ as the test function.
The subscript $\mathit{\Omega}$ denotes the material domain, $\mathit{\Gamma}_{\mathrm{t}}$ represents the traction boundary, while the dot denotes the time derivative. 
 And 
$\bm{\sigma}^{\mathrm{s}} = \bm{\sigma} / \rho$, $\overline{\bm{t}}^{\mathrm{s}} = \overline{\bm{t}} / \rho$
, as well as $\bm{b}$, are the stress, traction, and body force, per unit mass, respectively. 

As introduced in Section~\ref{sec2}, the MPM formulation is based on a particle-based representation of the continuum and the use of interpolation functions to map quantities between particles and a background grid. For clarity and to make this appendix self-contained, we restate these two fundamental equations here. The density field is expressed in terms of material points as:
\begin{equation}
    \rho (\bm{x}) = \sum_{p=1}^{n_{\mathrm{p}}} m_p \delta (\bm{x} - \bm{x}_p)  \, \text{,}
    \label{eq:rho_appendix} 
\end{equation}
and the interpolation of a field variable (e.g., displacement $\bm{u}$) from grid nodes to a material point is given by:
\begin{equation}
    \bm{u}_p = \sum_{I=1}^{n_{\mathrm{g}}} N_I(\bm{x}_p) \bm{u}_I  \,  \text{.}
    \label{eq:u_appendix} 
\end{equation}

It is important to highlight that Eq.~(\ref{eq:rho_appendix}) serves as a tool for discretizing the governing equations from the weak form rather than an approximation to the density field itself.
This is a key difference between MPM and the SPH method \citep{Ma2009,Liu2019}, where Eq.~(\ref{eq:rho_appendix}) acts as a smoothing kernel for interpolating densities.

The remainder of this appendix details how these expressions are used to discretize the governing equations of motion.

By substituting Eqs. (\ref{eq:rho_appendix}) and (\ref{eq:u_appendix}) into the momentum equation described in (\ref{eq:weakform}) and leveraging the properties of the Dirac function along with the arbitrariness of the virtual velocity, the weak form can be discretized at each grid node as follows
\begin{equation}
    \dot{\bm{P}}_I = \bm{F}_I^{\mathrm{int}} + \bm{F}_I^{\mathrm{ext}}  \, \text{,}
    \label{eq:pdot}
\end{equation}
in which $\bm{P}_I$, $\bm{F}_I^{int}$ and $\bm{F}_I^{ext}$represent the momentum, the internal force and the external force of the $I$-th grid node, respectively, taking the form of
\begin{equation}
    \bm{P}_I = M_{IJ}\dot{\bm{u}}_{J}
    \quad \text{where} \quad
    M_{IJ} = \sum_{p=1}^{n_{\mathrm{p}}} m_p N_I(\bm{x}_p) N_J(\bm{x}_p)  \,  \text{,}
    \label{eq:Mij}
\end{equation}
\begin{equation}
    \bm{F}_I^{\mathrm{int}} = - \sum_{p=1}^{n_{\mathrm{p}}} \frac{m_p}{\rho_p} \nabla N_I(\bm{x}_p) \cdot \bm{\sigma}_p  \,  \text{,}
    \label{eq:fi}
\end{equation}
and
\begin{equation}
    \bm{F}_I^{\mathrm{ext}} = \sum_{p=1}^{n_{\mathrm{p}}} m_p N_I(\bm{x}_p) \bm{b}_p + \int_{\mathit{\Gamma}_{\mathrm{t}}} N_I(\bm{x}_p) \overline{\bm{t}} \mathrm{d} \mathit{\Gamma}  \, \text{.}
    \label{eq:fe}
\end{equation}
in which $\bm{\sigma}_p$ represents the Cauchy stress of a material point $p$, so as $\bm{b}_p$.

Moreover, applying Eq.~(\ref{eq:u_appendix}) to velocity $\bm{v}$, multiplying both sides with $m_p N_{Jp}$ then summing over all the material points yields
\begin{equation}
    \sum_{p=1}^{n_{\mathrm{p}}} m_p N_J(\bm{x}_p) \bm{v}_p = \sum_{p=1}^{n_{\mathrm{p}}} \sum_{I=1}^{n_{\mathrm{g}}} m_p N_I(\bm{x}_p) N_J(\bm{x}_p) \bm{v}_I  \, \text{.}
    \label{eq:mapping}
\end{equation}
By defining the lumped grid mass matrix
\begin{equation}
    M_I = \sum_{J=1}^{n_{\mathrm{g}}} M_{IJ} = \sum_{p=1}^{n_{\mathrm{p}}} m_p N_I(\bm{x}_p)  \, \text{,}
    \label{eq:mass}
\end{equation}
using $M_{IJ}$ in Eq.~(\ref{eq:Mij}) \citep{Burgess1992}, Eq.~(\ref{eq:mapping}) can be further simplified to
\begin{equation}
    \bm{P}_I = M_{I} {\bm{v}}_{I} = \sum_{p=1}^{n_{\mathrm{p}}} m_p N_I(\bm{x}_p) \bm{v}_p  \, \text{.}
    \label{eq:momentum}
\end{equation}
Thus far, the mapping from the material points to the background grid nodes is completed, and the algorithm is brought to full closure.

\subsection{Explicit solution scheme}\label{secA_2}
\begin{figure}[pos=htbp]
\centering
  \includegraphics[width = 0.5\textwidth]{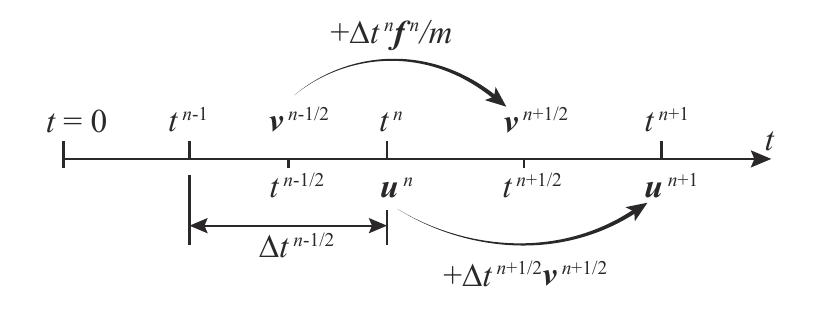}
  \caption{The leapfrog central difference explicit integration scheme.}
  \label{fig:timestep}
\end{figure}

To achieve the numerical calculation, the variable step size leapfrog central difference explicit integration scheme is used for the time advancement of the momentum equation, as shown in Fig.~\ref{fig:timestep}. 
The leapfrog scheme ensures second-order precision, and the explicit time integration is computationally highly efficient,
with the stability requirement of $\Delta t \leqslant \Delta t_{cr}$, where the critical time step $\Delta t_{cr}$ equals to
\begin{equation}
    \Delta t_{\mathrm{cr}} = \frac{d_{\mathrm{c}}}{\max_{p}\left(c_p + \left| \bm{v}_p \right|\right)}  \, \text{,}
    \label{eq:dt}
\end{equation}
in which $d_{\mathrm{c}}$ represents the edge length of the regular grid, and $c_p$ and $v_p$ represent the sound speed (given in Section~\ref{sec3_2}) and velocity of each material point $p$, respectively. 
This requirement restricts both the movement of points and the propagation of stress waves to no more than one grid in each time step.

Besides, it's necessary to choose one form of the different MPM schemes, meaning the different grid nodal velocity fields being employed when updating the stress state. 
The stress could be updated at the beginning of the $n$-th time step with the grid nodal momentum $\bm{P}_I^{n - 1/2}$ called the Update-Stress-First (USF) scheme \citep{Bardenhagen2002}, at the end of the time step with $\bm{P}_I^{n + 1/2}$ called the Update-Stress-Last (USL) \citep{Sulsky1994}, or at the end of the time step but using the grid nodal velocity obtained by mapping the updated particle momentum $\bm{P}_p^{n + 1/2}$ back to the grid nodes called the Modified Update-Stress-Last (MUSL) scheme \citep{Sulsky1995}. 
\citet{Nairn2003} and \citet{Ni2020} analyzed the energy conservation error and the simulation stability of each scheme, and MUSL performed well in most of the extreme cases. 
Therefore, the MUSL scheme is used in the present work, and it is implemented in the following process during the $n$-th time step:

\begin{enumerate}[1.]

\item According to Eqs.~(\ref{eq:mass}) and (\ref{eq:momentum}), map the particle mass and momentum to their related grid nodes, to obtain the grid nodal mass and momentum, as shown in Fig.~\ref{fig:MPMillu}(b)
\begin{equation*}
\begin{aligned}
    & M_I^n = \sum_{p=1}^{n_{\mathrm{p}}} m_p N_I^n(\bm{x}_p)   \,  \text{,} \\
    & \bm{P}_I^{n - 1/2} = \sum_{p=1}^{n_{\mathrm{p}}} m_p N_I^n(\bm{x}_p) \bm{v}_p^{n - 1/2}  \, \text{.}
\end{aligned}
\end{equation*}

\item Apply essential boundary conditions to the grid nodal momentum, for example, $P_{iI}^{n - 1/2} = 0$ for the boundary node $I$ that is fixed in the $i$-th coordinate direction.

\item According to Eqs. (\ref{eq:fi}) and (\ref{eq:fe}), calculate the grid nodal force
\begin{equation*}
    \bm{F}_I^n = \bm{F}_I^{{\mathrm{int}},n} + \bm{F}_I^{{\mathrm{ext}},n}
    = - \sum_{p=1}^{n_{\mathrm{p}}} \frac{m_p}{\rho_p^n} \nabla N_I^n(\bm{x}_p) \cdot \bm{\sigma}^n_p
    + \sum_{p=1}^{n_{\mathrm{p}}} m_p N_I^n(\bm{x}_p) \bm{b}^n_p + \int_{\mathit{\Gamma}_{\mathrm{t}}} N_I^n(\bm{x}_p) \overline{\bm{t}^n} \mathrm{d}\mathit{\Gamma}  \, \text{.}
\end{equation*}
Also apply essential boundary conditions, such as let $F_{iI}^n = 0$.

\item Integrate the momentum equation Eq.~(\ref{eq:pdot}) on the grid nodes
\begin{equation*}
    \bm{P}_I^{n + 1/2} = \bm{P}_I^{n - 1/2} + \bm{F}_I^n \Delta t^n  \, \text{.}
\end{equation*}

\item Integrate the velocity and position of the material points by interpolating the nodal properties to points, as shown in Fig.~\ref{fig:MPMillu}(c) and Fig.~\ref{fig:timestep}
\begin{equation*}
\begin{aligned}
    & \bm{x}_p^{n + 1} = \bm{x}_p^n + \Delta t^{n + 1/2} \sum_{I=1}^{n_{\mathrm{g}}} \frac {N_I^n(\bm{x}_p) \bm{P}_I^{n + 1/2}} {M_I^n}  \,  \text{,}\\
    & \bm{v}_p^{n + 1/2} = \bm{v}_p^{n - 1/2} + \Delta t^n \sum_{I=1}^{n_{\mathrm{g}}} \frac {N_I^n(\bm{x}_p) \bm{F}_I^n} {M_I^n}  \,  \text{.}
\end{aligned}
\end{equation*}

\item Map the updated momentum of points back to the grid nodes, recalculate the grid nodal momentum and reimpose essential boundary conditions
\begin{equation*}
    \bm{P}_I^{n + 1/2} = \sum_{p=1}^{n_{\mathrm{p}}} m_p N_I^n(\bm{x}_p) \bm{v}_p^{n + 1/2}  \, \text{.}
\end{equation*}

\item Calculate the grid nodal velocity
\begin{equation*}
    \bm{v}_I^{n + 1/2} = \frac{\bm{P}_I^{n + 1/2}} {M_I^n}   \, \text{,}
\end{equation*}
which determines the spatial gradient and divergence terms, and derives the strain increment and vorticity increment of points based on Eqs. (\ref{eq:D}) and (\ref{eq:Omega})
\begin{equation*}
\begin{aligned}
    & \Delta \bm{\varepsilon}_p^{n + 1/2} = \Delta t^{n + 1/2} \frac{1}{2} \sum_{I=1}^{n_{\mathrm{g}}} (\nabla N_I^n(\bm{x}_p) \otimes \bm{v}_I^{n + 1/2} + \bm{v}_I^{n + 1/2} \otimes \nabla N_I^n(\bm{x}_p))  \, \text{,} \\
    & \Delta \bm{\mathit{\Omega}}_p^{n + 1/2} = \Delta t^{n + 1/2} \frac{1}{2} \sum_{I=1}^{n_{\mathrm{g}}} (\nabla N_I^n(\bm{x}_p) \otimes \bm{v}_I^{n + 1/2} - \bm{v}_I^{n + 1/2} \otimes \nabla N_I^n(\bm{x}_p))  \, \text{.}
\end{aligned}
\end{equation*}
Then, update the density of the material points with
\begin{equation*}
    \rho_p^{n + 1} = \rho_p^n / (1 + \text{tr}(\Delta \bm{\varepsilon}_p^{n + 1/2}))  \, \text{,}
\end{equation*}
and the stress based on Eq.~(\ref{eq:stress}) and Section~\ref{sec3}
\begin{equation}
    \bm{\sigma}_p^{n + 1} = \bm{\sigma}_p^n + 
    (\bm{\mathit{\Omega}}_p^{n + 1/2} \cdot \bm{\sigma}_p^n + 
    \bm{\sigma}_p^n \cdot (\bm{\mathit{\Omega}}_p^{n + 1/2})^{\mathrm{T}}) \Delta t^{n + 1/2} + 
    \bm{\sigma}^{\triangledown} (\dot{\bm{\varepsilon}}^{n + 1/2}_p, \bm{\sigma}_p^n, \text{etc.}) \Delta t^{n + 1/2}  \, \text{.}
    \label{eq:stressupdate}
\end{equation}
The specific internal energy $E$ and pressure $p$ are also solved iteratively within this step.
Note that $E$ is given by the incremental form $E^{n+1} = E^{n} + \Delta \bm{\varepsilon}^{n+1/2} \colon \bm{\sigma}^{n+1/2} / \rho$.
In the implementation of the algorithm, it is again necessary to compute the trial solution of $E^{*n+1}$ to find the pressure $p^{n+1}$, and then update $E^{n+1}$.
Given the small time steps, a single iteration is typically sufficient for convergence.

\item At this point, all the properties of the material domain have been updated and stored on the material points, so that the deformed background grid can be discarded. 
Regenerate a new regular grid, go back to the $1$st step of the loop, as shown in Fig.~\ref{fig:MPMillu}(d).

\end{enumerate}

\section{Fundamentals of elastoplasticity and strength models}\label{secB}
\subsection{The plastic flow law}\label{secB_1}
Hooke's law gives the linear relation between strain and stress for isotropic elastic materials. However, materials generally exhibit elastoplastic behavior at higher loads.
The critical yield strength determined by a specific yield function $f(\bm{\sigma},\ \bm{q})$ is a common form of the strength models for elastoplastic materials, where the yield criterion requires that $f(\bm{\sigma},\ \bm{q}) \leqslant 0$.

According to Eq.~(\ref{eq:stressupdate}), the calculation of deviatoric stress follows
\begin{equation}
    \bm{s}_p^{n + 1} = \bm{s}_p^n + 
    (\bm{\mathit{\Omega}}_p^{n + 1/2} \cdot \bm{s}_p^n + 
    \bm{s}_p^n \cdot (\bm{\mathit{\Omega}}_p^{n + 1/2})^{\mathrm{T}}) \Delta t^{n + 1/2} + 
    \bm{s}^{\triangledown n + 1/2} \Delta t^{n + 1/2}  \,  \text{.}
    \label{eq:devstr}
\end{equation}
The Jaumann rate term, $\bm{s}^{\triangledown n + 1/2}$, is related to the deviatoric strain rate 
\begin{equation}
    \bm{s}^\triangledown = 2 G \left( \dot{\bm{\varepsilon}} - \frac{1}{3} \text{tr}(\dot{\bm{\varepsilon}}) \bm{I} \right)  \,  \text{,}
    \label{eq:elastic}
\end{equation}
for isotropic elastic materials conforming to Hooke's law.
Here, $G$ represents the shear modulus, related to the Young's modulus $E_{\mathrm{Y}}$ and the Poisson's ration $\nu$ in the form of $G = E_{\mathrm{Y}} / 2 (1 + \nu)$.
But for elastoplastic materials, Eq.~(\ref{eq:devstr}) and Eq.~(\ref{eq:elastic}) only gives the elastic trial solution $\bm{\sigma}^*$.

When $\bm{\sigma}^*$ results in $f(\bm{\sigma}^*,\ \bm{q}^n) > 0$ at time $t^{n+1}$ (the superscript $n+1$ is omitted without ambiguity), it signifies that the material is in a state of plastic loading.
Hence, the stress should be corrected back to the yield surface $f = 0$.
The correction process and the corresponding cumulative plastic strain $\bm{\varepsilon}^{\mathrm{p}}$ are established by the plastic flow law
\begin{equation}
    \dot{\bm{\varepsilon}}^{\mathrm{p}} = \dot{\lambda} \bm{r}
    \label{eq:plaflow}
\end{equation}
that for a given plastic flow potential $\psi$, $\bm{r} = \partial \psi / \partial \bm{\sigma}$ determines the direction of plastic flow which is perpendicular to the yield surface, and the loading parameter $\dot{\lambda}$ is calculated as below.
The yield function could be used as the plastic potential function, i.e., $\psi \equiv f$, which is called an associated flow law, otherwise non-associated.

Based on the plastic flow law, the plastic correctors of strain, stress, and internal state variables are obtained.
By associating the aforementioned variables with the yield condition, namely 
\begin{equation}
  \begin{cases}
    \Delta \bm{\varepsilon}^{\mathrm{p}} = \dot{\lambda} \bm{r} \Delta t \\
    \bm{\sigma} = \bm{\sigma}^* - \bm{C}^{\sigma J} \colon \Delta \bm{\varepsilon}^{\mathrm{p}} \\
    \bm{q}^{n+1} = \bm{q}^n + \dot{\lambda} \bm{h} \Delta t \\
    f(\bm{\sigma},\ \bm{q}) = 0
    \label{eq:returnmapping}
  \end{cases}
    \,  \text{,}
\end{equation}
the plasticity correction is accomplished.
Here $\bm{C}^{\sigma J}$ is the fourth-order elasticity tensor, and $\bm{h}$ represents the evolution equations of $\bm{q}$.
This process is called the return mapping algorithm, where the elastic trial stress returns to the yield surface along the plastic flow.
The above system of equations can be solved with an iterative scheme, or be simplified through linearization or a semi-implicit algorithm.
Cumulative plastic strain is also calculated during the solution process, allowing for the measurement of plastic heat generation and modeling of fatigue damage. 
Additionally, the increment of heat energy generated by plastic deformation per unit volume takes the form of
\begin{equation}
    \Delta Q = \bm{\sigma} \colon \Delta \bm{\varepsilon}^{\mathrm{p}}  \,  \text{,}
    \label{eq:heatenergy}
\end{equation}
which results in thermal softening.
And dedicated equations can be further derived using specific strength models.

\subsection{The von Mises yield criterion}\label{secB_2}
A simple strength model is the von Mises yield criterion, where the von Mises effective stress $\bar{\sigma}$ satisfies the condition
\begin{equation}
    f (\bm{\sigma}) = \bar{\sigma} - \sigma_{\mathrm{Y}} \leqslant 0  \,  \text{,}
    \label{eq:J2perfectpla}
\end{equation}
which defines the limit of yield strength.
$\sigma_{\mathrm{Y}}$ represents a single constant yield strength. $\bar{\sigma}$ is defined as $\sqrt{3 J_2}$ \citep{Mises1913}, where $J_2 = \begin{smallmatrix} \frac{1}{2} \end{smallmatrix} \bm{s} \colon \bm{s}$
is the second invariant of the deviatoric stress $\bm{s}$.
(Therefore, this criterion is also called the $J_2$ flow theory.)
Eq.~(\ref{eq:J2perfectpla}) illustrates a cylindrical yield surface in the $3$-D principal stress space that remains unaffected by the hydrostatic pressure (as shown in Fig.~\ref{fig:yield}(\textit{a})). If the associated $J_2$ flow theory is used, the return mapping algorithm is directed to the radial return method. When $f(\bm{\sigma}^*) > 0$, the plasticity correction goes to
\begin{equation}
    \Delta \bm{\varepsilon}^{\mathrm{p}} = \frac {f(\bm{\sigma}^*)} {2 G \bar{\sigma}^*} \bm{s}^*\, \text{,} \quad
    \bm{s} = \frac {\sigma_{\mathrm{Y}}} {\bar{\sigma}^*} \bm{s}^* \, \text{.}
\end{equation}

The mechanical properties reflected by the von Mises yield criterion accurately capture the elastic--perfectly plastic behavior.
By selecting different expressions for the yield stress $\sigma_{\mathrm{Y}}(\bar{\varepsilon}^{\mathrm{p}})$, the $J_2$ flow theory can be further extended to describe a wider range of material properties.
For the linear isotropic hardening model $\sigma_{\mathrm{Y}}^{n+1}(\bar{\varepsilon}^{\mathrm{p}}) = \sigma_{\mathrm{Y}}^n + E^{\mathrm{p}} \Delta \bar{\varepsilon}^{{\mathrm{p}}(n+1)}$ with a constant plastic modulus $E^{\mathrm{p}}$, the yield surface is still cylindrical but with an increasing diameter.
Hence, the plasticity correction of the deviatoric stress remains the same, yet the plastic strain increment
\begin{equation}
    \Delta \bm{\varepsilon}^{\mathrm{p}} = \frac 3 2 \frac {f(\bm{\sigma}^*)} {3 G + E^{\mathrm{p}}} \frac {\bm{s}^*} {\bar{\sigma}^*} \, \text{,}
    \label{eq:J2elapla}
\end{equation}
where the effective plastic strain increment $\Delta \bar{\varepsilon}^{\mathrm{p}} = \sqrt{\begin{smallmatrix} \frac{2}{3} \end{smallmatrix} \Delta \bm{\varepsilon}^{\mathrm{p}} \colon \Delta \bm{\varepsilon}^{\mathrm{p}}}$, and the accumulated effective plastic strain $\bar{\varepsilon}^{{\mathrm{p}}(n+1)} = \bar{\varepsilon}^{{\mathrm{p}}(n)} + \Delta \bar{\varepsilon}^{{\mathrm{p}}(n+1)}$.
For a more general function $\sigma_{\mathrm{Y}} = \sigma_{\mathrm{Y}}(\bar{\varepsilon}^{\mathrm{p}}, \, \dot{\bar{\varepsilon}}^{\mathrm{p}}, \, T)$, such as the Johnson-Cook flow stress model, the $E^{\mathrm{p}}$ in Eq.~(\ref{eq:J2perfectpla}) is given by $\mathrm{d} \sigma_{\mathrm{Y}}/\mathrm{d} \bar{\varepsilon}^{\mathrm{p}}$.

\subsection{Pressure-dependent strength models}\label{secB_3}
The $J_2$ flow theory successfully accounts for pressure-independent plastic deformation, which is widely used especially for metals and other non-porous ductile materials. However, the behavior of rocks, soils, and other geomaterials is always governed by friction, and the shear strength increases substantially with the confining pressure, i.e., pressure-dependent.

Drucker-Prager yield model \citep{Drucker1952} is a basic one.
By defining the effective shear stress $\tau = \sqrt{J_2}$, the yield surface is expressed as 
\begin{equation}
    f^{\mathrm{s}} (\bm{\sigma}) = \tau - q_{\phi} p - Y_0 \, \text{,}
\end{equation}
where $q_{\phi}$ is the internal friction coefficient of friction angle $\phi$, and $Y_0$ represents the cohesive shear strength, i.e., the maximum shear stress that the material can withstand in the pressureless state.
Considering the fact that rocks are generally brittle, a tensile strength $p^{\mathrm{t}}$ is also included, adding another yield surface
\begin{equation}
    f^{\mathrm{t}} (\bm{\sigma}) = p^{\mathrm{t}} - p  \, \text{.}
\end{equation}
With this tension cut-off, the Drucker-Prager yield surface becomes a truncated cone in the principal stress space, and the shear envelope could be sketched in the $p$--$\tau$ plane (see Fig.~\ref{fig:yield}). 

To properly embody the prevalent shear-dilatant behavior observed in geotechnical materials, the plastic potential function $\psi$ could be given by 
\begin{equation}
    \psi = \tau - q_{\psi} p  \, \text{,}
    \label{eq:potential}
\end{equation}
that $q_{\psi}$ , alike $q_{\phi}$, denotes the dilatancy coefficient with the dilatancy angle. 
And the plastic corrector could be derived based on Eqs.~(\ref{eq:returnmapping}).
It is important to note that the direction of plastic flow in the junction region may abruptly change due to the multi-surface model, which necessitates the determination of the appropriate flow rule and increases the computational complexity.

Another feature of the Drucker-Prager model is that the strength keeps increasing with pressure, which is incongruent with reality when the pressure approaches infinite.
According to experimental data, \citet{Lundborg1968} proposed a modified smooth model with an upper limit $Y_{\mathrm{m}}$. The yield stress is in the form of
\begin{equation}
    \sigma_{\mathrm{Yi}} = Y_0 + \frac {\mu_{\mathrm{i}} p} {1 + \mu_{\mathrm{i}} p / ( Y_{\mathrm{m}} - Y_0 )}  \, \text{,}
\end{equation}
and $\mu_{\mathrm{i}}$ is the same as $q_{\phi}$.
In this case the tensile strength $p^{\mathrm{t}}$ is equal to $-Y_0 (Y_{\mathrm{m}} - Y_0) / \mu_{\mathrm{i}} Y_{\mathrm{m}}$.
Since $Y_{\mathrm{m}}$ is always much larger than $Y_0$, $p^{\mathrm{t}}$ can be approximated as $-Y_0 / \mu_{\mathrm{i}}$.
\citet{Collins2004} supplements the yield strength of completely fragmented rock material as the Coulomb dry-friction law
\begin{equation}
    \sigma_{\mathrm{Yd}} = \mu_{\mathrm{d}} p  \, \text{,}
\end{equation}
and is constrained to $\sigma_{\mathrm{Yd}} \leqslant \sigma_{\mathrm{Yi}}$.
Here $\mu_{\mathrm{d}}$ is still the coefficient of internal friction but for the fully damaged material.
For partially damaged material, an interpolation is defined as
\begin{equation}
    \sigma_{\mathrm{Y}} = (1-D)\sigma_{\mathrm{Yi}} + D \sigma_{\mathrm{Yd}}  \, \text{,}
\end{equation}
which is based on the degree of damage $D$ that ranging from $0$ for intact to $1$ for totally damaged (as shown in Fig.~\ref{fig:yield}(\textit{b}) by the thiner dot--dash line).
A temperature modifier is also included.
Then \citet{Jutzi2015} utilized a linear melting coefficient $1 - u_{\mathrm{I}} / u_{\mathrm{I-melt}}$ to capture the softening, in which $u_{\mathrm{I}}$ relates to the specific internal energy and $u_{\mathrm{I-melt}}$ represents the specific melting energy.
The effect of the rate-dependent friction coefficient was also evaluated.
However, it displays insignificant influence.

\section*{Acknowledgements}
We express our sincere gratitude to the Editor, Dr. Stuart Robbins, and the anonymous reviewers for their constructive comments and careful review, which significantly improved the quality of this manuscript.
We would like to thank Yan Liu (Tsinghua University), Bin Cheng (Tsinghua University), Wenyue Dai (Tsinghua University) and Xian Shi (Shanghai Astronomical Observatory) for helpful discussions and constructive suggestions.
X.Y. would like to personally thank Dr. Yun Zhang (University of Michigan, Ann Arbor) for her invaluable inspiration and encouragement in implementing the MPM method.
We acknowledge support from Tsinghua University and Université Côte d'Azur.
X.Y. and J.L. acknowledge support from the National Natural Science Foundation of China under Grant 12372047.
X.Y. acknowledges support from the National Natural Science Foundation of China under Grant 62227901. 
P.M. acknowledges support from the French space agency CNES and from the French National Centre for Scientific Research (CNRS) through the exploratory research program of the Mission for Transversal and Interdisciplinary Initiatives.

\printcredits{
Xiaoran Yan: Methodology - MPM, Methodology - material model, Software, Validation, Formal analysis, Visualization, Writing - original draft, Writing - review and editing.
Patrick Michel: Conceptualization, Supervision, Writing - review and editing.
Ruichen Ni: Methodology - MPM, Writing - review and editing.
Yifei Jiao: Methodology - material model, Writing - review and editing.
Junfeng Li: Funding acquisition, Supervision, Writing - review and editing.
}

\bibliographystyle{cas-model2-names}

 \bibliography{refs.bib}



\end{document}